\documentclass[12pt]{article}

\pdfoutput=1

\usepackage[utf8]{inputenc}
\usepackage[T1]{fontenc}
\usepackage{amsmath,amssymb,mathtools}
\usepackage[separate-uncertainty=true,retain-unity-mantissa=false]{siunitx}
\usepackage{slashed}
\usepackage{graphicx}
\usepackage[compat=1.1.0]{tikz-feynman}
\usepackage{booktabs,multicol,multirow}
\usepackage[shortlabels]{enumitem}
\usepackage{hyperref}
\usepackage[capitalize]{cleveref}
\usepackage{xspace}
\usepackage[noblocks]{authblk}
\usepackage[dvipsnames]{xcolor}
\usepackage{soul}
\usepackage[textsize=tiny]{todonotes}
\usepackage{subcaption}
\usepackage{nicefrac}
\usepackage{ulem}
\usepackage{bbold}
\usepackage{bbm}
\usepackage{cite}
\usepackage{makecell}

\newcommand{\citere}[1]{Ref.~\cite{#1}}
\newcommand{\citeres}[1]{Refs.~\cite{#1}}

\newcommand{\cp}{\ensuremath{{\cal CP}}}
\newcommand{\lsim}{\;\raisebox{-.3em}{$\stackrel{\displaystyle <}{\sim}$}\;}
\newcommand{\gsim}{\;\raisebox{-.3em}{$\stackrel{\displaystyle >}{\sim}$}\;}

\newcommand{\gev}{\;\text{GeV}\xspace}
\newcommand{\tev}{\;\text{TeV}\xspace}
\newcommand{\sw}{s_\text{w}}
\newcommand{\cw}{c_\text{w}}

\newcommand{\mhh}{\ensuremath{m_{hh}}}
\newcommand{\mbbbb}{\ensuremath{m_{b\bar b b\bar b}}}
\newcommand{\lahhh}{\lambda_{hhh}}
\newcommand{\lahhH}{\lambda_{hhH}}
\newcommand{\lahHH}{\lambda_{hHH}}
\newcommand{\laHHH}{\lambda_{HHH}}
\newcommand{\laijk}{\lambda_{ijk}}
\newcommand{\laijkz}{\laijk^{(0)}}
\newcommand{\rlahhh}[1]{\hat{\lambda}_{hhh}^{(#1)}}
\newcommand{\rlahhH}[1]{\hat{\lambda}_{hhH}^{(#1)}}
\newcommand{\rlahHH}[1]{\hat{\lambda}_{hHH}^{(#1)}}
\newcommand{\rlaHHH}[1]{\hat{\lambda}_{HHH}^{(#1)}}
\newcommand{\rlaijk}[1]{\hat{\lambda}_{ijk}^{(#1)}}

\newcommand{\laSMz}{\lambda_{hhh}^{\SM, (0)}}
\newcommand{\kala}{\kappa_\lambda}
\newcommand{\sihh}{\sigma_{hh}^{\mathrm{RxSM}}}
\newcommand{\sihhSM}{\sigma_{hh}^{\mathrm{SM}}}
\newcommand{\siZhh}{\sigma_{Zhh}^{\text{RxSM}}}
\newcommand{\siHZhh}{\sigma_{H,Zhh}^{\text{RxSM}}}
\newcommand{\siZhhSM}{\sigma_{Zhh}^{\text{SM}}}

\newcommand{\SM}{{\text{SM}}}

\newcommand{\br}{\rm{BR}}

\newcommand{\order}[1]{\ensuremath{{\cal O}(#1)}}

\newcommand{\cA}{{\mathcal A}}
\newcommand{\cL}{{\mathcal L}}

\newcommand{\iab}{\rm{ab}^{-1}}

\renewcommand{\rm}{\mathrm}
\newcommand{\nn}{\nonumber}

\newcommand{\HB}{\texttt{HiggsBounds}}

\date{}

\evensidemargin \oddsidemargin
\begin{document}
\thispagestyle{empty}
\def\thefootnote{\fnsymbol{footnote}}

\begin{flushright} \texttt{DESY-25-092}\\
\texttt{IFT–UAM/CSIC-25-071}
\end{flushright}
\vspace{3em}
\begin{center}
{\Large{\bf Impact of one-loop corrections to trilinear\\[.3em] scalar couplings on di-Higgs production in the RxSM}}
\\
\vspace{3em}
{
Johannes Braathen$^{1}$\footnotetext[0]{\hspace{-0.65cm}\href{mailto:johannes.braathen@desy.de}{johannes.braathen@desy.de}\\ \href{mailto:sven.heinemeyer@cern.ch}{sven.heinemeyer@cern.ch}\\\href{mailto:andrea.parra@estudiante.uam.es}{andrea.parra@estudiante.uam.es}\\ \href{mailto:alain.verduras@desy.de}{alain.verduras@desy.de}},
Sven Heinemeyer$^{2}$,
Andrea Parra Arnay$^{2}$,
Alain Verduras Schaeidt$^{1}$
}\\[2em]
{\sl $^1$ Deutsches Elektronen-Synchrotron DESY, Notkestr.~85, 22607 Hamburg, Germany}\\[0.2em]
{\sl $^2$ Instituto de F\'isica Te\'orica (UAM/CSIC), Cantoblanco, 28049, Madrid, Spain}
\end{center}

\vspace{2ex}

\begin{abstract}\noindent
    We investigate di-Higgs production at the (HL-)LHC and possible high-energy future $e^+e^-$ colliders within the real Higgs singlet extension
of the Standard Model (SM), the RxSM. This model has two CP-even Higgs bosons, $h$ and $H$, for which we assume $m_h\sim 125 \gev < m_H$.
We analyse the effect of one-loop corrections to the two trilinear scalar couplings relevant for di-Higgs production, $\lahhh$ and $\lahhH$, by performing an extensive parameter 
scan within the RxSM. We find that the one-loop 
corrections have a strong impact on the total production cross-sections, as well as on the differential cross-sections with respect to the 
invariant di-Higgs mass, \mhh. We evaluate the sensitivity of the HL-LHC and a high-energy $e^+e^-$ collider with $\sqrt{s} = 1 \tev$, the ILC1000, to probe BSM physics effects in these processes.
We demonstrate that the RxSM can be distinguished from the SM for large parts of the sampled parameter space. The resonant $H$ structure in the 
\mhh\ distribution, on the other hand, can be observed only if the corresponding couplings, in particular $\lahhH$, are sufficiently large. 
Here the ILC1000 yields a substantially better sensitivity than the HL-LHC.

\end{abstract}
\def\thefootnote{\arabic{footnote}}
\setcounter{page}{0}
\setcounter{footnote}{0}

\newpage

\tableofcontents

\newpage


\section{Introduction}
\label{sec:intro}

The discovery of a scalar particle with a mass around $125 \gev$ in 2012 by the ATLAS and CMS 
collaborations~\cite{Aad:2012tfa,Chatrchyan:2012xdj,Khachatryan:2016vau} has established the Brout-Englert-Higgs mechanism 
as the origin of electroweak symmetry breaking (EWSB). However, our understanding of the fundamental interactions of Nature, 
currently taking the form of the Standard Model (SM), remains incomplete, as it does not provide explanations for several 
fundamental phenomena observed in Nature.
Current measurements of the couplings of the discovered Higgs boson leave significant room for beyond-the-Standard-Model (BSM) physics, 
with, for instance, precisions in the determination of the Higgs boson's couplings to gauge bosons or third-family fermions 
of \order{10\%}. 
While there are many different ways of extending the SM, in this work we focus on models that enlarge the SM Higgs sector,
since these models can provide explanations to several of the unexplained phenomena.

One of the most important unresolved fundamental problems of High-Energy Physics is the matter-antimatter asymmetry, 
or baryon asymmetry, of the Universe (BAU). 
A promising framework in which the BAU could be dynamically generated is the mechanism of electroweak (EW) 
baryogenesis~\cite{Kuzmin:1985mm}. A crucial ingredient in this scenario is the occurrence of a strong first-order electroweak phase transition 
(SFOEWPT), which allows fulfilling the Sakharov condition~\cite{Sakharov:1967dj} of departure from thermal equilibrium.
During such a transition, 
patches of space transition from a ``false'' 
vacuum to a deeper, and thus more stable
one~\cite{Morrissey:2012db,Ramsey-Musolf:2019lsf,Espinosa:1993bs,Profumo:2007wc,Espinosa:2007qk,Espinosa:2011ax,Curtin:2014jma,Kurup:2017dzf,Ham:2004cf,Barger:2008jx}, 
leading to the nucleation and expansion of bubbles of the ``true'' vacuum. 
Given the measured mass of the Higgs boson, the SM fails to undergo a SFOEWPT~\cite{Kajantie:1996mn}. 
On the other hand, models with extended Higgs sectors offer 
the possibility of a SFOEWPT in the early universe, thus satisfying the third Sakharov condition as required for 
successful EW baryogenesis~\cite{Sakharov:1967dj}.

A major goal of ongoing and future collider programmes is to explore whether the scalar boson 
discovered at the LHC is part of a more complex Higgs sector, and whether such a Higgs sector would accommodate a SFOEWPT
and thus facilitate EW baryogenesis to explain the observed BAU.

In the study of BSM models with extended Higgs sectors, the trilinear Higgs self-coupling $\lahhh$ plays a crucial role. 
It provides information about the shape of the scalar potential away from the EW minimum and is closely connected to the
dynamics of the EW phase transition (EWPT). 
The study of $\lahhh$ is therefore important for determining whether the EWPT occured as a SFOEWPT. 
As shown in early studies~\cite{Grojean:2004xa,Kanemura:2004ch} and confirmed in more recent analyses, 
like e.g.,~\citeres{Kakizaki:2015wua,Hashino:2016rvx,Hashino:2016xoj,Basler:2017uxn,Biekotter:2022kgf,Bittar:2025lcr}, a SFOEWPT 
is often linked to a deviation of $\lahhh$ from the SM value. Beyond its implications for the evolution of the early Universe,
the trilinear Higgs coupling also constitutes a sensitive tool to probe indirect signs of new physics, through 
the radiative corrections it can receive from additional scalars. Indeed, deviations of $\lahhh$ of up to several hundred
percent from its SM prediction can occur in models with additional scalars if there are significant mass splittings between different BSM scales.
This type of higher-order corrections can even be found for scenarios where other couplings of the Higgs boson are SM-like at tree-level,
such as scenarios with alignment~\cite{Gunion:2002zf}. These so-called ``mass-splitting'' effects were first found in the context of Two-Higgs-Doublet Models 
in \citeres{Kanemura:2002vm,Kanemura:2004mg}, but are now known to occur in a variety of BSM theories~\cite{Aoki:2012jj,Kanemura:2015fra,Kanemura:2015mxa,Arhrib:2015hoa,Kanemura:2016sos,Kanemura:2016lkz,He:2016sqr,Kanemura:2017wtm,Kanemura:2017gbi,Chiang:2018xpl,Basler:2018cwe,Kanemura:2019slf,Basler:2020nrq,Falaki:2023tyd,Bahl:2022jnx,Bahl:2022gqg,Bahl:2023eau,Cherchiglia:2024abx,Basler:2024aaf}. It should be noted that mass-splitting effects do not arise as a
perturbation of the tree-level prediction for $\lahhh$, but rather as a new type of corrections only entering at loop level. 
Moreover, explicit calculations of the two-loop corrections to $\lahhh$~\cite{Senaha:2018xek,Braathen:2019pxr,Braathen:2019zoh,Braathen:2020vwo,Aiko:2023nqj,Bahl:2025wzj} have confirmed the perturbative convergence of these BSM effects,
even in scenarios with large deviations from the SM. Following the same mechanism, BSM trilinear scalar couplings can also receive large 
radiative corrections, see e.g.\ \citeres{Arco:2025pgx,anyHH}.

Experimentally, the most direct probe of $\lahhh$ is via the process of di-Higgs production, which features a dependence on 
$\lahhh$ already at leading order. 
Examples are $gg \to hh$, which is the leading channel at the LHC, 
as well as $e^+e^- \to Zhh \,/\, \nu\bar\nu hh$ at high-energy $e^+e^-$ colliders (see e.g.\ Refs.~\cite{Barklow:2017awn,LinearColliderVision:2025hlt,Altmann:2025feg}). Here it is important that
the contribution involving $\lahhh$ interferes with ``background diagrams'', where the interference can be constructive or destructive, depending
on the process and on the value of $\lahhh$. As an important example, at the LHC the interference is destructive, 
leading to a very small cross-section for SM-like values of the Higgs couplings (including $\lahhh$). 
This also implies that BSM deviations in $\lahhh$ can significantly modify the di-Higgs production cross-section --- by several orders of magnitude. 
There have already been several searches for this process 
by the ATLAS and CMS collaborations at CERN, which have set bounds on the di-Higgs cross-section and on 
$\lahhh$~\cite{ATLAS:2022jtk,ATLAS:2022kbf,CMS:2022dwd,ATLAS:2024ish,CMS:2024awa,ATLAS:2025lbo}, and significant improvements are expected at the HL-LHC. In addition to sensitivity to BSM deviations in $\lahhh$, 
di-Higgs production at the LHC, as well as at future high-energy $e^+e^-$ colliders also offers access to BSM trilinear scalar couplings
that involve BSM Higgs bosons, see e.g.~\citere{Arco:2025nii}. Thus, di-Higgs production is an open window to new physics which is sensitive 
to the trilinear scalar couplings realised in Nature.

A minimal and well-studied extension of the SM Higgs sector is the addition of a real singlet scalar, 
the RxSM~\cite{Barger:2007im,Lerner:2009xg,Costa:2015llh,Li:2019tfd}%
\footnote{We note that this model is also sometimes called ``xSM'', ``SSM'' or ``HSM'' in the literature.}.  
The scalar sector of the RxSM contains two physical Higgs states: in what follows, we identify the lighter state, denoted $h$, 
with the $125\gev$ Higgs boson discovered at the LHC, while the heavier scalar, $H$, remains undetected so far. 
Different versions of real singlet models have been devised, depending in particular on whether a $\mathbb{Z}_2$ (or some other) symmetry 
acting on the singlet state is imposed, or not. In this work, we consider the most general variant, in which no additional
symmetry is considered. In particular, this adds to the model three (two) additional free parameters compared to the variants with an unbroken (spontaneously broken) $\mathbb{Z}_2$ symmetry. These extra parameters 
can have important phenomenological implications on the trilinear scalar couplings, which can receive large BSM contributions already at tree level, 
or at loop level, depending on the considered region of the RxSM parameter space. 
It has also been shown in \citere{Espinosa:1993bs} that the RxSM without a $\mathbb{Z}_2$ symmetry can give rise to a SFOEWPT in the early universe.
This makes it one of the most minimal Higgs-sector extensions capable of supporting a SFOEWPT, and the simplest allowing so with a one-step transition.

Large radiative corrections to $\lahhh$ as well as to BSM trilinear scalar couplings can have a significant impact on di-Higgs 
production, as studied e.g.\ in the 2HDM~\cite{Heinemeyer:2024hxa}. It is therefore crucial to properly take these corrections 
into account to obtain reliable interpretations of future collider searches for di-Higgs production  
in terms of constraints on BSM scenarios. In order to include loop-corrected trilinear scalar couplings in a calculation of di-Higgs 
production, a desirable property is that the corrected trilinear scalar couplings, entering the cross-section calculation, should be 
independent of the renormalisation scale $Q$. To ensure this, we define in this work --- for the first time --- a fully on-shell (OS) renormalisation scheme for the RxSM. In our analysis, we compute the OS one-loop corrections to the trilinear scalar couplings relevant for di-Higgs
production, namely $\lahhh$ and $\lahhH$, using the public tool \texttt{anyH3}~\cite{Bahl:2023eau,anyHH}. 
The resulting loop-corrected couplings are then used as inputs for the computation of total and differential di-Higgs production 
cross-sections, both at the High-Luminosity LHC (HL-LHC) and at a future 1~TeV $e^+e^-$ collider.
For the HL-LHC, we compute the leading-order (LO) cross-section of the process $gg \to hh$, 
using the program \texttt{HPAIR}~\cite{Abouabid:2021yvw,Arco:2022lai,Dawson:1998py,Nhung:2013lpa,Grober:2015cwa}.
In the case of the 1~TeV $e^+e^- $ collider, we compute the polarised cross-section of the process 
$e^+e^- \to Zhh $ using the public tool \texttt{MadGraph5\_aMC@NLO} v3.5.7~\cite{Alwall:2014hca}.

Our paper is organised as follows. In \cref{sec:model} we review the RxSM, fix our notations, and briefly summarise the theoretical and 
experimental constraints that we apply. In \cref{sec:ren} we describe our setup for the computation of the one-loop corrections to the 
couplings $\lambda_{hhh}$ and $\lambda_{hhH}$, using a complete on-shell renormalisation of the RxSM. In 
\cref{sec:one} we present numerical results for the one-loop OS-renormalised values of $\lahhh$ and $\lahhH$,
obtained with scans of the RxSM parameter space. Finally, in \cref{sec:dihiggs}, we investigate the impact of the one-loop
corrections to the trilinear scalar couplings on the total and differential di-Higgs production cross-section at the HL-LHC and
at a 1 TeV $e^+e^-$ collider. Our conclusions are given in \cref{sec:conclusions}.


\section{General singlet extension of the SM (RxSM)}
\label{sec:model}

\subsection{Model definitions}
\label{sec:model_def}
We consider in this work the most general real singlet extension of the SM, which we denote RxSM~\cite{Lerner:2009xg,Li:2019tfd,Gonderinger:2009jp,Arco:2025nii,Ramsey-Musolf:2019lsf}. This model adds a real singlet $S$ to the Higgs sector of the SM, and unlike other singlet extension variants, does not contain any $\mathbb{Z}_2$ symmetry of the Lagrangian. After EWSB, the SM-like doublet $\Phi$ and the singlet $S$ can be expanded as
\begin{equation}
{\Phi}=\frac{1}{\sqrt{2}}\left(
 \begin{matrix}
 \sqrt{2}G^+  \\
 v+\phi+i G^0
 \end{matrix}\right)\,,\qquad S=s+v_S, \label{singlet}
\end{equation}
where $s$ is the new scalar field and $v_S$ is the singlet VEV, while $\phi$ is the \cp-even component of the doublet, $G^0$ ($G^\pm$) is the neutral (charged) would-be Goldstone boson, and $v$ is the EW (or SM) VEV. The scalar potential of the RxSM reads, at the tree level,
\begin{align}
    V({\Phi},{ S})=\mu^2({\Phi}^{\dagger }{\Phi})+\frac{\lambda}{2}({\Phi}^{\dagger }{\Phi})^2+\kappa_{SH}({\Phi}^{\dagger }{\Phi}){ S}+\frac{\lambda_{SH}}{2}(\Phi^{\dagger }{\Phi}){S}^2+\frac{M_S^{2}}{2}{S}^2+\frac{\kappa_S}{3}{ S}^3+\frac{\lambda_S}{2}S^4. \label{potential}
\end{align}
Since no $\mathbb{Z}_2$ symmetry has been imposed on the model, there would in principle be terms linear and cubic in the singlet field. However, the RxSM offers the freedom to redefine the singlet VEV $v_S$ to absorb one of these terms~\cite{Espinosa:2011ax}, which we have used here to remove the linear (tadpole) term. We furthermore note that as the singlet $S$ is real, \cp-violation cannot be included in the scalar potential of the RxSM (as is the case for the SM), so that all the parameters in \cref{potential} are real.

Taking into account the two mass parameters and five couplings in the Lagrangian, as well as the VEVs of the doublet and the singlet, the model contains nine free parameters. We can reduce their number to seven using the minimisation conditions of the potential (i.e.\ the tadpole equations),
\begin{equation}
    \left.\frac{dV}{d\phi}\right|_{\phi=0,s=0}=t_{\phi}=0,\qquad\left.\frac{dV}{ds}\right|_{\phi=0,s=0}=t_S=0,
\end{equation}
which lead to two relations among the Lagrangian parameters
\begin{align}
    \mu^2&=-\frac{\lambda v^2}{2}-\kappa_{SH}v_S -\frac{\lambda_{SH}v_S^2}{2}+\frac{t_\phi}{v}\,,\nn\\
    M_S^{2}&=-2\lambda_Sv_S^2 -\kappa_Sv_S - \frac{\kappa_{SH}v^2}{2v_S}-\frac{\lambda_{SH}v^2}{2}+\frac{t_S}{v_S}\,.
\end{align}
Next, expanding the potential using the definitions from \cref{singlet}, the mass matrix can be computed as
\begin{equation}
    \mathcal{M}^2= 
    \begin{pmatrix}
\frac{d^2V}{d\phi^2} & \frac{d^2V}{d\phi ds} \\[0.5em]
 \frac{d^2V}{d\phi ds}  & \frac{d^2V}{ds^2}
\end{pmatrix}=
\begin{pmatrix}
\mathcal{M}_{\phi}^2 & \mathcal{M}_{\phi s}^2 \\[0.5em]
 \mathcal{M}_{\phi s}^2  &\mathcal{M}_s^2
\end{pmatrix}\,,
\end{equation}
where
\begin{align}
    \mathcal{M}_{\phi}^2&=\mu^2+\frac{3\lambda v^2}{2}+\kappa_{SH}v_S+\frac{\lambda_{SH}v_S^2}{2}\,, \nn\\ 
    \mathcal{M}_s^2&= M_S^{2}+\frac{\lambda_{SH}v^2}{2}+2v_S(\kappa_S+3\lambda_S v_S) \,,\nn\\
    \mathcal{M}_{\phi s}^2&=(\kappa_{SH}+\lambda_{SH}v_S)v\,. 
\end{align}

\noindent
The \cp-even gauge eigenstates can be rewritten in terms of mass eigenstates $h$ and $H$ through a mixing matrix $R_\alpha$, defined by,
\begin{equation}
    \begin{pmatrix}
  \phi \\
  s
\end{pmatrix}=
R_{\alpha}
  \begin{pmatrix}
  h \\
  H
\end{pmatrix}=
\begin{pmatrix}
\cos\alpha & -\sin\alpha \\
 \sin\alpha & \cos\alpha
\end{pmatrix}
  \begin{pmatrix}
  h \\
  H
\end{pmatrix}\,.
\label{mixingmatrix}
\end{equation}
For the rest of this work, we will assume that $h$ corresponds to the detected Higgs boson with a mass around 125 GeV, while $H$ is a BSM Higgs boson, assumed to be always heavier than $h$ (as we want to allow the decay $H\to hh$). 
After the diagonalisation of the mass matrix, the \cp-even scalar mass eigenvalues and corresponding mixing angle $\alpha$ are found to be
\begin{align}
    m_h^2 &= \mathcal{M}_{\phi}^2 \cos^2\alpha + \mathcal{M}_s^2\sin^2\alpha + \mathcal{M}_{\phi s}^2 \sin2\alpha\,,\nn\\
    m_H^2 &= \mathcal{M}_{\phi}^2\sin^2\alpha+ \mathcal{M}_s^2\cos^2\alpha - \mathcal{M}_{\phi s}^2 \sin2\alpha\,,\nn\\
    \tan2\alpha &= \frac{2\mathcal{M}_{\phi s}^2}{\mathcal{M}_{\phi}^2  - \mathcal{M}_s^2}\,.
\end{align}

For our calculational setup, and in particular the renormalisation of the RxSM Higgs sector, it is especially helpful to employ quantities defined in terms of the following seven parameters
\begin{equation}
   m_h^2,\,m_H^2,\,\alpha,\,v,\,v_S,\,\kappa_S,\,\kappa_{SH},
\end{equation}
which we will refer to as ``\textit{mass basis}''. Among these parameters, $m_h\simeq 125\gev$ and $v\simeq 250 \gev$) are fixed, i.e., one is left with five free BSM parameters. 

We therefore re-express the Lagrangian mass parameters and quartic couplings in terms of $m_h$, $m_H$, $\alpha$, and the tadpole parameters $t_\phi$ and $t_S$. We note that while $t_\phi=t_S=0$ in the minimum of the potential, it is useful to keep the dependence on these parameters for the sake of properly including tadpole counterterms in our radiative calculations, see \cref{sec:ren}. We find 
\begin{align}
\label{eq:rep_lambdas_and_masses}
\lambda &= \frac{c_\alpha^2m_h^2+s_\alpha^2m_H^2}{v^2} - \frac{t_\phi}{v^3} \,,\nn\\
\lambda_{SH}&=\frac{(m_h^2-m_H^2)c_{\alpha}s_{\alpha}}{vv_S}-\frac{\kappa_{SH}}{v_S}\,,\nn\\
\lambda_S&=\frac{(m_h^2+m_H^2)v_S+(m_H^2-m_h^2)v_Sc_{2\alpha}-2t_S-2\kappa_Sv_S^2+\kappa_{SH}v^2}{8v_S^3}\,,\nn\\
M_S^{2}&=\frac{6t_S-\kappa_{SH}v^2-2\kappa_Sv_S^2-(m_h^2+m_H^2)v_S+(m_H^2-m_h^2)(vs_{2\alpha}-v_Sc_{2\alpha})}{4v_S}\,,\nn\\
\mu^2&=\frac{3}{2}\frac{t_\phi}{v}-\frac{1}{2}\kappa_{SH}v_S-\frac{1}{2}\big(m_h^2c_\alpha^2+m_H^2s_\alpha^2\big)-\frac{v_S}{2v}\big(m_h^2-m_H^2\big)c_\alpha s_\alpha\,,
\end{align}
where we have used the shorthand notations $c_x\equiv\cos x$ and $s_x\equiv \sin x$. %

Finally, using the relations of \cref{eq:rep_lambdas_and_masses}, the following expressions can be obtained at the tree level for the trilinear Higgs couplings in the mass basis (keeping once again the parametric dependence on the tadpoles, for use in the next section),
{\allowdisplaybreaks
\begin{align}
\label{eq:trilinears_0L}
    \lambda_{hhh}=&\ \frac{1}{4 v^2 v_S^2} \Big\{ -3 v_S \left[ \kappa_{SH} v^3 + 3 (t_\phi - m_h^2 v) v_S \right] c_{\alpha} + 3 v_S \left[ \kappa_{SH} v^3 - t_\phi v_S + m_h^2 v v_S \right] c_{3\alpha}  \nn\\
    &\hspace{1.4cm}+2 v^2 \left[ -6 t_S + 3 \kappa_{SH} v^2 + 6 m_h^2 v_S - 2 \kappa_S v_S^2 \right] s^3_{\alpha} \Big\}\,,\nn\\
    \lambda_{hhH}=&\ \frac{1}{4 v^2 v_S^2} s_{\alpha} \Big\{ -2 v_S \left[ \kappa_{SH} v^3 - 3 t_\phi v_S +(2 m_h^2 + m_H^2) v v_S \right]\nn\\
    &\hspace{1.9cm}- 2 v_S [ 3 \kappa_{SH} v^3 - 3 t_\phi v_S +(2 m_h^2 + m_H^2) v v_S ] c_{2\alpha} \nn\\
    &\hspace{1.9cm}+ v^2 \left[ -6 t_S + 3 \kappa_{SH} v^2 + 2 v_S \left( 2 m_h^2 + m_H^2 - \kappa_S v_S \right) \right] s_{2\alpha} \Big\}\,,\nn\\
    \lambda_{hHH}=&\ \frac{1}{4 v^2 v_S^2}c_{\alpha} \Big\{ 2 v_S \left[ \kappa_{SH} v^3 - 3 t_\phi v_S + (m_h^2 + 2 m_H^2) v v_S \right]\nn\\
    &\hspace{1.9cm}- 2 v_S [ 3 \kappa_{SH} v^3 - 3 t_\phi v_S +(m_h^2 + 2 m_H^2) v v_S ] c_{2\alpha} \nn\\
    &\hspace{1.9cm}+ v^2 \left[ -6 t_S + 3 \kappa_{SH} v^2 + 2 v_S \left( m_h^2 + 2 m_H^2 - \kappa_S v_S \right) \right] s_{2\alpha} \Big\}\,,\nn\\
    \lambda_{HHH}=&\ \frac{1}{8 v^2 v_S^2} \Big\{ 3 v^2 \left[ -6 t_S + 3 \kappa_{SH} v^2 + 6 m_H^2 v_S - 2 \kappa_S v_S^2 \right] c_{\alpha} \nn\\
    &\hspace{1.4cm}+ v^2 [ -6 t_S + 3 \kappa_{SH} v^2 + 6 m_H^2 v_S -2 \kappa_S v_S^2 ] c_{3\alpha} \\
    &\hspace{1.4cm}+ 12 v_S \left[ \kappa_{SH} v^3 + t_\phi v_S - m_H^2 v v_S + \left( \kappa_{SH} v^3 - t_\phi v_S + m_H^2 v v_S \right) c_{2\alpha} \right] s_{\alpha} \Big\}\,.\nn
\end{align}
}


\subsection{Theoretical and experimental constraints}

In this section, we briefly summarise the various theoretical and experimental constraints on the RxSM considered in our analysis. It should be noted that constraints arising from di-Higgs measurements at the LHC are not checked when generating scan points, but are instead a central part of our phenomenological investigations in \cref{sec:dihiggs}.

First, on the theoretical side, we have to ensure that boundedness-from-below of the potential, perturbativity of the Lagrangian quartic couplings, 
and perturbative unitarity are fulfilled. Firstly, for the scalar potential to be bounded from below, the quartic couplings must be positive
in all field directions. To ensure this, we demand that the determinant of the Hessian matrix of the potential is positive, which leads to
\begin{equation}
    \lambda > 0,\quad \lambda_S > 0,\quad \text{and}\quad\lambda_{SH} > -2\sqrt{\lambda\;\lambda_S}\,.
    \label{eq:bfb}
\end{equation}
Second, to ensure perturbativity of the couplings, we require that
\begin{equation}
    \frac{|\lambda|}{2},\;\frac{|\lambda_S|}{2},\;\frac{|\lambda_{SH}|}{2}<4\pi.
    \label{eq:pert}
\end{equation}
Finally, to ensure pertubative unitarity, we compute the scalar $2 \rightarrow 2$ scattering amplitude matrix in the high-energy limit, using results from Ref.~\cite{Braathen:2017jvs}, and we demand that the eigenvalues are below 1. 

Next, we have to check compatibility with experimental measurements. To do so, we consider on the one hand constraints from direct Higgs-boson searches at colliders. The exclusion limits at the $95\%$ C.L.\ of relevant BSM Higgs boson searches (including Run~2 data from the LHC) are implemented in  \HB~\texttt{v.6}~\cite{Bechtle:2008jh,Bechtle:2011sb,Bechtle:2013wla,Bechtle:2015pma,Bechtle:2020pkv,Bahl:2022igd}, which is included in the public code \texttt{HiggsTools}~\cite{Bahl:2022igd}. 
On the other hand, agreement with measurements at the LHC of the mass and signal strengths of the detected Higgs boson is checked via the public tool \texttt{HiggsSignals}~\texttt{v.3}~\cite{Bechtle:2013xfa,Bechtle:2014ewa,Bechtle:2020uwn,Bahl:2022igd} (also now part of \texttt{HiggsTools}).


\section{Calculational setup and on-shell renormalisation of the RxSM}
\label{sec:ren}

In this section, we describe our setup to obtain predictions for the trilinear scalar couplings that enter in the calculation of the di-Higgs production cross-section, namely $\lambda_{hhh}$ and $\lambda_{hhH}$, with a complete on-shell (OS) renormalisation of the RxSM. 

Following the discussion in \cref{sec:model_def}, and taking into account the tadpole parameters, the RxSM contains nine parameters, in the mass basis, given by,
\begin{equation}
   m_h^2,\,m_H^2,\,v,\,\alpha,\,v_S,\,\kappa_S,\,\kappa_{SH},\,t_{\phi},\,t_{S}\,,
\end{equation}
As can be seen from \cref{eq:trilinears_0L}, all nine parameters enter the tree-level expressions of $\lambda_{hhh}$ and $\lambda_{hhH}$,
i.e.\ a calculation at the one-loop level requires a renormalisation of all of them. 

To fix our notation, in the rest of this section, for a given parameter $x$, we denote its bare value as $x_B$ 
and its counterterm as $\delta^\text{CT}x$. The tree-level value of $x$ is denoted as $x^{(0)}$ and its one-loop renormalised value as $\hat{x}^{(1)}$. Lastly, $\delta^{(1)}x$ corresponds to the genuine (possibly divergent) one-loop correction to $x$, i.e.\ $\delta^{(1)}x\equiv \hat{x}^{(1)}-x^{(0)}-\delta^\text{CT}x$. 

In the following, we begin with the renormalisation of the tadpoles, as the choice of treatment of the tadpoles affects all subsequent steps.  
Next, we will discuss the renormalisation of the \cp-even scalar masses and mixing angles, which relate to scalar two-point functions, before turning to the case of the VEVs.  
Finally, we will go beyond standard works on the renormalisation of extended Higgs sectors, and we will consider the renormalisation of the Lagrangian trilinear couplings 
$\kappa_S$ and $\kappa_{SH}$ as a necessary ingredient for the calculation of the one-loop corrections to the trilinear Higgs couplings.


\subsection{Tadpole renormalisation}
\label{sec:ren:tads}
We employ in this work the standard tadpole scheme~\cite{Denner:1991kt} (sometimes also referred to as ``parameter-renormalised tadpole scheme'') and define the tadpole counterterm via an OS prescription. At one loop, the tadpole equations take the form
\begin{align}
    \hat{t}_i^{(1)}=t^{(0)}_i+\delta^{(1)} t_i+\delta^\rm{CT} t_i=0\quad \text{for }i=\phi,\ S\,. 
\end{align}
Our scheme choice for the tadpoles results in
\begin{align}
\label{eq:OStad}
    \delta^\text{CT} t_i=-\delta^{(1)} t_i\quad \text{for }i=\phi,\ S\,.
\end{align}
This implies that the relations between parameters dictated by the tree-level tadpole equations $t^{(0)}_i=0$ remain valid at the one-loop level, and the values of the VEVs at the minima of the tree-level and one-loop corrected potentials coincide. As a side remark, we note that for scenarios with heavy BSM scalars but very small values of the singlet VEV, unphysical enhancements can occur in the standard tadpole scheme, as pointed out in Ref.~\cite{Braathen:2021fyq}. However, we will not consider such scenarios in our phenomenological investigations. 

Finally, while the tadpole parameters $t_\phi$ and $t_S$ and their corresponding counterterms are defined in the gauge eigenstate basis, the diagrammatic calculation of one-loop corrections to the tadpoles is most conveniently performed 
in the mass basis, i.e.\ $\delta^{(1)}t_h$ and $\delta^{(1)}t_H$. Therefore we need to perform the rotation 
\begin{align}
    \delta^\text{CT}t_{\phi}&=\sin\alpha\;\delta^{(1)} t_H-\cos\alpha\;\delta^{(1)} t_h\,,\nn\\
    \delta^\text{CT}t_{S}&=-\cos\alpha\;\delta^{(1)} t_H-\sin\alpha\;\delta^{(1)} t_h\,.
\end{align}

\subsection{Two-point function renormalisation}
In our work, we renormalise the two-point scalar functions on shell, 
following Refs.~\cite{Kanemura:2004mg,Kanemura:2017wtm}. The first step is to renormalise the scalar fields, which are defined in the mass basis in \cref{mixingmatrix}, by introducing 
the field renormalisation constant matrix $\sqrt{Z}$ that leads, at the one-loop order, to
\begin{equation}
     \begin{pmatrix}
h_B \\[.5em]
H_B
\end{pmatrix}=
\sqrt{Z}
 \begin{pmatrix}
\hat{h} \\[.5em]
 \hat{H}
\end{pmatrix}=
 \begin{pmatrix}
1+\frac{1}{2}\delta^{\mathrm{CT}} Z_{hh} & \frac{1}{2}\delta^{\mathrm{CT}} Z_{hH} \\[.5em]
\frac{1}{2}\delta^{\mathrm{CT}} Z_{Hh} & 1+\frac{1}{2}\delta^{\mathrm{CT}} Z_{HH}
\end{pmatrix}
 \begin{pmatrix}
\hat{h} \\[.5em]
 \hat{H}
\end{pmatrix}\,, \label{fieldren}
\end{equation}
where $\hat{h}$ ($h_B$) and $\hat{H}$ ($H_B$) are the renormalised (bare) \cp-even scalar fields. For the renormalisation of the two-point function $\Gamma$ this yields,
\begin{align}
    \hat{\Gamma}(p^2)=\left(
\begin{matrix}
        \hat{\Gamma}_{hh}(p^2) & \hat{\Gamma}_{hH}(p^2)\\[.5em]
        \hat{\Gamma}_{Hh}(p^2) & \hat{\Gamma}_{HH}(p^2)
\end{matrix}\right) =&\ 
    i\sqrt{Z}^{\dagger}[p^2\mathbbm{1}_{2\times 2}-\hat{M}_{\phi}^2+\Sigma_{\phi,0}(p^2)-\delta M_{\phi}^2]\sqrt{Z} \nonumber\\
    \approx &\ i[p^2\mathbbm{1}_{2\times 2}-\hat{M}^2_{\phi}+\hat{\Sigma}_{\phi}(p^2)]\,,
\end{align}
where in the second line we have only kept terms of one-loop order (hence the ``$\approx$''). In this equation, $\hat{\Sigma}_{\phi}(p^2)$ denotes the $2\times 2$ symmetric matrix of renormalised self-energies, defined as
\begin{equation}
        \hat{\Sigma}_{\phi}(p^2)\equiv\left(
\begin{matrix}
        \hat{\Sigma}_{hh}(p^2) & \hat{\Sigma}_{hH}(p^2)\\
        \hat{\Sigma}_{Hh}(p^2) & \hat{\Sigma}_{HH}(p^2)
\end{matrix}\right),
\end{equation}
while $\hat{M}^2_{\phi}$ is the renormalised mass matrix. 
The renormalisation conditions in the OS scheme are as follows:
\begin{enumerate}
  \item The mixing of particles with the same quantum numbers vanishes at $p^2=m_{h,H}^2$, so that
\begin{equation}
    \rm{Re}[\hat{\Sigma}_{hH}(m_h^2)]=\rm{Re}[\hat{\Sigma}_{Hh}(m_H^2)]=0. \label{con1}
\end{equation}  
  \item The mass parameters $m_{h,H}^2$ are defined as being the real parts of the poles of the renormalised propagator $\hat{G}_{\phi}(p^2)$ (the inverse of the two-point function). This implies that
\begin{equation}
    \rm{Re}[\hat{\Sigma}_{hh}(m_h^2)]=\rm{Re}[\hat{\Sigma}_{HH}(m_H^2)]=0.\label{con2}
\end{equation}  
  \item The physical fields are properly normalised through fixing the residue of the propagator at its pole to $i$. This gives
\begin{equation}
    \rm{Re}\left[\frac{\partial\hat{\Sigma}_{hh}(p^2)}{\partial p^2}\biggm|_{p^2=m_h^2} \right]=\rm{Re}\left[\frac{\partial\hat{\Sigma}_{HH}(p^2)}{\partial p^2}\biggm|_{p^2=m_H^2} \right]=0. \label{con3}
\end{equation}  
\end{enumerate}

\noindent From \cref{con2} 
the mass counterterms can be derived as
\begin{align}
    \delta^{CT} m^2_{h}=&\ \rm{Re}[\Sigma_{hh}(m_h^2)]\,, \nn\\ 
    \delta^{CT} m^2_{H}=&\ \rm{Re}[\Sigma_{HH}(m_H^2)]\,.
\end{align}
In turn from \cref{con1} and \cref{con3} we can obtain the field counterterms
\begin{align}
\label{eq:fieldCT_OS}
    \delta^{\mathrm{CT}} Z_{hh}&=-\rm{Re}\left[\frac{\partial \Sigma_{hh}(p^2)}{\partial p^2}\right]_{p^2=m_h^2} \,,\nn\\
    \delta^{\mathrm{CT}} Z_{HH}&=-\rm{Re}\left[\frac{\partial \Sigma_{HH}(p^2)}{\partial p^2}\right]_{p^2=m_H^2} \,,\nn\\
    \delta^{\mathrm{CT}} Z_{hH}&=\frac{\rm{Re}[\Sigma_{hH}(m_H^2)]}{m_h^2-m_H^2} \,,\nn\\ 
    \delta^{\mathrm{CT}} Z_{Hh}&=\frac{\rm{Re}[\Sigma_{Hh}(m_h^2)] }{m_H^2-m_h^2}\,.
\end{align}

\subsection{Mixing angle renormalisation}

In order to renormalise the mixing angle, we choose in this work to use the KOSY scheme \cite{Kanemura:2004mg,Kanemura:2015fra}. The idea behind this scheme is to renormalise the rotation matrix by temporarily switching to the gauge basis, performing a renormalisation transformation of the fields (in the gauge basis) and of the mixing angle, and then relating this to the renormalisation of the fields in the mass basis --- c.f.\ \cref{fieldren}. We begin with
\begin{gather}
\label{kan}
\left(
    \begin{matrix}
        h_B\\
        H_B
    \end{matrix}\right)=
    \rm{R}_{\alpha^{0}}^\rm{T} \left(
    \begin{matrix}
        s_B\\
        \phi_B
    \end{matrix}\right) \rightarrow \rm{R}^\rm{T}_{\delta^{\mathrm{ CT}} \alpha}\rm{R}^\rm{T}_{\hat{\alpha}}\left(
    \begin{matrix}
        s_B\\
        \phi_B
    \end{matrix}\right)= \rm{R}^\rm{T}_{ \delta^{\mathrm{ CT}} \alpha}\rm{R}^\rm{T}_{\hat{\alpha}}\sqrt{Z_{\phi,s}}\left(
    \begin{matrix}
        \hat{s}\\
        \hat{\phi}
    \end{matrix}\right)=  \nonumber \\
    =\rm{R}^\rm{T}_{\delta^{\mathrm{ CT}} \alpha}\rm{R}^\rm{T}_{\hat{\alpha}}\sqrt{Z_{\phi,s}}\rm{R}_{\hat{\alpha}}\rm{R}^\rm{T}_{\hat{\alpha}}\left(
    \begin{matrix}
         \hat{s}\\
        \hat{\phi}
    \end{matrix}\right) 
    =\rm{R}^\rm{T}_{ \delta^{\mathrm{ CT}} \alpha}\rm{R}^\rm{T}_{\hat{\alpha}}\sqrt{Z_{\phi,s}}\rm{R}_{\hat{\alpha}}\left(
    \begin{matrix}
        \hat{h}\\
        \hat{H}
    \end{matrix}\right)=\sqrt{\tilde{Z}}\left(
    \begin{matrix}
       \hat{h}\\
        \hat{H}
    \end{matrix}\right)\,.
\end{gather}
Expanding in the mass basis the WFR matrix $\rm{R}^\rm{T}_{\hat{\alpha}}\sqrt{Z_{\phi,s}}\rm{R}_{\hat{\alpha}}$ and the counterterm rotation matrix $\rm{R}^\rm{T}_{ \delta^{\mathrm{ CT}} \alpha}$, and then equating this equation to \cref{fieldren}, we obtain
\begin{align}
    \sqrt{\tilde{Z}}=&\rm{R}^\rm{T}_{ \delta^{\mathrm{ CT}} \alpha}\rm{R}^\rm{T}_{\hat{\alpha}}\sqrt{Z_{\phi,s}}\rm{R}_{\hat{\alpha}}=\rm{R}^\rm{T}_{ \delta^{\mathrm{ CT}} \alpha}\left(
    \begin{matrix}
        1+\frac{\delta^{\mathrm{CT}} Z_{hh}}{2}&\delta^{\mathrm{ CT}} C_{hH} \\
        \delta^{\mathrm{ CT}} C_{Hh}&1+\frac{\delta^{\mathrm{CT}} Z_{HH}}{2}
    \end{matrix}\right)\approx\nonumber\\
    &\approx\left(
    \begin{matrix}
        1+\frac{\delta^{\mathrm{CT}} Z_{hh}}{2}&\delta^{\mathrm{ CT}} C_{hH}- \delta^{\mathrm{ CT}} \alpha \\
        \delta^{\mathrm{ CT}} C_{Hh}+ \delta^{\mathrm{ CT}} \alpha&1+\frac{\delta^{\mathrm{CT}} Z_{HH}}{2}
    \end{matrix}\right)\equiv\sqrt{Z}\,,
\end{align}
where $\delta^\text{CT}C_{hH}$ and $\delta^\text{CT}C_{Hh}$ 
come from the expansion of the off-diagonal terms of the WFR matrix $\text{R}_{\hat\alpha}^\text{T}\sqrt{Z_{\phi,s}}\text{R}_{\hat\alpha}$. In the second line of the equation above, we have again only kept terms of one-loop order. 
We can observe that the diagonal terms are identical, while for the non-diagonal terms we find,
\begin{align}
    \delta^{\mathrm{ CT}} C_{hH}-\delta^{\mathrm{ CT}} \alpha&=\frac{\delta^{\mathrm{CT}} Z_{hH}}{2} \,, \nn\\
    \delta^{\mathrm{ CT}}  C_{Hh}+\delta^{\mathrm{ CT}} \alpha&=\frac{\delta^{\mathrm{CT}} Z_{Hh}}{2} \,.
\end{align}
To extract an expression for $\delta^\text{CT}\alpha$ from the equations above, at least one more relation between $\delta^\text{CT}C_{hH}$ and $\delta^\text{CT}C_{Hh}$  is necessary. Different choices for this have been discussed in the literature: while earlier works~\cite{Kanemura:2004mg} imposed that the matrix $Z_{\phi,s}$ be symmetric --- thus leading to $\delta^\text{CT}C_{hH}=\delta^\text{CT}C_{Hh}$ --- subsequent work~\cite{Kanemura:2015fra} demonstrated that this choice can lead to gauge dependences in renormalised mixing angles. Moreover, these works showed that one can instead use the additional degree of freedom from keeping $\delta^\text{CT}C_{hH}$ and $\delta^\text{CT}C_{Hh}$ non equal to absorb the gauge dependence. It should however be noted that this issue of gauge dependences only occurs in the renormalisation of extended \cp-odd scalar sector, see Ref. \cite{Kanemura:2015fra}. Because the RxSM only extends the \cp-even scalar sector, but not the \cp-odd one, we do not require separate $\delta^\text{CT}C_{hH}\neq\delta^\text{CT}C_{Hh}$, and we can simply impose that
\begin{equation}
    \delta^{\mathrm{ CT}} C_ {hH}=\delta^{\mathrm{ CT}}C_{Hh}\equiv\delta^{\mathrm{ CT}}  C \label{c}\,.
\end{equation}
Using the expressions of the field strength counter terms $\delta^\text{CT}Z_{hH}$ and $\delta^\text{CT}Z_{Hh}$ from \cref{{eq:fieldCT_OS}}, the following expression can finally be obtained for the mixing angle counterterm
\begin{equation}
    \delta^{\mathrm{ CT}} \alpha = \frac{1}{2(m_H^2-m_h^2)}\rm{Re}[\Sigma_{hH}(m_h^2)+\Sigma_{hH}(m_H^2)]\,.
\end{equation}

\subsection{Renormalisation of the VEVs}
\subsubsection*{Electroweak VEV}
Since the electroweak sector of the RxSM is identical to the SM one, we use the same procedure to renormalise the EW VEV as in the SM. 
Following the convention in \texttt{anyH3}~\cite{Bahl:2023eau} (which we will use below for the calculation of trilinear scalar couplings), we relate the EW VEV to the $W$- and $Z$-boson masses as well as $\alpha_\text{EM}$ (in the Thompson limit),
\begin{equation}
    v^2=\frac{m_W^2}{\pi \alpha_\text{EM}}\left(1-\frac{m_W^2}{m_Z^2}\right)\,.
\end{equation}
It follows that the EW VEV counterterm can be expressed, like in the SM, as (see also Ref.~\cite{Kanemura:2015fra})
\begin{align}
    \frac{\delta^\text{CT} v}{v}=&\ \frac{1}{2}\Bigg[\frac{\delta^\text{CT}m_W^2}{m_W^2}+\frac{\cw^2}{\sw^2}\left(\frac{\delta^\text{CT}m_Z^2}{m_Z^2}-\frac{\delta^\text{CT}m_W^2}{m_W^2}\right)-\frac{\delta^\text{CT}\alpha_\text{EM}}{\alpha_\text{EM}}\Bigg]\,.
\end{align}
where $\sw$ and $\cw$ are the sine and cosine of the weak mixing angle. Adopting an OS renormalisation of the EW input parameters, we have
\begin{align}
    \delta^\text{CT}m_V^2 = \mathrm{Re}\big[\Sigma^T_{VV}(p^2=m_V^2)\big]\,,\quad \text{for }V=W,\,Z\,,
\end{align}
where $\Sigma^T_{VV}(p^2)$ is the transverse part of the one-loop gauge boson self-energy, defined through the standard decomposition $\Sigma_{VV}^{\mu\nu}(p^2)=(g^{\mu\nu}-p^\mu p^\nu/p^2)\;\Sigma_{VV}^T(p^2)\;+\;p^\mu p^\nu/p^2\; \Sigma_{VV}^L(p^2)$. Moreover, 
\begin{align}
    \delta^\text{CT}\alpha_\text{EM}=\frac{1}{2}\Pi_{\gamma\gamma}(p^2=0)+\frac{2\sw}{\cw}\frac{\Sigma_{\gamma Z}^T(p^2=0)}{m_Z^2}\,,
\end{align}
where $\Pi_{\gamma\gamma}$ is defined from the (one-loop) photon self-energy as $\Sigma_{\gamma\gamma}^{\mu\nu}(p^2)=(p^2g^{\mu\nu}-p^\mu p^\nu)\Pi_{\gamma\gamma}(p^2)$. Combining all these expressions, we obtain
\begin{align}
\frac{\delta^\text{CT} v}{v}= \frac{1}{2}&\left[\frac{\sw^2-\cw^2}{s^2_W}\frac{\rm{Re}\big[\Sigma_{WW}^T(m_W^2)\big]}{m_W^2}+\frac{\cw^2}{\sw^2}\frac{\rm{Re}\big[\Sigma_{ZZ}^T(m_Z^2)\big]}{m_Z^2}-\Pi_{\gamma\gamma}(0)
-\frac{2\sw}{\cw}\frac{\Sigma_{\gamma Z}^T(0)}{m_Z^2}\right]\,.
\end{align}

\subsubsection*{Singlet VEV}
The one-loop RGE of the singlet VEV can be shown to vanish in the RxSM, which implies that the corresponding counterterm, $\delta^\text{CT}v_S$, contains no UV divergence. A theoretical understanding for this can be obtained from, e.g., Ref.~\cite{Sperling:2013eva}, where it was shown that, 
when the Lagrangian is invariant under a rigid gauge transformation of the field that acquires a VEV, 
the counterterm of said VEV can at most contain UV-finite contributions. 
This is precisely the case for the $SU(2)_L$ gauge singlet $S$ in the RxSM. Consequently, 
$\delta^\text{CT} v_S$ is UV-finite and, we choose (as we work in the standard tadpole scheme) to set the finite part of this counterterm to zero, i.e.
\begin{equation}
    \delta^\text{CT} v_S=0\,.
\end{equation}


\subsection{Diagrammatic calculation of trilinear scalar couplings}
Before we turn to the last part of our renormalisation procedure, we have to introduce the approach that we have followed to compute the corrections to the trilinear Higgs couplings. We perform full one-loop diagrammatric calculations of both $\lambda_{hhh}$ and $\lambda_{hhH}$, employing the public code \texttt{anyH3}~\cite{Bahl:2023eau,anyHH}. 
We note that, throughout this work, we compute predictions for the trilinear scalar couplings with all external momenta set to zero, given that these momentum-independent quantities are passed as modified couplings to \texttt{HPAIR}\footnote{\texttt{HPAIR} is the code that we use to perform the di-Higgs production cross-section computations in the HL-LHC and is introduced in Sec. 5.}~ \cite{Abouabid:2021yvw,Arco:2022lai,Dawson:1998py,Nhung:2013lpa,Grober:2017gut,Grober:2015cwa} to calculate total cross-sections and differential distributions for di-Higgs production.\footnote{The impact of the momentum dependence in $\lambda_{hhh}$ and $\lambda_{hhH}$ on the total di-Higgs cross-section and corresponding distributions is investigated in Ref.~\cite{anyHH} in the context of various models, including a 2HDM. The effect on distribution shapes is found to be typically moderate~\cite{Arco:2025pgx}, while modifications of the total cross-sections can reach up to $\mathcal{O}(20\%)$ --- this is to a large extent due to the modification of the height of the resonant peak from the $H$ scalar. } 
The different contributions to the renormalised trilinear Higgs couplings at the one-loop level, which we denote $\hat\lambda_{ijk}^{(1)}$ (where $i,j,k=h$ or $H$), can be divided between 
the tree level contribution $\lambda_{ijk}^{(0)}$, the one-loop one-particle-irreducible (1PI) vertex correction diagrams 
$\delta^{(1)}_{\mathrm{gen}}\lambda_{ijk}$, the external-leg corrections 
$\delta^{(1)}_{\mathrm{wfr}}\lambda_{ijk}$,  
and the one-loop counterterm $\delta^{(1)}_{\mathrm{CT}}\lambda_{ijk}$. 
Diagrams with insertions of one-loop tadpoles do not appear in our setup, because we employ an OS renormalisation of tadpoles (as discussed in \cref{sec:ren:tads}). The different contributions can be represented in terms of Feynman diagrams as 
\begin{equation}
    \hat{\lambda}_{ijk}^{(1)}=-\hat{\Gamma}^{(1)}_{h_i,h_j,h_k}(0,0,0)=\lambda_{ijk}^{(0)}+\delta^{(1)}_{\mathrm{gen}}\lambda_{ijk}+\delta^{(1)}_{\mathrm{wfr}}\lambda_{ijk}+\delta^{\mathrm{CT}}\lambda_{ijk}=\nonumber
\end{equation}
\begin{equation}
=
\begin{tikzpicture}[baseline=0]
    \begin{feynman}
        \vertex (X);
        \vertex[left=1cm of X] (e1) {${h}_i$};
        \vertex[right=1cm of X] (tmp);
        \vertex[above=1cm of tmp] (e2) {${h}_j$};
        \vertex[below=1cm of tmp] (e3) {${h}_k$};
        \diagram*{(e1)  -- [scalar] (X),
                  (e2)  -- [scalar] (X),
                  (e3)  -- [scalar] (X)};
    \end{feynman}
\end{tikzpicture}
+
\begin{tikzpicture}[baseline=0]
    \begin{feynman}
        \vertex (in) {${h}_i$ };
        \vertex[right=1cm of in](a);
        \vertex[right=1.cm of a](b);
        \vertex[right=1cm of b](tmp);
        \vertex[above=0.56cm of tmp](out1) {${h}_j$};
        \vertex[below=0.56cm of tmp](out2) {${h}_k$};
        \diagram*{(in)  -- [scalar] (a),
                  (out1) -- [scalar] (b),
                  (out2) -- [scalar] (b),
                  (a)  -- [half left,looseness=1.5] (b),
                  (b)  -- [half left,looseness=1.5] (a)
                 };
    \end{feynman}
\end{tikzpicture}
+
\begin{tikzpicture}[baseline=(a)]
    \begin{feynman}
      \vertex at (0,1.75) (a) {${h}_i$};
      \vertex at (3.5,0) (b) {${h}_j$};
      \vertex at (3.5,3.5) (c) {${h}_k$};

      \vertex at (1.16,1.75) (a1);
      \vertex at (2.62,0.58) (b1);
      \vertex at (2.62,2.917) (c1);

      \diagram* {
      (a) -- [scalar] (a1),
      (b) -- [scalar] (b1),
      (c) -- [scalar] (c1),
      (a1) -- [] (c1) -- [] (b1) -- [] (a1)
      };

    \end{feynman}
\end{tikzpicture}
+\nonumber
\end{equation}
\begin{equation}
+
\begin{tikzpicture}[baseline=(X.base)]
    \begin{feynman}
        \vertex (X);
        \vertex[left=1cm of X] (e1) {${h}_i$};
        \vertex[right=1cm of X] (tmp);
        \vertex[right=0.5cm of X] (tmp0);
        \vertex[above=0.5cm of tmp0] (x);
        \vertex[above=1cm of tmp] (e2) {${h}_j$};
        \vertex[below=1cm of tmp] (e3) {${h}_k$};
        \diagram*{(e1)  -- [scalar] (X),
                  (e3)  -- [scalar] (X),
                  (X)  -- [] (x),
                  (e2)  -- [scalar] (x)};
        \draw[] (x) arc [start angle=-45, end angle=315, radius=0.4cm] node[label={[xshift=-0.5cm,yshift=0.5cm]}] {};
    \end{feynman}
\end{tikzpicture}
+
\begin{tikzpicture}[baseline=(X.base)]
    \begin{feynman}
        \vertex (X);
        \vertex[left=1cm of X] (e1) {${h}_i$};
        \vertex[right=1.5cm of X] (tmp);
        \vertex[right=0.5cm of X] (tmp0);
        \vertex[right=1cm of X] (tmp1);
        \vertex[above=0.5cm of tmp0] (x1);
        \vertex[above=1cm of tmp1] (x2);
        \vertex[above=1.3cm of tmp] (e2) {${h}_j$};
        \vertex[below=1.3cm of tmp] (e3) {${h}_k$};
        \diagram*{(e1)  -- [scalar] (X),
                  (e3)  -- [scalar] (X),
                  (X)  -- [] (x1),
                  (x1)  -- [ half left] (x2),
                  (x2)  -- [ half left] (x1),
                  (e2)  -- [scalar] (x2)};
    \end{feynman}
\end{tikzpicture}
 +
\begin{tikzpicture}[baseline=0]
    \begin{feynman}
        \vertex[crossed dot] (X) {};
        \vertex[left=1.5cm of X] (e1) {${h}_i$};
        \vertex[right=1cm of X] (tmp);
        \vertex[above=1cm of tmp] (e2) {${h}_j$};
        \vertex[below=1cm of tmp] (e3) {${h}_k$};
        \diagram*{(e1)  -- [scalar] (X),
                  (X)  -- [scalar] (e2),
                  (X)  -- [scalar] (e3)};
    \end{feynman}
\end{tikzpicture}
\,,\label{one_diagrams}
\end{equation}
where $\hat\Gamma_{ijk}(p_1^2,p_2^2,p_3^2)$ denotes the renormalised scalar three-point function for external states $ijk$, evaluated for external momenta $p_1^2$, $p_2^2$ and $p_3^2$ (here all taken to zero) respectively. Furthermore, $\lambda_{ijk}^{(0)}$ corresponds to the first diagram, $\delta^{(1)}_\text{gen}\lambda_{ijk}$ to the second and third ones (first line), $\delta^{(1)}_\text{wfr}\lambda_{ijk}$ to the fourth and fifth diagrams (second line), and $\delta^\text{CT}\lambda_{ijk}$ to the last diagram. In the above diagrams the solid lines do not only indicate fermions, but serve to represent \textit{any} possible particle of the RxSM --- scalar, fermion, gauge boson or ghost.

The counterterms $\delta^\text{CT}\lambda_{ijk}$ can be computed in terms of the counterterms for each of the parameters on which the couplings depend, as
\begin{align}
    &\delta^{\mathrm{CT}}\lambda_{ijk}=\sum_x\delta^{\mathrm{CT}}x\frac{\partial\lambda_{ijk}^{(0)}}{\partial x}\,,
\end{align}
where $x\in \{m_h^2,m_H^2, v, \alpha, \kappa_S, \kappa_{SH}, t_\phi, t_S\}$. 

\subsection{Renormalisation of the $\mathbb{Z}_2$-breaking couplings}
At this point, we have already summarised the OS \footnote{We note that among our chosen renormalisation conditions, all are OS conditions except the one for the singlet VEV $v_S$. Given that $v_S$ does not require any renormalisation in the RxSM (even the UV divergence of its counterterm vanishes), we will thus refer to our overall renormalisation scheme as an OS scheme, although this constitutes a slight abuse of naming concerning $v_S$.} conditions that are commonly used for singlet extensions of the SM (or similarly for the 2HDM). However, two more parameters still need to be renormalised, namely the Lagrangian trilinear couplings $\kappa_S$ and $\kappa_{SH}$. These two additional degrees of freedom constitute a significant difference with the case of the $\mathbb{Z}_2$-symmetric singlet extension of the SM, in which a complete OS renormalisation of the model can be defined exclusively in terms of one- and two-point functions. 
In contrast, in the RxSM, we have now exhausted the scalar two-point functions that could be use to define OS renormalisation conditions. While an $\overline{\text{MS}}$ renormalisation of $\kappa_S$ and $\kappa_{SH}$ is in principle an option (and this is what is generally done in studies of the RxSM \cite{Kanemura:2015fra,Bahl:2025wzj}), this would leave a renormalisation scale dependence in our results for $\rlahhh1$ and $\rlahhH1$. Such a renormalisation scale dependence could of course be mitigated by including the renormalisation group running of the $\overline{\text{MS}}$ couplings in our setup. We prefer, however, to avoid this entirely by defining OS renormalisation conditions for $\kappa_S$ and $\kappa_{SH}$ in terms of scalar three-point functions, which constitutes one of the main new theoretical aspects of this paper.\footnote{Different options for the renormalisation of Lagrangian trilinear couplings were discussed e.g.\ in the context of the MSSM in Ref.~\cite{Bahl:2022kzs}. } 

The main objective of this calculation is to obtain the leading BSM contributions to the di-Higgs production process at NLO, which only involves the first two out of the four trilinear scalar couplings that exist in the RxSM, namely $\lahhh$, $\lahhH$, $\lahHH$ and $\laHHH$. Therefore, we can use the two trilinear Higgs couplings that do not enter in the di-Higgs production process, i.e.\ $\lahHH$ and $\laHHH$, in order to define two on-shell renormalisation conditions 
for 
$\kappa_S$ and $\kappa_{SH}$. We choose to impose 
\begin{align}
    \rlahHH1 &\overset{!}{=} \; \lahHH^{(0)}\,, \label{conh}\\
    \rlaHHH1 &\overset{!}{=} \; \laHHH^{(0)}\,. \label{conH}
\end{align}
The two renormalised couplings can be expanded as
\begin{align}
    \rlahHH1 &= \lahHH^{(0)}+\delta^{(1)}_{\text{gen}}\lahHH + \delta_{\mathrm{wfr}}^{(1)}\lahHH 
    + \delta^{\mathrm{CT}}_{m_{\varphi}^2}\lahHH + \delta^{\mathrm{CT}}_{v}\lahHH + \delta^{\mathrm{CT}}_{t_{\varphi}}\lahHH + \nonumber\\
    &\quad+\delta_\alpha^\text{CT}\lambda_{hHH}+\delta^{\mathrm{CT}}\kappa_S\frac{\partial\lahHH^{(0)}}{\partial\kappa_S}
    + \delta^{\mathrm{CT}}\kappa_{SH}\frac{\partial\lahHH^{(0)}}{\partial\kappa_{SH}} 
    \overset{!}{=} \; \lahHH^{(0)}\,,\nn\\
    \rlaHHH1 &= \laHHH^{(0)}+\delta^{(1)}_{\text{gen}}\laHHH+\delta_{\mathrm{wfr}}^{(1)}\laHHH+\delta^{\mathrm{CT}}_{m_{\varphi}^2}\laHHH+\delta^{\mathrm{CT}}_{v}\laHHH+\delta^{\mathrm{CT}}_{t_{\varphi}}\laHHH+\nonumber\\
    &\quad+\delta_\alpha^\text{CT}\lambda_{HHH}+\delta^{\mathrm{CT}}\kappa_S\frac{\partial\laHHH^{(0)}}{\partial\kappa_S}+\delta^{\mathrm{CT}}\kappa_{SH}\frac{\partial\laHHH^{(0)}}{\partial\kappa_{SH}}\overset{!}{=}\laHHH^{(0)}\,,
\end{align}
where $\delta^\text{CT}_x\lambda_{ijk}\equiv \delta^{\mathrm{CT}}x\frac{\partial\lambda_{ijk}^{(0)}}{\partial x}$ represents the contribution to the total one-loop counterterm for the coupling $\lambda_{ijk}$ that comes from the parameter $x$. Furthermore, we have used $\delta^{\mathrm{CT}}_{m_{\varphi}^2}\lambda_{ijk}$ and $\delta^{\mathrm{CT}}_{t_\varphi}\lambda_{ijk}$ as shorthand notations for the sum of all the contributions from mass and tadpole counterterms, respectively. 
The tree level parts on both sides of the equalities cancel with each other, and then we can rewrite the above equations as 
\begin{align}
    \delta^{(1)}_{\text{gen}+\text{wfr}}\lahHH+\sum_{x\in\{m_\varphi^2,v,t_\varphi,\alpha\}}\delta^\text{CT}_x\lahHH+\delta^\text{CT}\kappa_S\frac{\partial\lahHH^{(0)}}{\partial\kappa_S}+\delta^\text{CT}\kappa_{SH}\frac{\partial\lahHH^{(0)}}{\partial\kappa_{SH}}&=0\,,\nn\\
\delta^{(1)}_{\text{gen}+\text{wfr}}\laHHH+\sum_{x\in\{m_\varphi^2,v,t_\varphi,\alpha\}}\delta^\text{CT}_x\laHHH+\delta^\text{CT}\kappa_S\frac{\partial\laHHH^{(0)}}{\partial\kappa_S}+\delta^\text{CT}\kappa_{SH}\frac{\partial\laHHH^{(0)}}{\partial\kappa_{SH}}&=0\,.
\end{align}
We thus have a system of two equations with two unknowns, $\delta^\text{CT}\kappa_S$ and $\delta^\text{CT}\kappa_{SH}$, which we can solve 
to obtain our on-shell counterterms. These read 
\begin{align}
  \delta^{\mathrm{CT}}\kappa_S&= \frac{\frac{\partial\laHHH^{(0)}}{\partial\kappa_{SH}}(\delta^{(1)}_{\text{gen}+\text{wfr}}\lahHH+\sum_x\delta^\text{CT}_x\lahHH)-\frac{\partial\lahHH^{(0)}}{\partial\kappa_{SH}}(\delta^{(1)}_{\text{gen}+\text{wfr}}\laHHH+\sum_x\delta^\text{CT}_x\laHHH)}{\frac{\partial\lahHH^{(0)}}{\partial\kappa_{SH}}\frac{\partial\laHHH^{(0)}}{\partial\kappa_{S}}-\frac{\partial\lahHH^{(0)}}{\partial\kappa_{S}}\frac{\partial\laHHH^{(0)}}{\partial\kappa_{SH}}}\,,\nn\\
  \delta^{\mathrm{CT}}\kappa_{SH}&= \frac{\frac{\partial\lahHH^{(0)}}{\partial\kappa_{S}}(\delta^{(1)}_{\text{gen}+\text{wfr}}\laHHH+\sum_x\delta^\text{CT}_x\laHHH)-\frac{\partial\laHHH^{(0)}}{\partial\kappa_{S}}(\delta^{(1)}_{\text{gen}+\text{wfr}}\lahHH+\sum_x\delta^\text{CT}_x\lahHH)}{\frac{\partial\lahHH^{(0)}}{\partial\kappa_{SH}}\frac{\partial\laHHH^{(0)}}{\partial\kappa_{S}}-\frac{\partial\lahHH^{(0)}}{\partial\kappa_{S}}\frac{\partial\laHHH^{(0)}}{\partial\kappa_{SH}}}\,.
\end{align}

As a final remark, it should be noted that \cref{conh,conH} are of course not the only possible choices of renormalisation conditions that could be employed for $\kappa_S$ and $\kappa_{SH}$. In principle, we could have imposed similar conditions on $\lambda_{hhh}$ and $\lambda_{hhH}$. With such a choice, the BSM corrections to the di-Higgs production processes $gg\to hh$ and $e^+e^-\to Zhh$, investigated in \cref{sec:dihiggs}, would be entirely absorbed into the finite parts of the counterterms $\delta^\text{CT}\kappa_S$ and $\delta^\text{CT}\kappa_{SH}$. In turn, the investigation of these processes would be identical to the analysis with tree-level couplings in Ref.~\cite{Arco:2025nii}.  

\subsection{Dependence on the renormalisation scale \boldmath{$Q$}}

Having now determined a complete set of counterterms, we can compute the one-loop corrections to $\lahhh$ and $\lahhH$ in our OS scheme.\footnote{The \texttt{schemes.yml} implementation of our OS scheme for use in \texttt{anyH3} can be provided upon request.} 
A powerful check of the consistency of our scheme definition is offered by the renormalisation scale independence of our results, as these should only depend on OS-renormalised quantities that are, by definition, scale independent. 
To check this we have considered two benchmark points from Ref.~\cite{Arco:2025nii}, 
which are given in \cref{points1}.

\begin{table}[h]
\centering
\begin{tabular}{@{}ccccccc@{}}
\toprule
    & $m_h$ [GeV] & $m_H$ [GeV] & $\cos\alpha$ & $v_S$ [GeV]  & $\kappa_S$ [GeV] & $\kappa_{SH}$ [GeV] \\ \midrule
BPI & 125.1 & 327.0 & 0.974        & 60.9 & $-361.0$     & $-245.0$                          \\
BPII & 125.1 & 511.0 & 0.986        & 40.7 & $-618.0$     & $-372.0$                          \\ \bottomrule 
\end{tabular}
\caption{Two benchmark points of the RxSM, taken from Ref.~\cite{Arco:2025nii}. }
\label{points1}
\end{table}

\begin{figure}[h]
    \centering
    \includegraphics[width=0.483\linewidth]{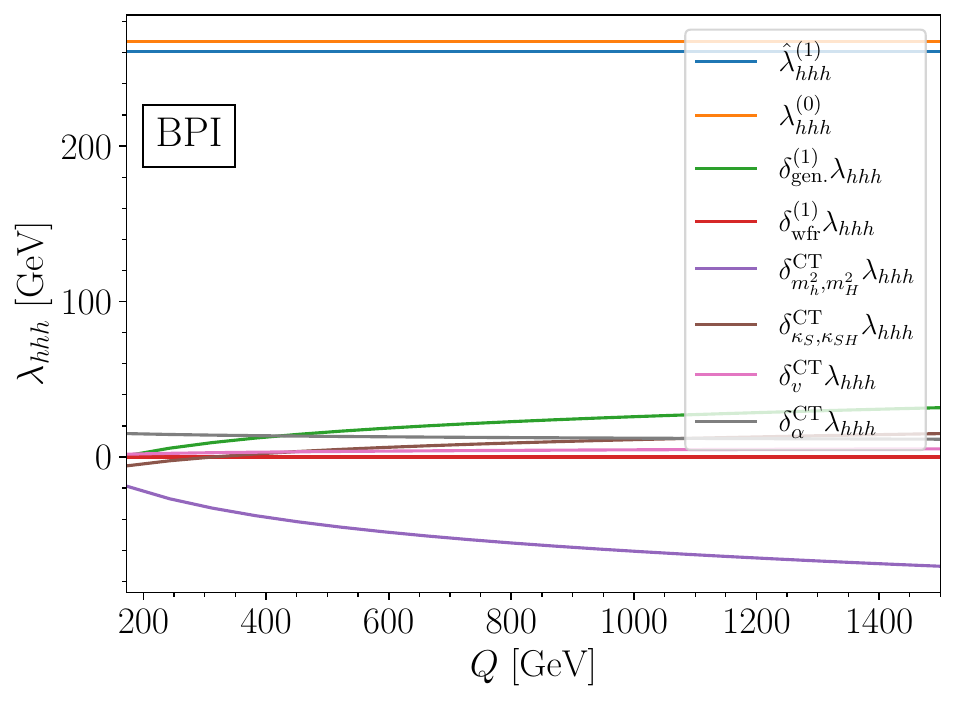}
    \includegraphics[width=0.497\linewidth]{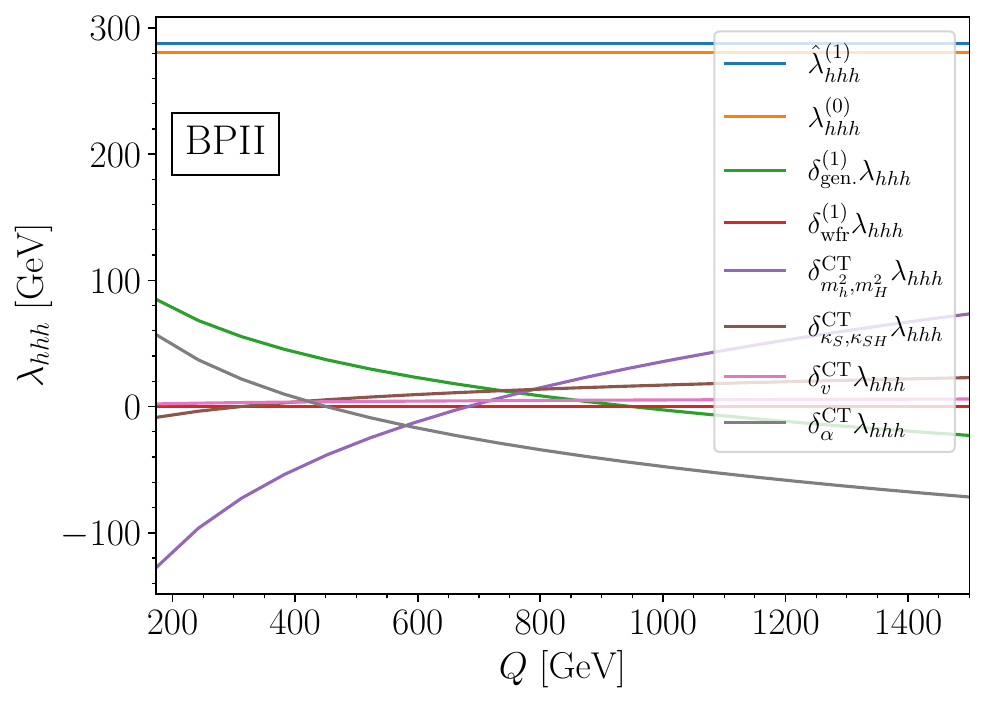}
    \includegraphics[width=0.485\linewidth]{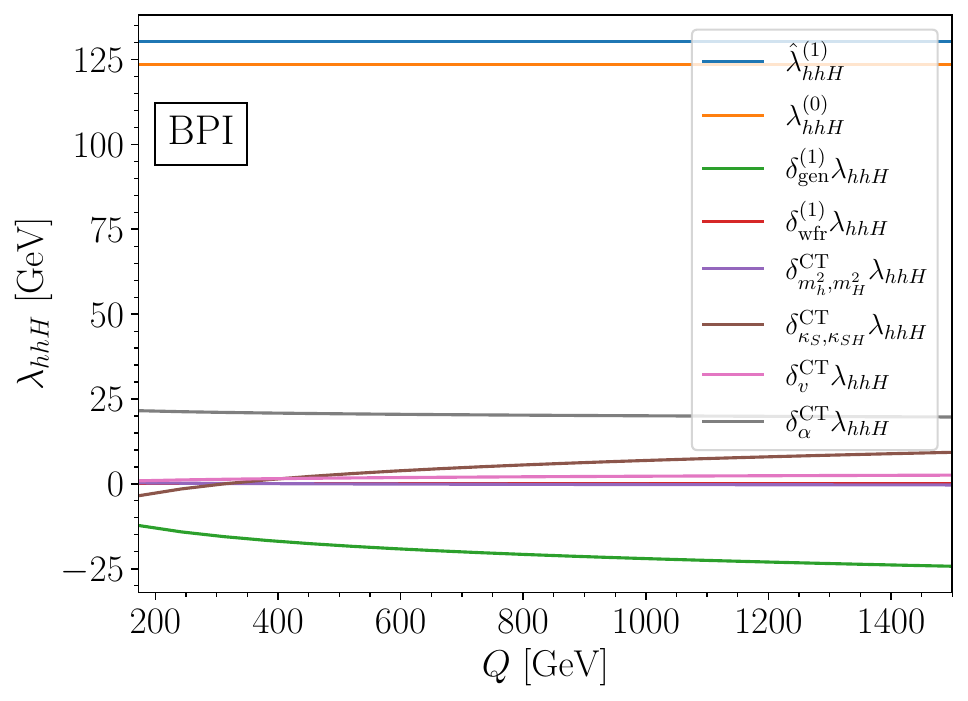}
    \includegraphics[width=0.495\linewidth]{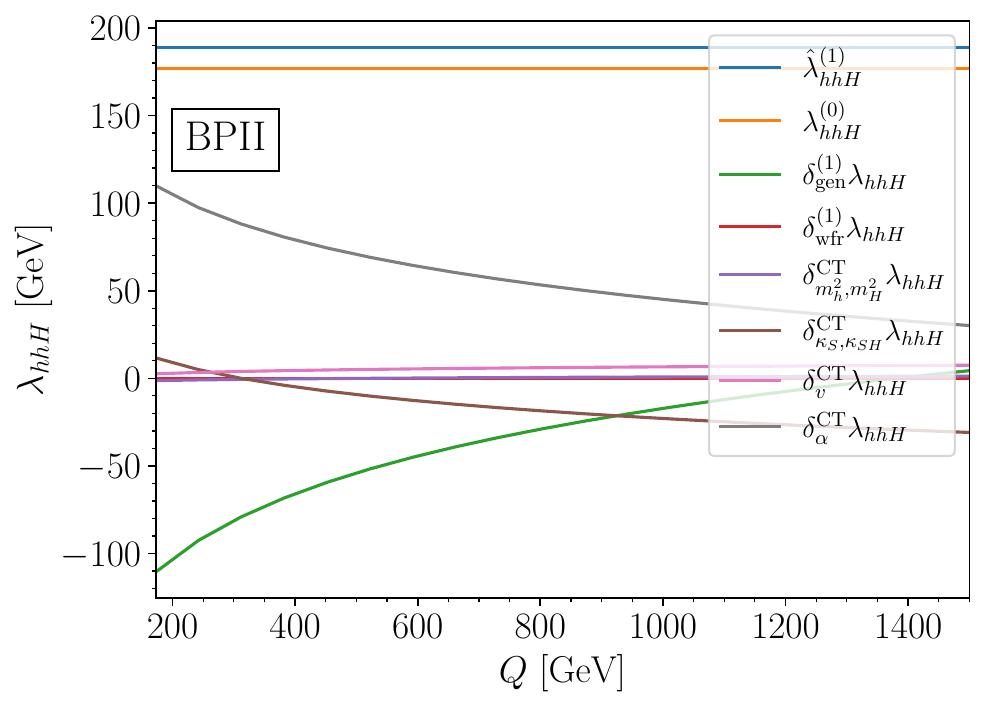}
    \caption{
   Individual contributions --- see \cref{one_diagrams} --- and total one-loop results for $\lambda_{hhh}$ and $\lambda_{hhH}$, using the OS renormalisation scheme defined in \cref{sec:ren}. 
   \textit{Upper row}: results for BP1 from \cref{points1}; \textit{bottom row}: results for BP2 from \cref{points1}; \textit{left column}: results for $\lambda_{hhh}$; \textit{right column}: results for $\lambda_{hhH}$. 
    The colour coding is as follows: the blue line is the full one-loop result $\rlaijk{1}$, the orange line is the tree-level contribution $\lambda^{(0)}_{ijk}$, the green curve corresponds to the one-loop 1PI contributions $\delta^{(1)}_\text{gen}\laijk$, the red curve represents the external-leg corrections $\delta^{(1)}_\text{wfr}\laijk$, the purple curve is the contribution from the mass renormalisation $\delta^\text{CT}_{m_h^2,m_H^2}\laijk$, the brown curve is the contribution from the renormalisation of $\kappa_S$ and $\kappa_{SH}$ $\delta^\text{CT}_{\kappa_S,\kappa_{SH}}\laijk$, the pink curve shows the contribution from the renormalisation of the EW VEV $\delta^\text{CT}_{v}\laijk$, and finally the grey curve is the contribution from the $\alpha$ counterterm $\delta^\text{CT}_\alpha\lambda_{ijk}$. 
   }
    \label{checkq}
\end{figure}

We have computed the dependence of each of the contributions entering the calculation of both couplings, as well as their sum, on the renormalisation scale $Q$, in the range $Q\in[m_t,1500 \gev]$. 
The results are shown in \cref{checkq}, where the blue lines show
our full one-loop results for $\lambda_{hhh}$ and $\lambda_{hhH}$ (for the other colors: see caption). One can observe that they are independent of $Q$, as expected. (Numerically, this holds up to \texttt{Python}'s working accuracy.)

\clearpage
\newpage


\section{One-loop corrections to \boldmath{$\laijk$}}

\label{sec:one}

In this section, we investigate the numerical results for the trilinear Higgs couplings $\lahhh$ and $\lahhH$ that can be obtained at one loop with the OS renormalisation scheme defined in the previous section. The scalar sector of the RxSM contains five free parameters, namely $m_H$, $\alpha$ $v_S$, $\kappa_S$, and $\kappa_{SH}$. To explore this five-dimensional 
parameter space, we perform parameter scans employing  
the following scan ranges
\begin{align}
\label{eq:paramscan_ranges}
    m_H&\in[260,1000]~\mathrm{GeV},\nonumber\\
    \cos\alpha&\in[0.95,1],\nonumber\\
    v_S&\in[1,800]~\mathrm{GeV},\nonumber\\
    \kappa_S&\in[-1000,1000]~\mathrm{GeV},\nonumber\\
    \kappa_{SH}&\in[-1000,0]~\mathrm{GeV}.
\end{align}
We note that we choose as lower bound for the heavy Higgs mass $m_H>260$~GeV in order to allow kinematically the decay channel of a heavy Higgs boson into two light ones, i.e.\ $H\rightarrow hh$. We do not start right at the threshold of this decay to avoid numerical instabilities in the calculation of the di-Higgs cross-section (see \cref{sec:dihiggs}). The requirement of $\kappa_{SH}$ to be negative comes from the theoretical constraint of boundedness-from-below of the potential, as discussed in \cref{sec:model}. 

\subsection{Corrections to \boldmath{$\kala$}}

\begin{figure}[htb!]
    \centering
    \includegraphics[width=0.49\linewidth]{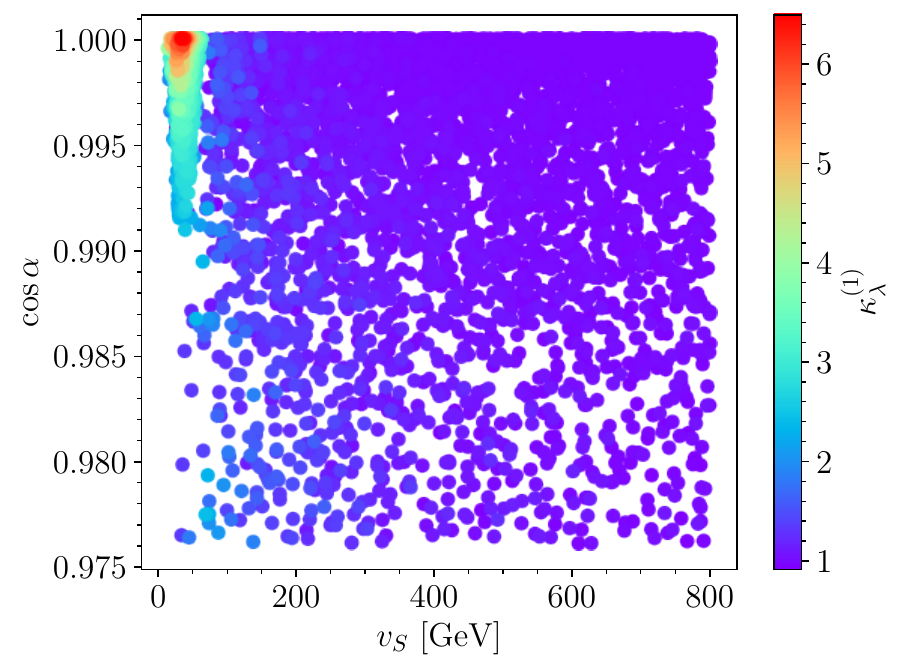}
    \includegraphics[width=0.49\linewidth]{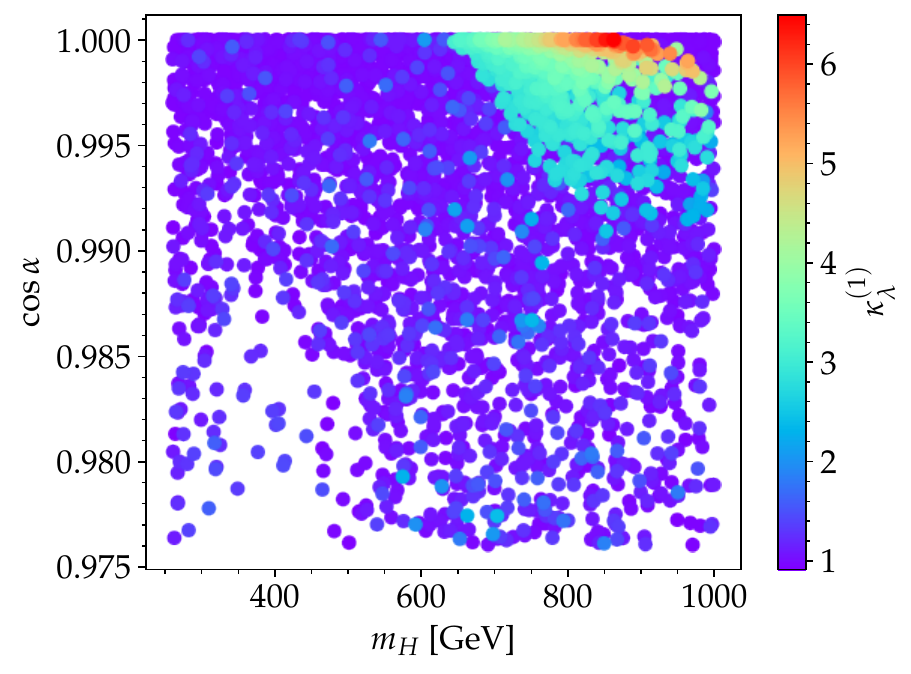}
    \caption{Parameter scan results in the RxSM, using the ranges given in \cref{eq:paramscan_ranges}. The colour coding 
     indicates $\kala^{(1)}$.
    \textit{Left}: results in the $\{\cos \alpha,v_S\}$ plane; \textit{right}: results in the $\{\cos \alpha,m_H\}$ plane. }
    \label{hhh}
\end{figure}

We begin with the case of the trilinear Higgs coupling of the detected Higgs boson, $\lambda_{hhh}$. We present our results in terms of the coupling modifier $\kala$, which is defined as the value of the coupling divided by its prediction in the SM at tree level,
\begin{equation}
    \kala = \frac{\lahhh}{\laSMz}\,.
\end{equation}
In \cref{hhh}, we show as colour coding the one-loop predictions of $\kala$ (which we denote $\kala^{(1)}$ to indicate the loop order) for our scan points. These results are presented in the $\{\cos \alpha,v_S\}$ plane (left panel) as well as in the $\{\cos \alpha,m_H\}$ plane (right panel). At first, 
we find that for most of the parameter space of the RxSM the one-loop corrected values of $\kala$ are positive with an overall range of 
$\kappa^{(1)}_{\lambda}\in[0.91,6.5]$, 
which is still fully allowed by the latest experimental constraints on $\kala$~\cite{ATLAS:2024ish,CMS:2024awa}.  We also observe that the large values of $\kala$ (red and orange points) are correlated with small values of the singlet VEV, $v_S$, and large values of the heavy Higgs mass $m_H$, as well as with values of $\cos\alpha$ close to 1 (i.e.\ near the alignment limit). The reason being that in these regions of the parameter space the one-loop corrections are enhanced due to large quartic coupling $\lambda_{SH}$. From \cref{eq:rep_lambdas_and_masses} one can see that $\lambda_{SH}$ is inversely proportional to $v_S$ and at the same time grows with $m_H$. 

\begin{figure}[htb!]
    \centering
    \includegraphics[width=0.49\linewidth]{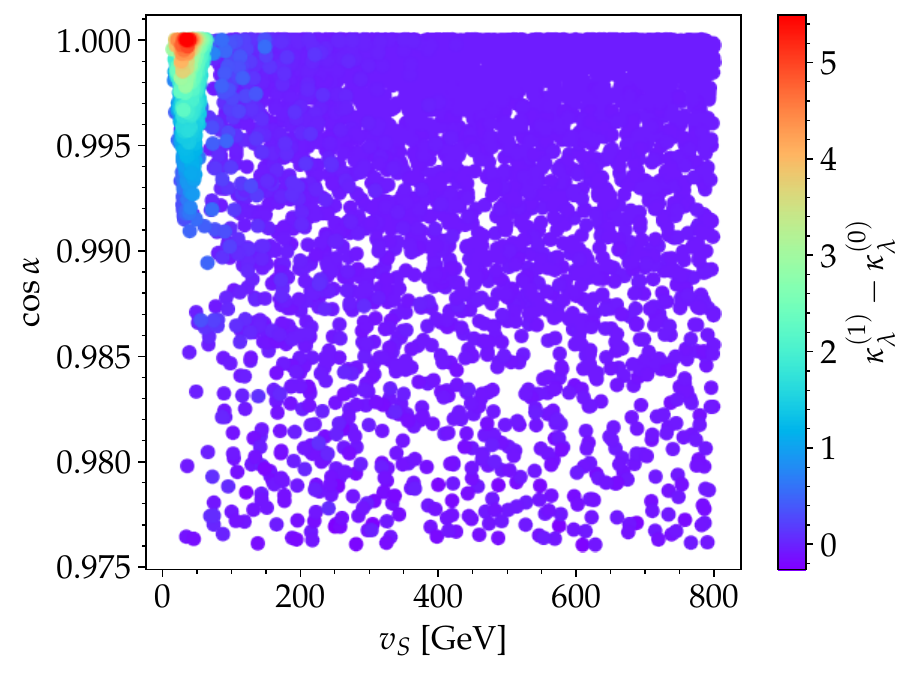}
    \includegraphics[width=0.49\linewidth]{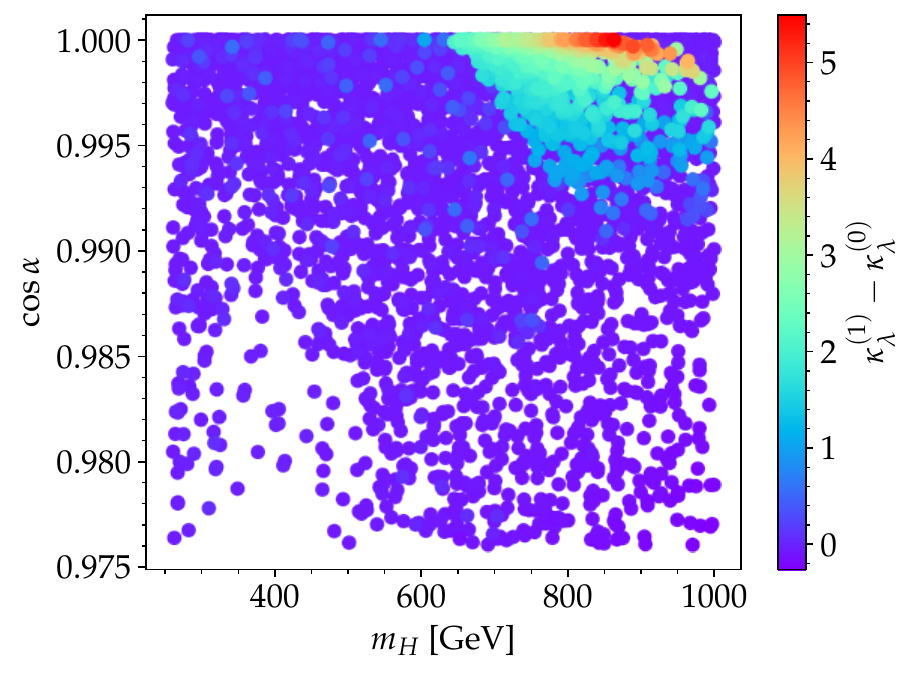}
    \caption{Difference between the one-loop and tree-level values of $\kala$, for the same RxSM parameter scan points as in \cref{hhh}. \textit{Left}: results in the $\{\cos \alpha,v_S\}$ plane; \textit{right}: results in the $\{\cos \alpha,m_H\}$ plane. }
    \label{difhhh}
\end{figure}

Next, in order to understand the origin of BSM deviations in $\kala$, between tree-level effects or one-loop corrections, we investigate separately the size of the one-loop corrections to $\kala$. 
We therefore show in \cref{difhhh} the difference between the one-loop and tree-level values of $\kala$, i.e.\ $\kala^{(1)} - \kala^{(0)}$.
The results are shown, as above, in the two planes: $\{\cos \alpha,v_S\}$ (left panel) and $\{\cos \alpha,m_H\}$ (right panel). 
We find the possible range of $\kala^{(0)} \in[0.97,2.3]$ and $\kala^{(1)} - \kala^{(0)} \in[-0.26, 5.49]$, where
large positive values of the latter are strongly correlated with the values of $\kala^{(1)}$ shown in \cref{hhh}. This indicates that, while BSM deviations are possible already at the tree level, the largest BSM deviations arise from loop corrections.  
By comparing \cref{hhh} and \cref{difhhh}, we find that for the points with the largest deviations from the SM, i.e.\ $\kala^{(1)}\in[3,6.5]$, found in the region with small $v_S$, large $m_H$ and closer to the alignment limit, the deviation is predominantly due to loop corrections, while the tree-level value is close to the SM, i.e.\ $\kala^{(0)} \approx 1$. On the other hand for those points with smaller BSM deviations, i.e.\ $\kala^{(1)} \in [0.91,3]$ (light blue points in \cref{hhh}), which are also further away from the alignment limit $\cos \alpha\in[0.975,0.990]$, the largest part of the deviation arises already at the tree-level.

\subsection{Corrections to \boldmath{$\lahhH$}}

\begin{figure}[htb!]
    \centering
    \includegraphics[width=0.49\linewidth]{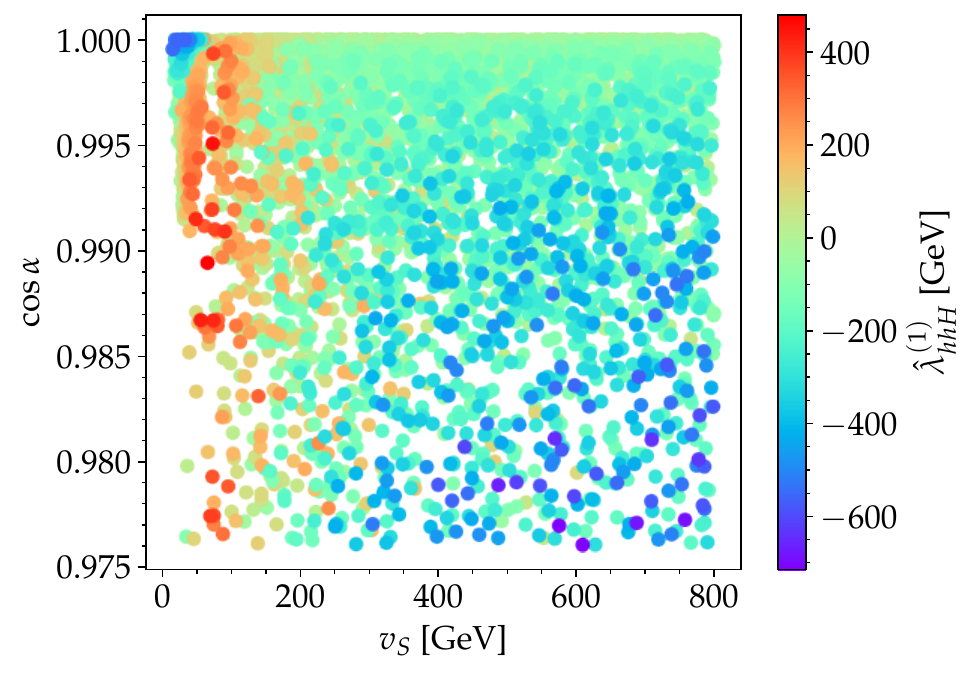}
    \includegraphics[width=0.49\linewidth]{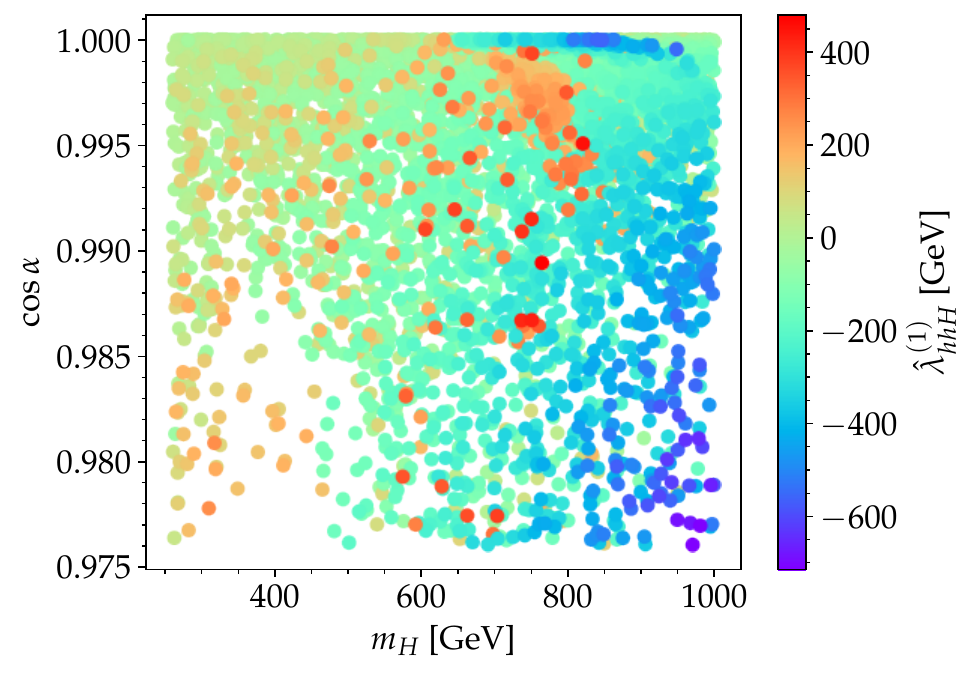}
    \caption{Results for $\rlahhH1$ of our RxSM parameter scan.
    \textit{Left}: results in the $\{\cos \alpha,v_S\}$ plane; \textit{right}: results in the $\{\cos \alpha,m_H\}$ plane.}
    \label{hhH}
\end{figure}

We turn in this section to the BSM trilinear Higgs coupling 
involved in di-Higgs production, $\lahhH$. 
In \cref{hhH} we present as colour coding the one-loop values of $\rlahhH1$ for the RxSM scan points, in the $\{\cos \alpha,v_S\}$ plane (left) and in the $\{\cos \alpha,m_H\}$ plane (right). The two main differences compared to the case of $\lambda_{hhh}$ is the existence of points with large negative 
predictions, reaching even larger absolute values than the points with positive predictions. 
For points close to the alignment limit, i.e.\ $\cos \alpha\in[0.995,1]$, 
large negative values of $\hat\lambda_{hhH}^{(1)}$ only occur for points with small\footnote{While it is known that the standard tadpole scheme can suffer from numerical instabilities stemming from terms $\sim t_S/v_S$ (see e.g.\ \cref{eq:rep_lambdas_and_masses}), as pointed out in Ref.~\cite{Braathen:2021fyq}, we have checked that no such instability occurs for the points considered in this study.} singlet VEVs, $v_S\lesssim 20 \gev$ and heavy Higgs masses, $m_H \gtrsim 750 \text{ GeV}$. 
On the other hand, for points further from the alignment limit, i.e.\ $\cos \alpha\in[0.975,0.995]$, significant positive (red points) and negative (blue points) values of $\rlahhH1$ are possible, depending on the values of $v_S$ and $m_H$. Smaller (larger) values of both $v_S$ and $m_H$ correlate with positive (negative) $\rlahhH1$. 

\begin{figure}[htb!]
    \centering
    \includegraphics[width=0.49\linewidth]{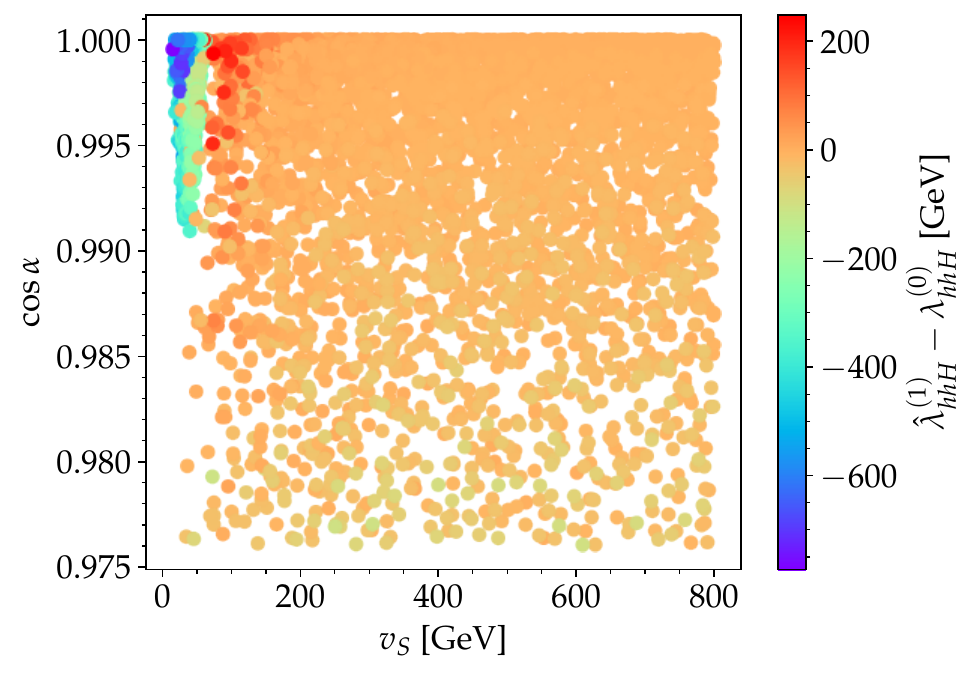}
    \includegraphics[width=0.49\linewidth]{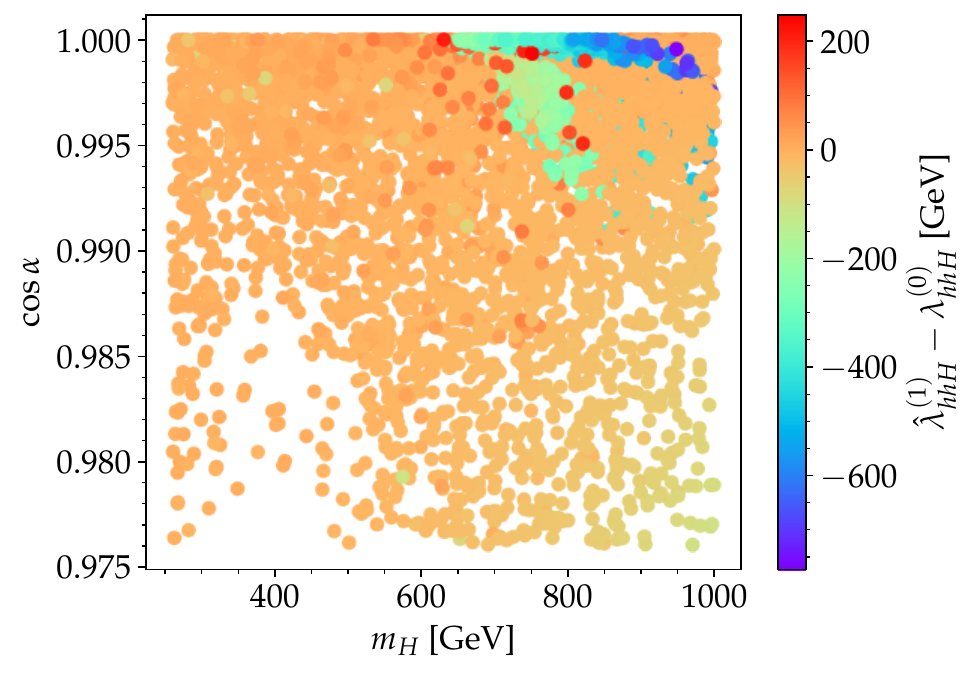}
    \caption{Difference between the one-loop and tree-level values of $\lahhH$ for our RxSM parameter scan.
    \textit{Left}: results in the $\{\cos \alpha,v_S\}$ plane; \textit{right}: results in the $\{\cos \alpha,m_H\}$ plane.}
    \label{difhhH}
\end{figure}

As in the previous section, it is important to understand whether this behaviour stems from tree- or loop-level effects. 
In \cref{difhhH} we present the difference between the one-loop and tree-level values of $\lahhH$ for our scan points in the same two planes as the previous figures.  
It should be kept in mind that $\lahhH^{(0)} = 0$ in the alignment limit. 
One can observe that in the region of parameter space closer to the alignment limit and for small values of $v_S$ and large values of $m_H$, the large results for $\rlahhH1$ (in absolute values) arise for the most part from radiative corrections driven by 
large values of $\lambda_{SH}$. Inversely, for the points farther from the alignment limit as well as for $v_S\gtrsim 200 \gev$, the tree-level contribution to $\lahhH$ dominates and loop effects are moderate.

As a general conclusion, we find that the largest loop corrections occur for points close to the alignment limit, i.e.\ $\cos \alpha\in[0.99,1]$.

\clearpage

\section{Predictions for di-Higgs production}
\label{sec:dihiggs}

In this section we turn to the investigation of the phenomenological implications for di-Higgs production of the potentially large loop corrections 
in the trilinear Higgs couplings found  
in the previous section. For this purpose, 
we consider two possible collider settings, namely the HL-LHC and a $e^+e^-$ collider with centre-of-mass energy of $1 \tev$.  
As a concrete example, we use the specifics for the ILC~\cite{Bambade:2019fyw}.


\subsection{HL-LHC}

\begin{figure}
    \centering
    \includegraphics[width=\linewidth]{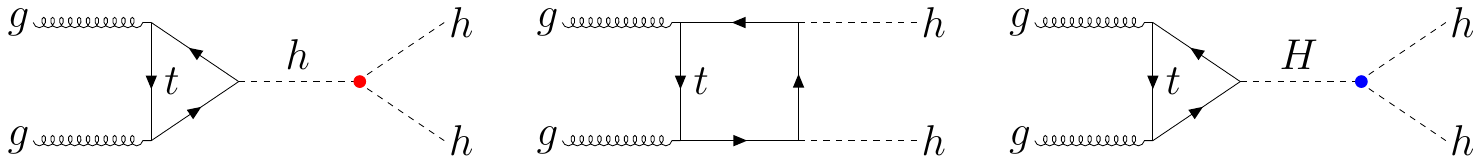}
    \caption{Leading-order diagrams contributing to the di-Higgs production process at the (HL-)LHC.The red and blue dots indicate the trilinear scalar couplings $\lahhh$ and $\lahhH$, respectively.}
    \label{fig:dihiggsLOdiags}
\end{figure}

To compute the di-Higgs production at the HL-LHC, we employ a modified version of \texttt{HPAIR} \cite{Abouabid:2021yvw,Arco:2022lai,Dawson:1998py,Nhung:2013lpa,Grober:2017gut,Grober:2015cwa}, which was already used in Ref.~\cite{Arco:2025nii}. In this code, the three leading-order (LO) diagrams contributing to the $gg\to hh$ process, shown in \cref{fig:dihiggsLOdiags}, as well as the 
interference between them, are taken into account. The code also computes the NLO QCD corrections for the total cross-section, but not for the differential distributions. In order to consistently provide results for the total cross-section and the differential distributions at the same order, we present results for both at LO in QCD only. However, one should remember that these predictions are modified by a QCD $K$ factor close to 2. 
\texttt{HPAIR} takes as inputs the trilinear Higgs couplings $\lahhh$ and $\lahhH$, so that by providing one-loop corrected versions of these two couplings, we obtain a prediction for the di-Higgs cross-section (and differential distributions) including leading NLO BSM contributions.  
When comparing results in the RxSM with the SM, one should have in mind that the LO value for the total di-Higgs production cross-section in the SM is $\sihhSM = 19.76$~fb~\cite{Abouabid:2021yvw}.

\subsubsection{Total cross-section}

\begin{figure}[htb!]
    \centering
    \includegraphics[width=0.49\linewidth]{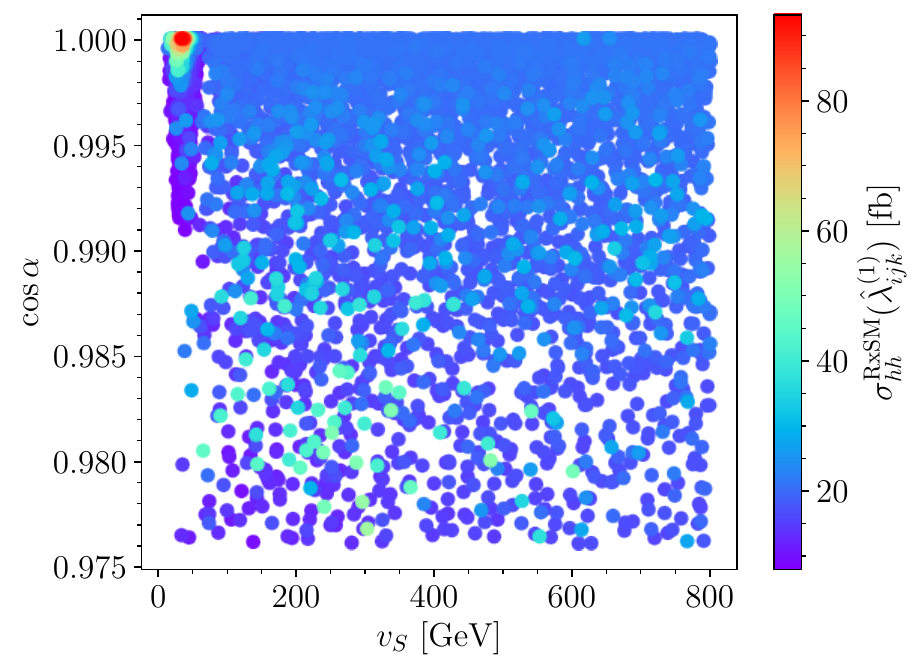}
    \includegraphics[width=0.49\linewidth]{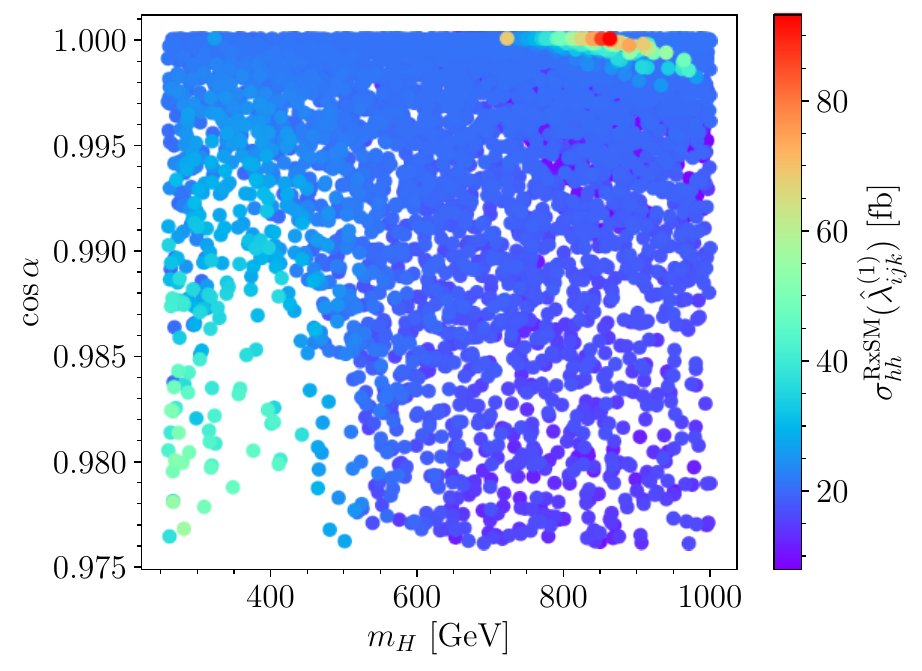}
    \caption{Predictions for $\sihh$ in our RxSM parameter scan. 
    \textit{Left}: results in the $\{\cos \alpha,v_S\}$ plane; \textit{right}: results in the $\{\cos \alpha,m_H\}$ plane. }
    \label{crossHL-LHC}
\end{figure}

Our predictions for the total di-Higgs production cross-section, including one-loop corrections to the trilinear Higgs couplings, for the RxSM scan points are shown in the colour coding of \cref{crossHL-LHC}, 
for our two benchmark plane projections, the $\{\cos \alpha,v_S\}$ plane (left) and the $\{\cos \alpha,v_S\}$ plane (right). 
In \cref{crossHL-LHC}, one can see that there is a region, indicated by red points, exhibiting a large enhancement of the cross-section $\sihh \approx 4.5\,\sihhSM$. They are found for small values of $v_S\lesssim 50 \gev$, large values of $m_H\simeq 850 \gev$, and close to the alignment limit $\cos\alpha>0.99$. However, we can also observe for this same 
region points (in purple) with a decrease in the cross-section w.r.t.\ the SM, with values as low as 
$\sihh \approx 0.5\,\sihhSM$.
The region with significant deviations in the di-Higgs cross-section corresponds exactly to the region for which we
observed large deviations in $\kala$ in \cref{hhh}, which were due to loop contributions. The change in the behaviour of the cross-section,
encompassing increase and decrease w.r.t. $\sihhSM$ is well known from the SM with a free $\lahhh$ (see e.g.~\citeres{Baglio:2012np,LHCHiggsCrossSectionWorkingGroup:2016ypw} and references therein), due to the negative interference of 
the $h$-exchange contribution and the box diagram, see \cref{fig:dihiggsLOdiags}.
Values smaller than the SM are found for $1 \le \kala \lsim 3.5$, whereas larger values are found for $\kala \gsim 3.5$, 
with a minimum around $\kala \approx 2.5$.

In \cref{crossHL-LHC}, there are also light blue points indicating smaller deviations in the cross-section, reaching maximal values of $\sihh \approx 2.5\,\sihhSM$.
These points are further from the alignment limit, i.e.\ $\cos \alpha\in[0.975,0.99]$, and also spread over the entire range of $v_S$. However, they have a dependence on the heavy Higgs mass $m_H$, where most of these points are found for $m_H \lesssim 500 \gev$. 
The fact that these points depend only on $m_H$, but not on changes in the trilinear Higgs couplings, which vary substantially over this part of the parameter space, see \cref{hhh,difhhh,hhH,difhhH}, indicates that this effect originates from the $s$-channel heavy Higgs-boson exchange already at the tree-level.

\begin{figure}[htb!]
    \centering
    \includegraphics[width=0.49\linewidth]{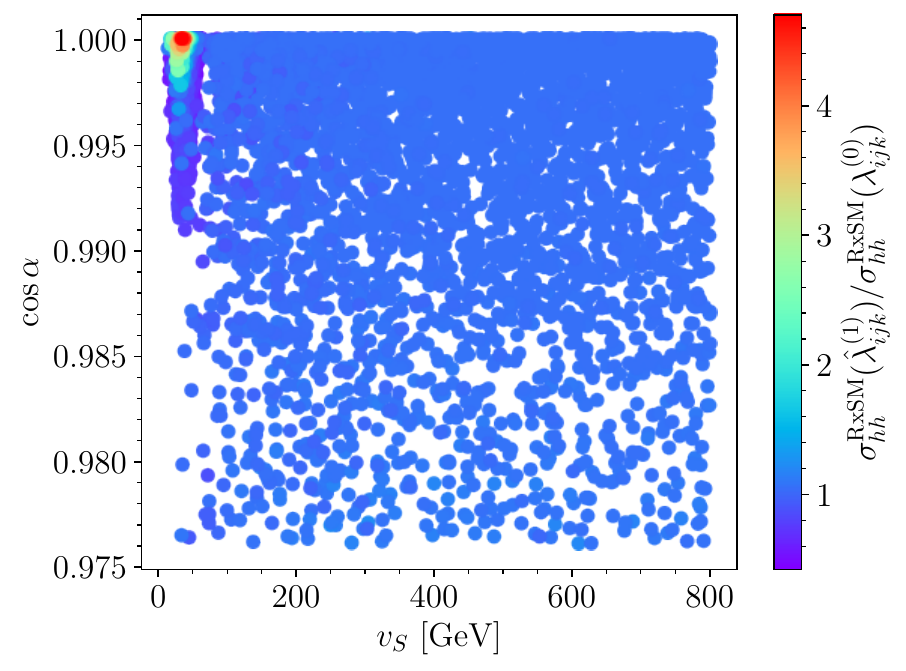}
    \includegraphics[width=0.49\linewidth]{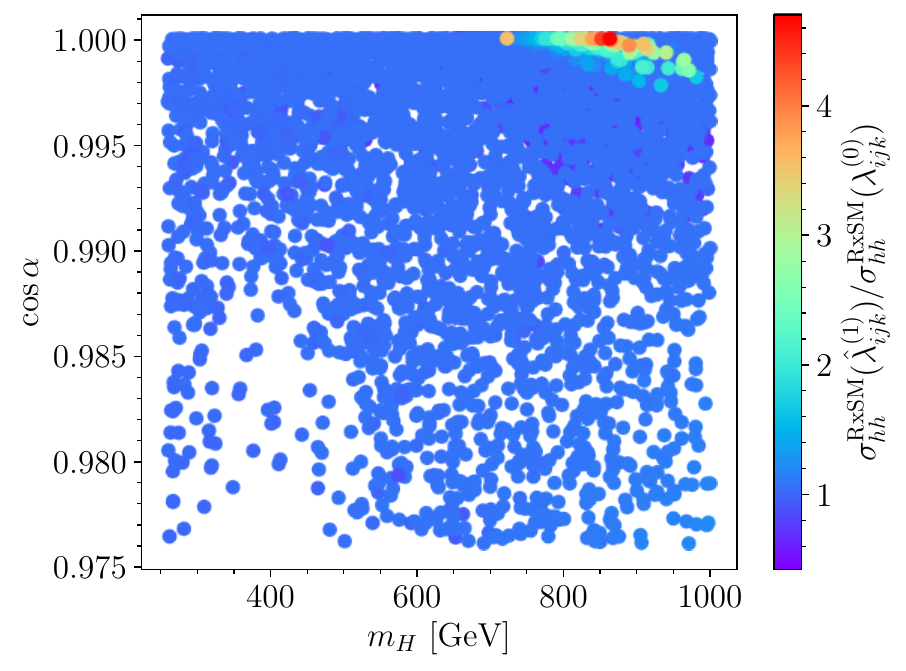}
    \caption{The ratio between the total cross-section for di-Higgs production at the (HL-)LHC including one-loop corrections to the trilinear Higgs couplings and using the tree-level value for the points from our parameter scan in the RxSM. \textit{Left}: Results in the $\{\cos \alpha,v_S\}$ plane; \textit{right}: results in the $\{\cos \alpha,m_H\}$ plane.}
    \label{ratiocrossHL-LHC}
\end{figure}

In \cref{ratiocrossHL-LHC} we present for our RxSM scan points the ratio between the total di-Higgs production cross-section including one-loop corrections to the trilinear Higgs couplings, which we denote $\sihh(\rlaijk{1})$, and the same cross-section using the tree-level values, $\sihh(\laijk^{(0)})$. As for the previous figures, the value of the ratio is given by the colour coding of the points and shown for 
the $\{\cos \alpha,v_S\}$ plane (left panel) and  
the $\{\cos \alpha,m_H\}$ plane (right panel). From \cref{crossHL-LHC} we concluded that the large deviations from the SM in the cross-section close to the alignment limit is a loop-induced effect and this is indeed confirmed by \cref{ratiocrossHL-LHC} as the same pattern of deviations can be observed in the ratio $\sihh(\rlaijk{1})/\sihh(\laijk^{(0)})$. 
For the other region where we observe deviations from the SM, further from the alignment limit ($\cos \alpha\in[0.975,0.99]$), we concluded that this is a tree-level effect coming from the propagator of the heavy Higgs boson, and \cref{ratiocrossHL-LHC} also supports this conclusion, as the ratio $\sihh(\rlaijk{1})/\sihh(\laijk^{(0)})$ remains close to 1 for these points.

 \subsubsection{Differential cross-section}
\label{sec:mhh-HLLHC}

\begin{table}[ht]
\centering
\begin{tabular}{cccccccccc}
\toprule
BP  & $m_H$ & $\cos\alpha$ & $v_S$ & $\kappa_S$ & $\kappa_{SH}$ & $\kappa_{\lambda}^{(0)}$ & $\kappa_{\lambda}^{(1)}$ & $\lambda_{hhH}^{(0)}$ & $\hat\lambda_{hhH}^{(1)}$ \\
&  [GeV] &  & [GeV] & [GeV] & [GeV] & & & [GeV] & [GeV] \\ \midrule
1 & 659.4     & 0.9999        & 56.1      & $-880$           & $-880$              & 1.01                      & 3.02                      & 96.3                      & $-119.7$                    \\
2 & 777.6     & 1.0000          & 44.7      & $-931$           & $-931$              & 1.00                      & 4.87                      & 2.2                       & $-303.6$                    \\
3 & 594.5     & 0.9987         & 110.7     & 146            & $-931$              & 1.08                      & 1.41                      & 127.6                     & 181.9                     \\
4 & 891.1     & 0.9957         & 37.0      & $-993$           & $-993$              & 1.62                      & 3.28                      & 543.2                     & 131.2                     \\
5 & 411.4     & 0.9878         & 92.6      & $-605$           & $-380$              & 1.29                      & 1.31                      & 153.9                     & 165.1                     \\
6 & 625.6     & 0.9976         & 96.4      &$ -851$           & $-826$              & 1.14                      & 1.46                      & 163.5                     & 263.8                     \\ \bottomrule
\end{tabular}
\caption{Definitions of the RxSM benchmark points for the study of di-Higgs production, in terms of the five free BSM parameters of the model ($m_H,\ \cos\alpha,\ v_S,\ \kappa_S$ and $\kappa_{SH}$). Additionally, for each benchmark point, tree-level and one-loop predictions for the trilinear self-coupling modifier $\kappa_\lambda$ as well as for the BSM trilinear coupling $\lahHH$ are included. We note that the numbers in this table are rounded; the actual input values to 10 digits, necessary to reproduce the quoted results for the trilinear scalar couplings, are provided in the ancillary file \texttt{BPs.csv}. }
\label{bps}
\end{table}
 
The objective of this section is to investigate the impact of the loop corrections to the trilinear Higgs couplings on the differential 
di-Higgs production cross-section. For this analysis,  we have selected six different benchmark points, defined in \cref{bps},
corresponding to different phenomenological scenarios. We begin by calculating with \texttt{HPAIR} theoretical predictions for the di-Higgs invariant mass distributions for the process $gg \to hh$, comparing the results using tree-level or one-loop corrected trilinear 
couplings. 
In a second step, we determine for these benchmark points whether the di-Higgs process could be observed at the HL-LHC, taking into account different experimental uncertainties. In order to obtain quantitative results, incorporating also the possibility of statistical fluctuations in experimental signals, we define a statistical significance for the RxSM deviation from the expected SM result. 
In order to include experimental effects in our analysis, we start by multiplying our cross-section results by the luminosity $\mathcal{L}$ to compute the number of events. We use here the value of $\mathcal{L}=6\iab$, obtained by combining the luminosities expected to be collected by ATLAS and CMS at the end of the HL-LHC. 
Next, following the procedure in \citere{Frank:2025zmj}, we take into account the decays of the Higgs bosons, 
choosing the channel with the largest branching ratio, $h \to b\bar b$. We furthermore take into account the corresponding detector efficiencies,
following \citere{ATLAS:2022hwc}. The latter are given by the product of $\epsilon=\epsilon_{\mathrm{TOT}}\epsilon_{\mathrm{SR}}$, where 
$\epsilon_{\mathrm{TOT}}$ is the efficiency factor of the pre-selection of the events, and $\epsilon_{\mathrm{SR}}$ is the efficiency of the 
reconstruction of the $b\bar{b}$ pairs. The total number of events is then given by, 
\begin{equation}
    N = \sigma(gg\to hh) \times\mathcal{L}\times\big(\mathrm{BR}(h\to b\bar{b})\big)^2
    \times\epsilon_{\mathrm{TOT}}\times\epsilon_{\mathrm{SR}} \,.\label{nevents}
\end{equation}
Additionally, we have to take into account the smearing and binning of the distributions, which estimate,
respectively, the experimental error in the measurement of the four-bottom invariant mass $m_{b\bar{b}b\bar{b}}$ and the finite resolution in $m_{b\bar{b}b\bar{b}}$ of the detector (see e.g.\ Refs.~\cite{Arco:2022lai, Arco:2025nii}). We employ, as was done in Refs.~\cite{Arco:2022lai, Arco:2025nii}, experimentally motivated values of 15\% for the smearing and 50 GeV for the binning.  

One aim of this analysis is to investigate whether there is a realistic chance to distinguish experimentally the di-Higgs distributions in the RxSM, using either tree-level or one-loop trilinear scalar couplings, from the SM. In addition to the experimental effects discussed above, we also include here the statistical uncertainty. For the discrimination of the RxSM (the considered hypothesis to be tested) from the SM (the null hypothesis), we define a statistical significance following \citere{Cowan:2010js}. 
We define our signal and background event numbers for each bin (labelled by the index $i$) respectively as
\begin{align}
        s_i&=  N_i^{\mathrm{RxSM}}-N_i^{\mathrm{SM}},\nonumber\\
        b_i &=N_i^{\mathrm{SM}}\,,
\end{align}
where $N_i^{\mathrm{RxSM (SM)}}$  is the number of events in the RxSM (SM) in bin $i$. Assuming Poisson statistics for the distributions, we can use a likelihood ratio method to define a statistical significance \cite{Cowan:2010js}
\begin{equation}
    Z=\sqrt{\sum_i2\left[(s_i+b_i)\log\left(1+\frac{s_i}{b_i}\right)-s_i\right]},\label{z}
\end{equation}
summing over all the bins of the distribution. 
The results for the significances for the RxSM distributions calculated with tree-level and one-loop trilinear scalar couplings are denoted as $Z^{(0)}$ and $Z^{(1)}$, respectively.

\begin{figure}[ht!]
    \centering
    \includegraphics[width=0.99\linewidth]{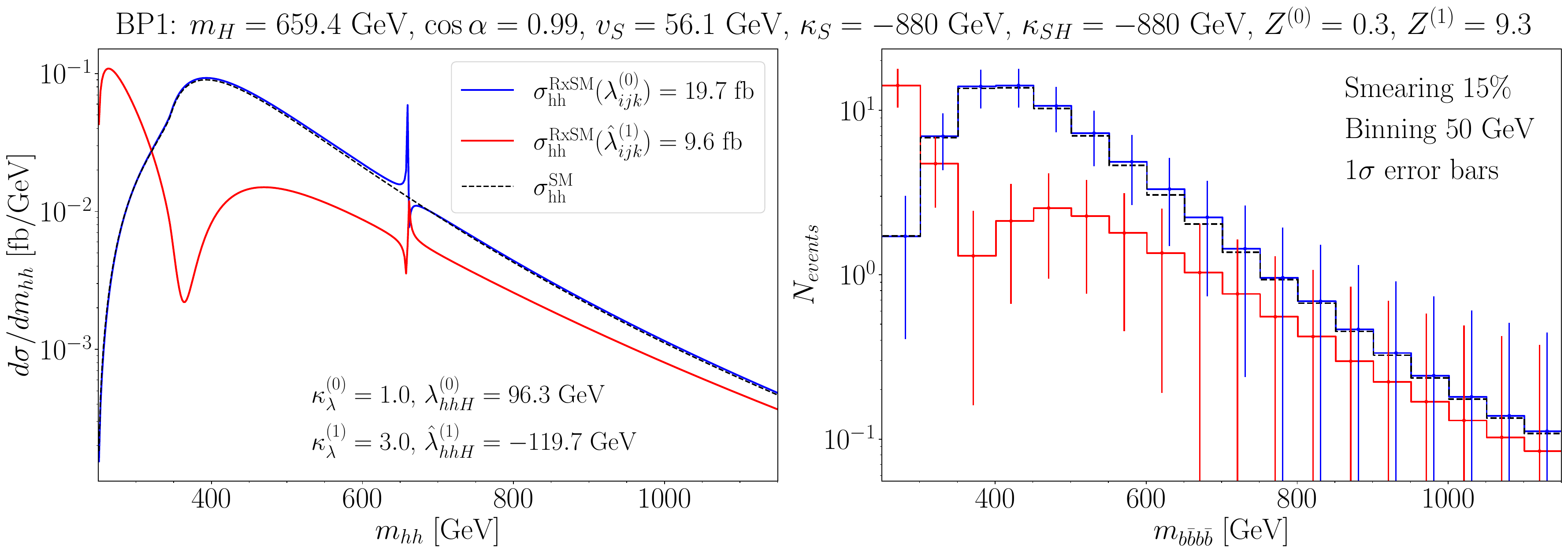}\\[1em]
    \includegraphics[width=0.99\linewidth]{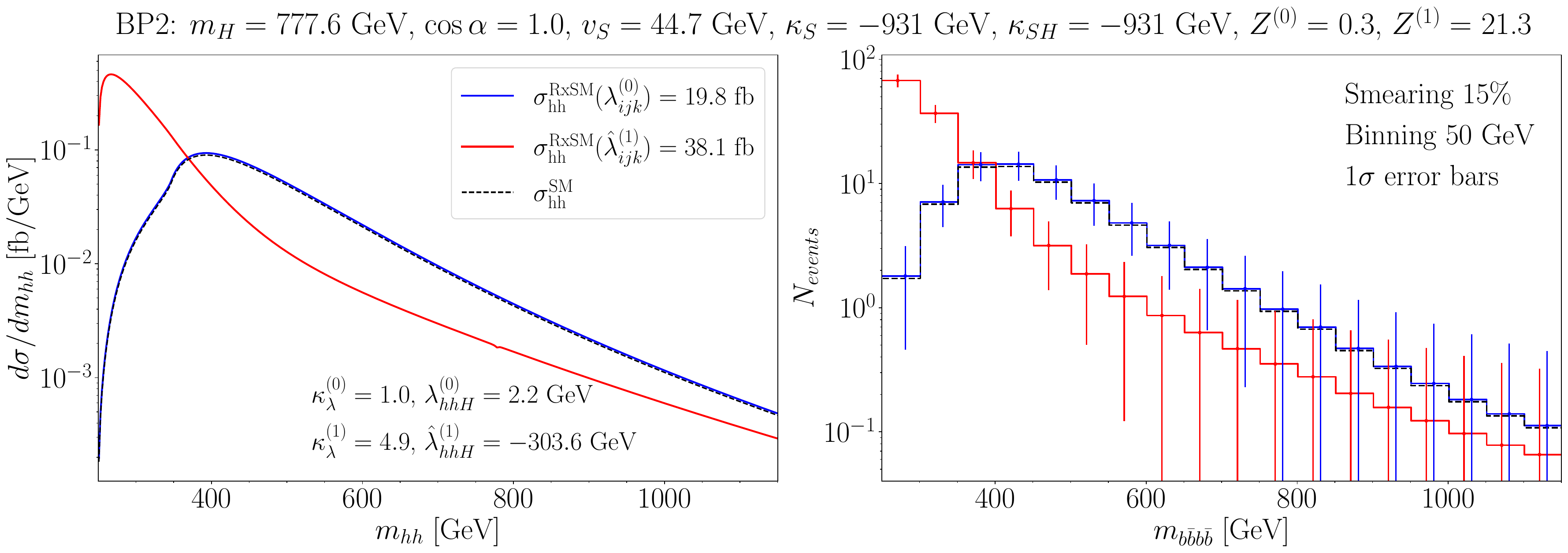}\\[1em]
    \includegraphics[width=0.99\linewidth]{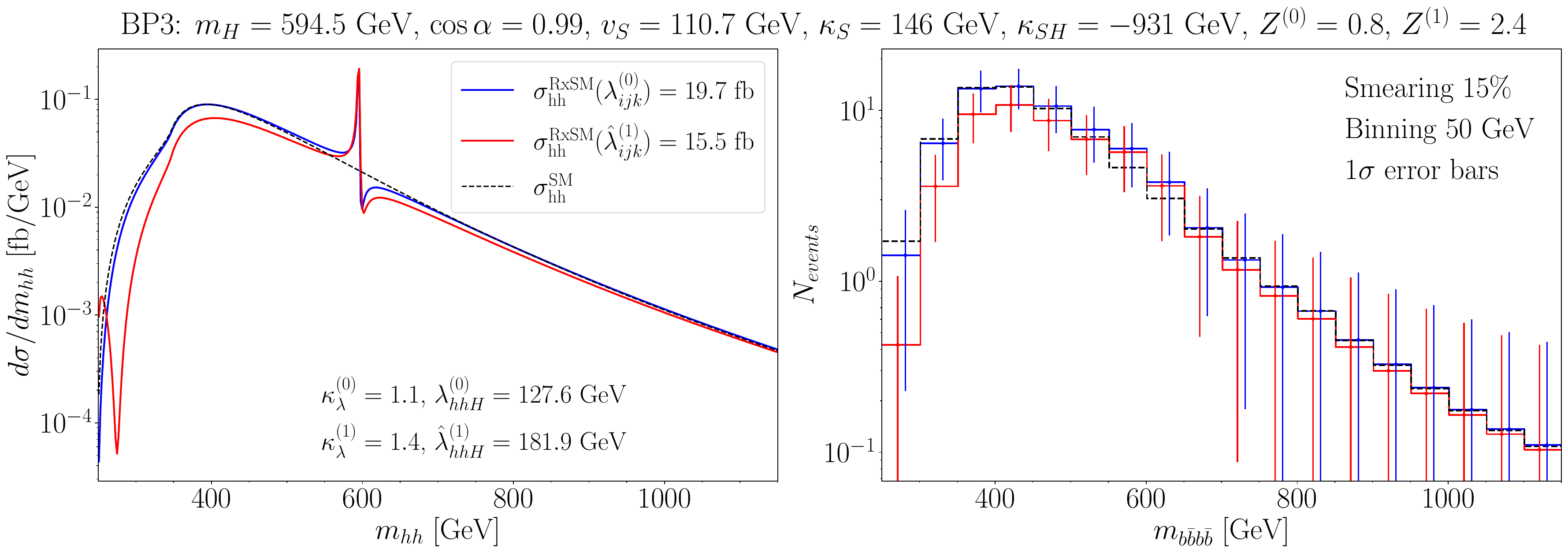}
    \caption{
Differential $\mhh$ di-Higgs production cross-section distributions;
blue (red) curves show results using the tree-level (one-loop corrected) trilinear Higgs couplings; the black dashed line shows the
corresponding SM result.
\textit{Left:} Distribution without experimental uncertainties;
\textit{right:} distribution of the number of di-Higgs events in the $b\bar b$ final state, taking into account 
smearing, binning and the experimental efficiencies (see text). Error bars indicate 
Poisson statistical uncertainties in each bin.
\textit{Top:} Results for BP1; \textit{centre:} results for BP2; \textit{bottom:} results for BP3 from \cref{bps}.
    }
    \label{bp123}
\end{figure}

\begin{figure}[ht!]
    \centering
    \includegraphics[width=0.95\linewidth]{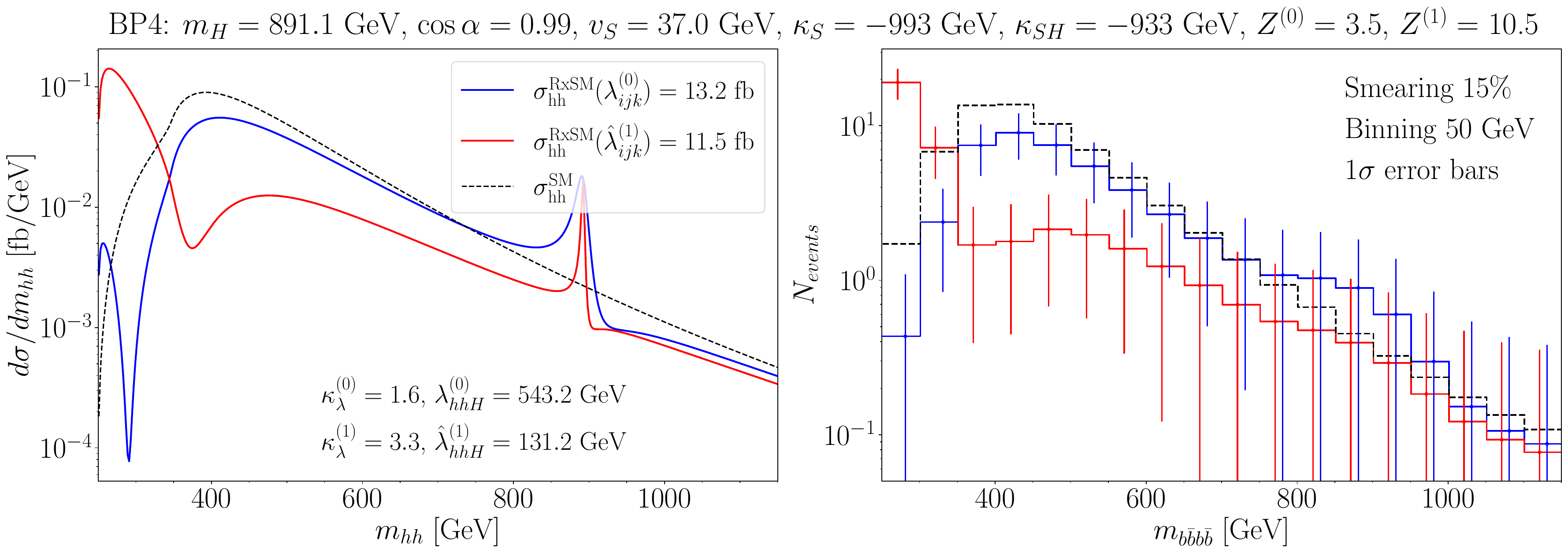}\\[1em]
    \includegraphics[width=0.95\linewidth]{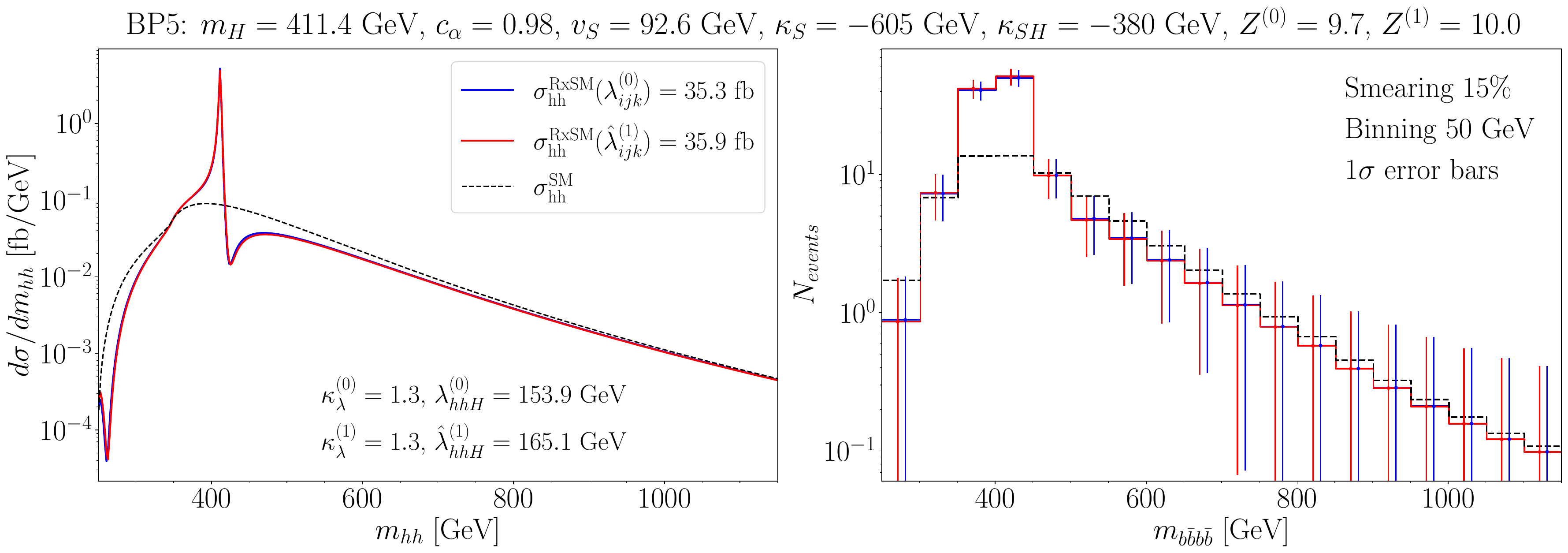}\\[1em]
    \includegraphics[width=0.95\linewidth]{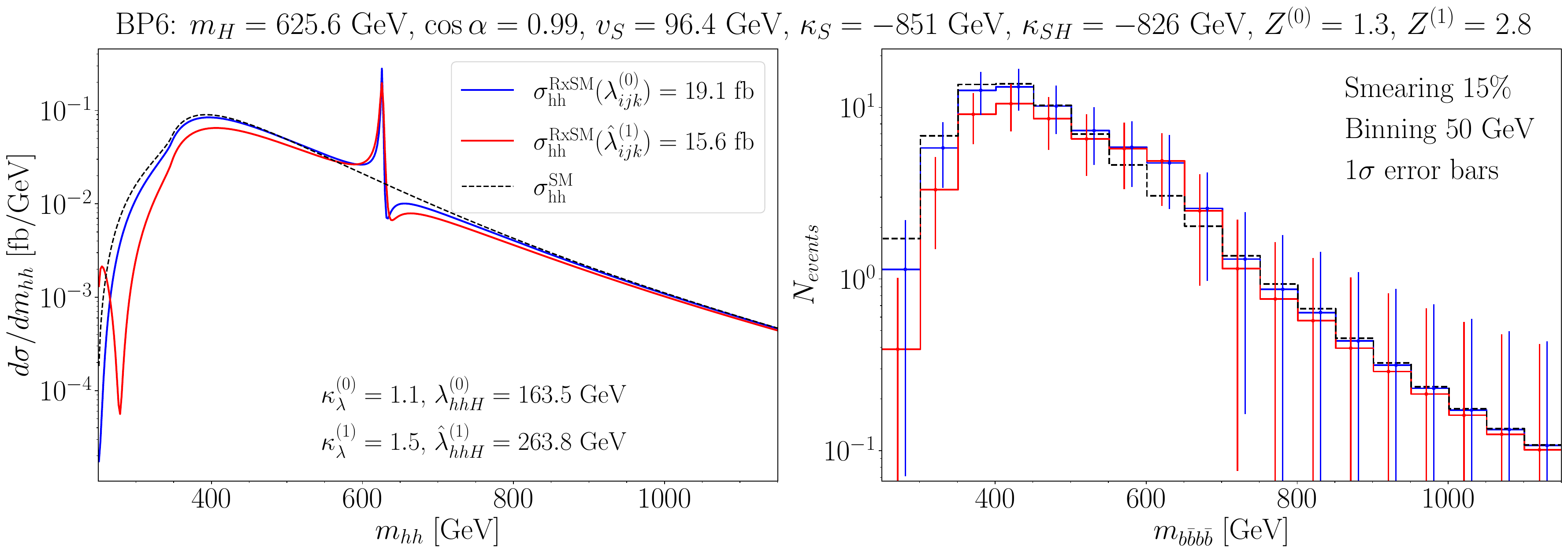}
    \caption{
    Differential $\mhh$ di-Higgs production cross-section distributions; line styles and colour coding as in 
    \protect\cref{bp123}.
\textit{Top:} Results for BP4; \textit{centre:} results for BP5; \textit{bottom:} results for BP6 from \cref{bps}.}
    \label{bp456}
\end{figure}

In \cref{bp123,bp456}, we show the results for the theoretical distributions and the distributions including the experimental effects discussed above for the six benchmark points of \cref{bps}. 
The blue (red) curves show results using the tree-level (one-loop corrected) trilinear Higgs couplings; the black dashed line shows the
corresponding SM result.\footnote{We note that for the SM we employ the tree-level value of $\kappa_\lambda=1$, as the use of an effective trilinear coupling is known not to be a good approximation in the SM~\cite{Muhlleitner:2022ijf}. The main reason for this is that momentum effects are significant for $\kappa_\lambda\simeq 1$, while they can be safely neglected for points with BSM deviations in $\kappa_\lambda$ (see e.g.\ the discussion in Ref.~\cite{Bahl:2023eau}). Nevertheless, we emphasise that the SM distributions obtained with $\kappa_\lambda^{(0)}=1$ or $\kappa_\lambda^{(1)}\simeq 0.94$ are extremely similar, and once experimental effects are included, they are undistinguishable.}
In the left column we show the distributions without experimental uncertainties;
the distributions in the right columns give the number of di-Higgs events in the $b\bar b$ final state, taking into account 
smearing, binning and the experimental efficiencies as discussed above. The error bars indicate 
Poisson statistical uncertainties in each bin for the signal.

In \cref{bp123}, we observe that for BP1 (upper row) the theoretical RxSM distribution with tree-level trilinear scalar couplings (blue curve) is essentially indistinguishable from the SM value for almost the entire range of $m_{hh}$. At the level of the theoretical curves, only a resonance in the RxSM, for $m_{hh}=m_H$, yields a visible peak-dip structure.  After experimental effects are included, the resonance is too small to be distinguished from the SM, which results in a significance $Z^{(0)}=0.3$. 
When including loop calculations, the value of $\kala$ is increased to $\kala^{(1)}\simeq 3$, 
while the value of $\lahhH$ effectively switches sign. While the former
leads to an important modification of the interference pattern for low values of $\mhh/\mbbbb$, the latter switches the peak-dip
structure to a dip-peak structure. However, after taking into account the experimental uncertainties, this has no visible effect
in the distributions due to the very narrow width of the resonance structure. The inclusion of the one-loop effects, particularly in $\lahhh$,
results in a significance of $Z^{(1)}=9.3$.
BP1 is therefore an example scenario for which an analysis using tree-level trilinear scalar couplings would not have allowed distinguishing the RxSM from the SM, while the inclusion of one-loop corrections renders this possible. 

BP2 is another example in which one would not expect to be able to distinguish the RxSM from the SM in an analysis using only tree-level trilinear couplings. Indeed, for BP2 $\kala^{(0)} \simeq 1$ and $\lahhH^{(0)}\simeq 0$, leading to total and differential cross-sections essentially identical to those of the SM. However, when taking into account the one-loop corrections in $\rlaijk{1}$, we find $\kala^{(1)} \sim 4.9$ and $\lahhH^{(1)} \sim -300 \gev$. Using these values,
one can already distinguish the RxSM from the SM at the level of the total cross-section, as $\sihh$ is increased by $\sim100\%$ compared to $\sihhSM$ --- unlike in BP1, where $\sihh$ decreased by $\sim50\%$. This increase in $\sihh$ is due to the one-loop value of $\kala^{(1)}$. This can be understood well at the level of differential distribution: indeed the interference pattern for such large values of $\kala$ yields a large peak for low values of $\mhh/\mbbbb$, which results in an increase of the total cross-section.
On the other hand, the resonance structure of the heavy Higgs boson remains unobservable, also including the one-loop corrections to 
$\lahhH$. The significance at the one-loop level between the RxSM and the SM is found to be $Z^{(1)} = 21.3$.

In BP3, the radiative corrections to the trilinear scalar couplings are moderate.  
In the case of a tree-level analysis, we are in a similar situation as with BP1, with no deviation in $\kala$ and a resonance that is relatively suppressed by experimental effects. 
This leads to a significance of $Z^{(0)} = 0.8$. 
At the one-loop level, we have a small correction to $\lahhh$, yielding $\kala^{(1)}= 1.4$, which modifies slightly 
the interference for low values of $\mhh/\mbbbb$. Nevertheless, the value of $\kala$ is still close to~1, and 
therefore we only observe a dip in the distribution around $\mhh \sim 280 \gev$, but no peak (unlike e.g.\ BP2). 
The modest correction to $\lambda_{hhH}$ yields a slightly larger resonance peak with respect to the tree-level case.
Combining these two effects, we find an increase in significance to $Z^{(1)}=2.4$. This remains insufficient to conclude 
that both curves can be distinguished, but it provides an indication of new physics, which we do not observe at tree level. 
Possible future improvements in experimental analysis techniques may offer the chance to further improve the significance for scenarios
like BP3.

In BP4, with $\kala^{(0)} = 1.6$ there is already a large deviation from the SM in the theoretical distributions 
employing tree-level trilinear scalar couplings, and this deviation persists even after considering 
the effects of smearing and binning, leading to $Z^{(0)} = 3.5$, 
which provides indications of new physics. 
However, the resonance occurs at large masses, and thus once statistical errors are taken into account, the peak in the
differential distribution in the RxSM cannot be distinguished from the continuum (i.e.\ the SM distribution). 
After including one-loop corrected trilinear scalar couplings, the resonant peak is not modified significantly.
However the large loop corrections to $\lahhh$, leading to $\kala^{(1)} = 3.3$, 
have a strong impact on the interference pattern at low values of $\mhh/\mbbbb$.
This yields a large enhancement just above threshold and a dip around $\sim 375 \gev$. 
This results in a large deviation from the SM distribution that is not erased by experimental effects, 
giving a significance of $Z^{(1)}=10.5$. On the other hand, the increase in $\kala$ does not have a significant impact on the total cross-section because the contributions from the peak and the dip cancel each other out. An analysis at the level of differential distributions (and including one-loop corrections to $\lahhh$) is therefore required to be able to 
distinguish the RxSM from the SM in BP4. 

Turning next to BP5, the tree-level and one-loop distributions are very similar to each other, and, once experimental 
uncertainties are taken into account, the two would not be distinguishable from each other. BP5 features a small deviation in $
\kala^{(0)}\simeq\kala^{(1)}\sim 1.3$, which causes a slight decrease in the differential cross-section for low values of 
$\mhh/\mbbbb$. At the same time, there is also a very large resonant peak allowing to differentiate the RxSM from the SM no matter 
the order at which the analysis is performed, with significances of $Z^{(0)}\simeq Z^{(1)}\sim 10$. This is 
due to a combination of a non-negligible value of the trilinear Higgs coupling, $\lahhH^{(0)} \sim \lahhH^{(1)} \sim 150 \gev$, and of the low
value of the BSM Higgs mass, $m_H \sim 400 \gev$, which enhances the interference with the non-resonant contributions.
As a general observation from our parameter scans, we find that, for points with a large resonant peak, found in the range $350\gev\lesssim m_H\lesssim 500\gev$, there are no significant corrections to $\lahhH$. Consequently, for this type of scenarios, the one-loop corrections in the RxSM do not enhance the sensitivity to the BSM trilinear Higgs coupling.

Finally, our last scenario, BP6, is similar to BP3 in terms of the trilinear scalar couplings and total di-Higgs cross-sections at tree level and one loop, as well as of the statistical significances. On the other hand, BP6 features a larger resonant peak at the 
tree level, which is somewhat suppressed by loop corrections to $\kappa_\lambda$. The dip for low values of $m_{hh}$ due to the deviation of $\kappa_\lambda$ also pushes down the continuum distribution in the invariant mass region around the resonant peak, therefore the resonant peak is also affected by this interference. 
To better understand to what extent the 
resonant peak and, consequently, $\lahhH$ can be resolved experimentally, we have defined a statistical significance, which 
we denote $Z_{\mathrm{peak}}$, for discriminating the resonant peak (the signal hypothesis) 
from the continuum (the null hypothesis, $\lahhH = 0$) for BP5 and BP6.%
\footnote{Further details about the definition of $Z_\text{peak}$ can be found in \cref{app:peak}.}
For BP5, we find for the peak significance $Z^{(0)}_{\mathrm{peak}}=11.5$ and $Z^{(1)}_{\mathrm{peak}}=11.8$. 
In both cases, we can distinguish the resonant peak from the continuum, and therefore, we have sensitivity to the $\lahhH$ coupling.
However, as discussed above, the one-loop corrections do not enhance the significance.
In the case of BP6, we find $Z^{(0)}_{\mathrm{peak}}=1.2$ and $Z^{(1)}_{\mathrm{peak}}=0.5$. 
In this case, the resonant peak cannot be resolved from the continuum in either case, and the negative one-loop 
corrections to $\lahhH$ reduce $Z_\text{peak}$.
All significances are summarised in \cref{bpscrosslhc} for our benchmark points. 

\begin{table}[htb!]
\centering
\begin{tabular}{@{}ccccccccccc@{}}
\toprule
BP  & $\kappa_{\lambda}^{(0)}$ & $\kappa_{\lambda}^{(1)}$ & $\lambda_{hhH}^{(0)}$ & $\hat\lambda_{hhH}^{(1)}$ & $\sigma^{\mathrm{RxSM}}_{\mathrm{hh}}(\lambda_{ijk}^{(0)})$ & $\sigma^{\mathrm{RxSM}}_{\mathrm{hh}}(\hat\lambda_{ijk}^{(1)})$ &$Z^{(0)}$ &$Z^{(1)}$& $Z^{(0)}_{\mathrm{peak}}$&$Z^{(1)}_{\mathrm{peak}}$\\
 & & & [GeV] & [GeV] & [fb] & [fb] & & \\ \midrule
1 & 1.0                      & 3.0                      & 96.3                      & $-119.7$                    & 19.7                                                          & 9.6                      & 0.3&9.3 &0.0 &0.0                                      \\
2 & 1.0                      & 4.9                      & 2.2                       & $-303.6$                    & 19.8                                                        & 38.1                            & 0.3 & 21.3  &0.0 &0.0                                      \\
3 & 1.1                      & 1.4                      & 127.6                     & 181.9                     & 19.7                                                          & 15.5                              &0.8&2.4&0.0 &0.0                                      \\
4 & 1.6                      & 3.3                      & 543.2                     & 131.2                     & 13.2                                                         & 11.5                         &3.5 & 10.5 &0.0 &0.0                                      \\
5 & 1.3                      & 1.3                      & 153.9                     & 165.1                     & 35.3                                                        & 35.9           &9.7&10.0  &11.5 &11.8                                      \\
6 & 1.1                      & 1.5                      & 163.5                     & 263.8                     & 19.1                                                      & 15.6                                  &    1.3&2.8&1.2 &0.5                                      \\ \bottomrule
\end{tabular}
\caption{Predictions for $\kala$ and $\lahhH$ at tree level and one loop, for the total di-Higgs production cross-section at the (HL-)LHC using tree-level or one-loop trilinear couplings, statistical significances to distinguish the di-Higgs invariant mass differential distributions (with tree-level or one-loop trilinear couplings) from the SM and  statistical significances to distinguish the di-Higgs invariant mass differential distributions resonant peak (with tree-level or one-loop trilinear couplings) from the continuous distribution, for the benchmark points defined in \cref{bps}.} 
\label{bpscrosslhc}
\end{table}

 
\subsection{\boldmath{$e^+e^-$} colliders}
\label{sec:ilc}

\begin{figure}[htb!]
    \centering
    \includegraphics[width=0.7\linewidth]{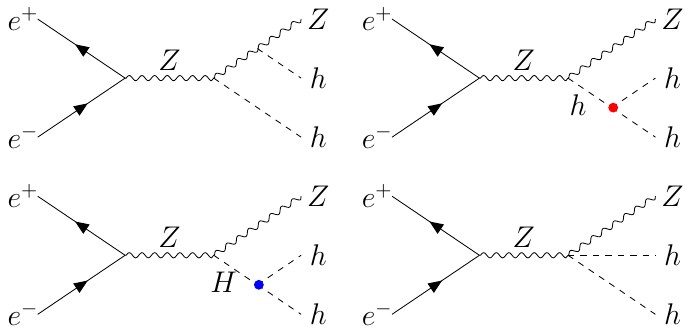}
    \caption{Example diagrams contributing to the $e^+e^-\to Zhh$ process at high-energy $e^+e^-$ colliders. The red and blue dots represent, 
    respectively, the trilinear scalar couplings $\lahhh$ and $\lahhH$. }
    \label{fig:diHiggs_epem}
\end{figure}

In this section, we present our results for di-Higgs production at future high-energy $e^+e^-$ colliders. We consider the 
di-Higgs-strahlung channel 
$e^+e^-\to Zhh$, which is the dominant production channel of two SM-like Higgs bosons up to centre-of-mass energies slightly above 1~TeV.
The contributing Feynman diagrams are shown in \cref{fig:diHiggs_epem}.
Similarly to the HL-LHC study, we investigate the impact of loop corrections to the trilinear Higgs couplings on the total and 
differential cross-sections and the experimental sensitivity to possible deviations from the SM. Contributions involving $\lahhh$ 
arise due to a non-resonant diagram, shown as the upper right diagram in \cref{fig:diHiggs_epem}. Consequently, the strongest 
effects of BSM modifications of $\lahhh$ are expected at low values of $\mhh$, close to the kinematic threshold. On the other hand, the 
contributions proportional to $\lahhH$ come from a (potentially) resonant diagram (lower left diagram in \cref{fig:diHiggs_epem}) 
mediated by the heavy Higgs boson $H$. This contribution is expected to have the largest effect around $\mhh = m_H$. Consequently,
the loop corrections to the two trilinear Higgs couplings will have a different impact for different values of $\mhh$. 
This makes it crucial to accurately measure the differential distributions over the whole allowed range.

\subsubsection{Calculation of \boldmath{$e^+e^- \to Zhh$}}
\label{sec:eeZhh}

We compute the leading-order unpolarised $e^+e^-\to Zhh$ production cross-section, which we denote $\sigma_{Zhh}$, using the public 
code \texttt{Madgraph5\textunderscore aMC v3.5.7}~\cite{Alwall:2014hca}. The input \texttt{UFO} model files for the RxSM required 
by \texttt{Madgraph} were obtained with the \texttt{Mathematica} package \texttt{SARAH-4.15}~\cite{Staub:2008uz,Staub:2009bi,Staub:2010jh,Staub:2012pb,Staub:2013tta}. We 
compute the cross-section for future $e^+e^-$ collider at a centre-of-mass energy of $\sqrt{s} = 1000\gev$, where we take as a concrete 
example the ILC1000~\cite{ILC:2013jhg,Moortgat-Pick:2015lbx,Bambade:2019fyw}. We choose a high centre-of-mass energy due to the large 
masses of our heavy Higgs boson. In this work, we assume an integrated luminosity of 8~ab$^{-1}$, as projected for the 
ILC1000~\cite{Bambade:2019fyw}. 
After computing the unpolarised cross-section we can apply a simple correction factor following \citere{Arco:2025pgx} to take into 
account polarised beams as possible at the ILC1000. The optimised 
polarisation of ILC1000 is an opposite sign polarisation of $80\%$ for electrons and $30\%$ for positrons giving as a result:
\begin{align}
\label{sigma-pol-mp}
    \sigma(-80\%,+30\%) &\simeq 1.476\;\sigma_{\mathrm{unpol}}\,, \\
    \sigma(+80\%,-30\%)
    &\simeq 1.004 \;\sigma_{\mathrm{unpol}}\,.
    \label{sigma-pol-pm}
\end{align}

With our calculational setup, we obtain for the SM di-Higgs production cross-sections of $\siZhhSM \simeq 0.236\text{ fb}$ 
for $\sqrt{s} = 500 \gev$ and $\siZhhSM \simeq 0.177\text{ fb}$ for $\sqrt{s} = 1 \tev$. The observation of the di-Higgs-strahlung
process at $\sqrt{s} = 500 \gev$ is expected at the $8\sigma$ level for an integrated luminosity of $4 \text{ ab}^{-1}$ (combining several polarisation runs), corresponding to a relative experimental uncertainty of 16.8\% on $\siZhhSM$ obtained in Ref.~\cite{Durig:2016jrs}, and which was recently improved to 12.8\% in Refs.~\cite{Altmann:2025feg,Berggren:2025fpw}. 
Applying a simple scaling of the number of events, this yields a relative uncertainty of the cross-section at $\sqrt{s} = 1000$~GeV of 
$\sim 10\%$ for an integrated luminosity of $8\text{ ab}^{-1}$, which corresponds to a discovery significance of close to 
$13\sigma$.


\subsubsection{Total cross-section}

\begin{figure}[h]
    \centering
    \includegraphics[width=0.49\linewidth]{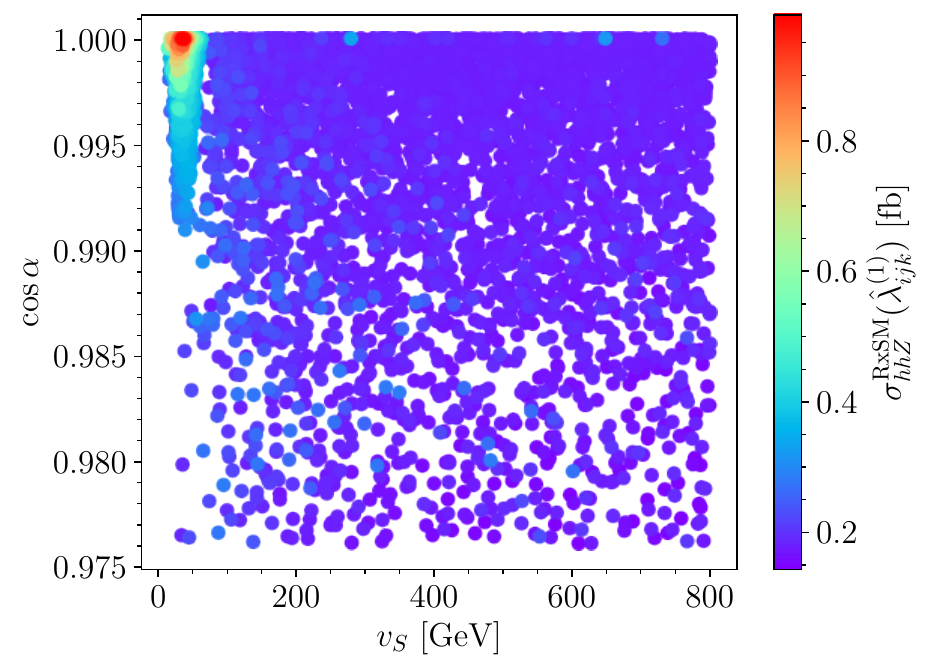}
    \includegraphics[width=0.49\linewidth]{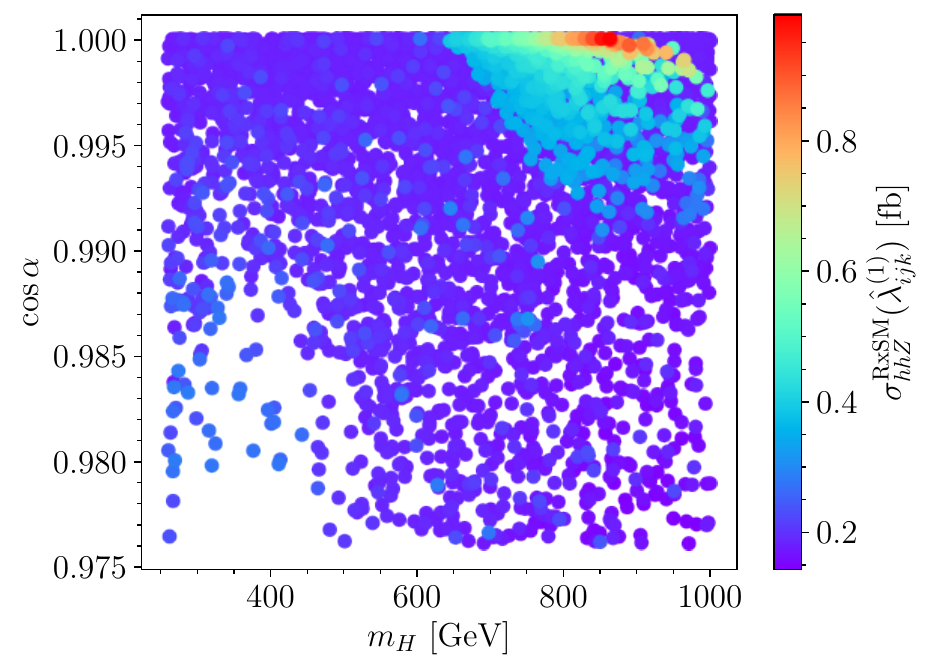}
    \caption{Total cross-section for the di-Higgs production process $e^+e^-\to Zhh$ at $\sqrt{s} = 1 \tev$
    including one-loop corrections to the trilinear scalar couplings for the RxSM scan points. \textit{Left}: Results in the $\{\cos \alpha,v_S\}$ plane; \textit{right}: results in the $\{\cos \alpha,m_H\}$ plane. }
    \label{crossee}
\end{figure}

Our predictions for the total di-Higgs production cross-section in the RxSM at $\sqrt{s} = 1 \tev$ 
including one-loop corrections to the trilinear Higgs couplings, which we denote $\siZhh(\rlaijk1)$, 
are shown in the colour coding of \cref{crossee}. Results are projected in two different planes: in the left panel in the 
$\{\cos \alpha,v_S\}$ plane, and in the right panel in the $\{\cos \alpha,m_H\}$ plane. We find first that values of the 
total cross-section up to six times larger than the SM result, $\siZhh \sim 6 \siZhhSM$, are possible. 
This would clearly allow  to distinguish the corresponding parameter points from the SM. 
The enhancement of $\siZhh$ with respect to the SM occurs in the same parameter region as for the $gg\to hh$ process,
namely in the region of low values of $v_S$, high values of $m_H$, and close to the alignment limit. 
This is due to the large corrections to $\kala$ at the one-loop level. However, comparing \cref{hhh}  
and \cref{crossee}, one can observe a continuous enhancement of  $\siZhh$ with $\kappa_{\lambda}$ --- in contrast to what was found for the $gg\to hh$ process in \cref{crossHL-LHC}. This can be explained by the monotonous increase of the $e^+e^-\to Zhh$ cross-section with $\kappa_\lambda$, which is unlike the $gg\to hh$ cross-section that is non-monotonous, with a minimum around $\kala\sim2.5$. The total di-Higgs production cross-section of $e^+e^-\to Zhh$ is therefore more sensitive to small changes of $\kappa_\lambda$ than that one of $gg\to hh$.

\begin{figure}[h]
    \centering
    \includegraphics[width=0.49\linewidth]{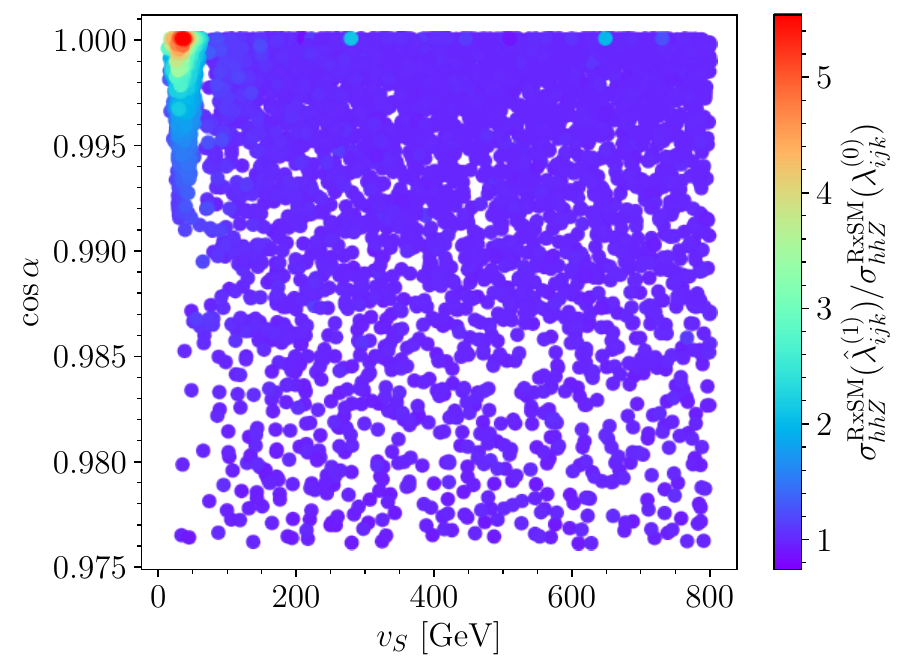}
    \includegraphics[width=0.49\linewidth]{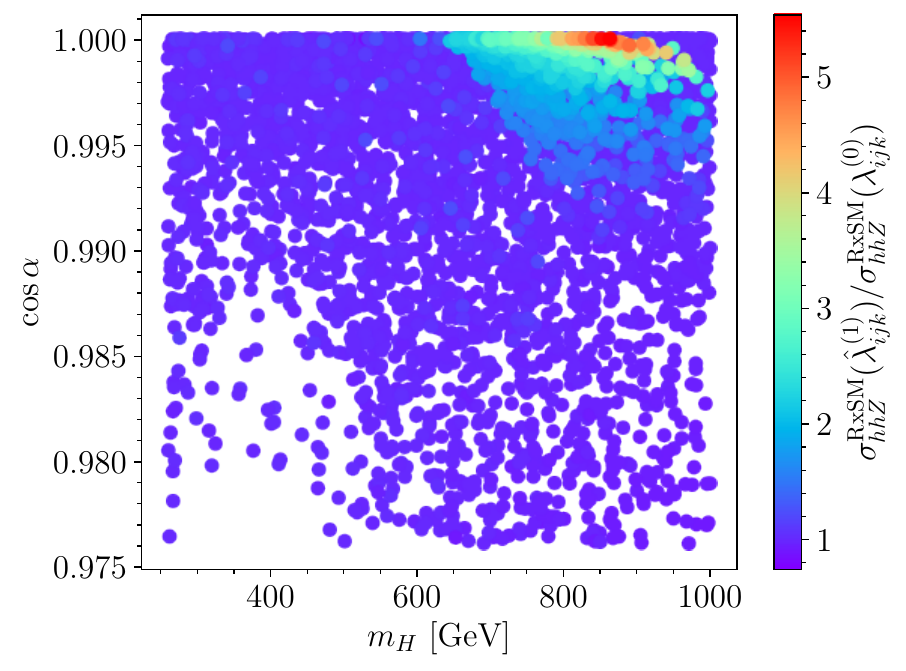}
    \caption{$\siZhh(\rlaijk1)/\siZhh(\laijk^{(0)})$ (see text). 
    \textit{Left}: Results in the $\{\cos \alpha,v_S\}$ plane; \textit{right}: results in the $\{\cos \alpha,m_H\}$ plane.}
    \label{ratiocrossee}
\end{figure}

In \cref{crossee} we can also find points with smaller deviations from the SM, with $\siZhh \sim 2 \siZhhSM$, 
and that are further from the alignment limit and are not correlated with the loop corrections to $\kala$. 
In order to understand the behaviour of these points, we present in \cref{ratiocrossee} the ratio between $\siZhh(\rlaijk1)$ 
and the same cross-section using the tree-level values of the trilinear Higgs couplings, $\siZhh(\laijk^{(0)})$. 
The ratio is indicated by the colour coding, shown for the $\{\cos \alpha,v_S\}$ plane (left panel) and the 
$\{\cos \alpha,m_H\}$ plane (right panel). As can be seen from \cref{ratiocrossee}, the large enhancement of $\siZhh$ 
for points near the alignment limit arises from $\kala^{(1)}$. 
On the other hand, for the points with a smaller increase, i.e.\ $\siZhh \sim 2 \siZhhSM$, 
the increase in the cross-section stems from tree-level effects, e.g. from the heavy Higgs-boson resonance 
contribution, as $\siZhh(\rlaijk1)/\siZhh(\laijk^{(0)})$ is close to unity.

\subsubsection{Differential cross-section}

In this subsection, 
we analyse the differential di-Higgs production cross-sections with respect to \mhh\ 
for the six benchmark points defined in \cref{bps} at an $e^+e^-$ collider of $\sqrt{s}=1 \tev$, 
using additionally polarised beams. As in our study for the HL-LHC case, we want to take into account
experimental uncertainties, although in the case of $e^+e^-$ colliders the events are much cleaner than in $pp$ collisions.
For this reason, we present in this subsection distributions taking into account the Higgs-boson decays and applying realistic 
experimental cuts. 

\begin{figure}
    \centering
    \includegraphics[width=0.7\linewidth]{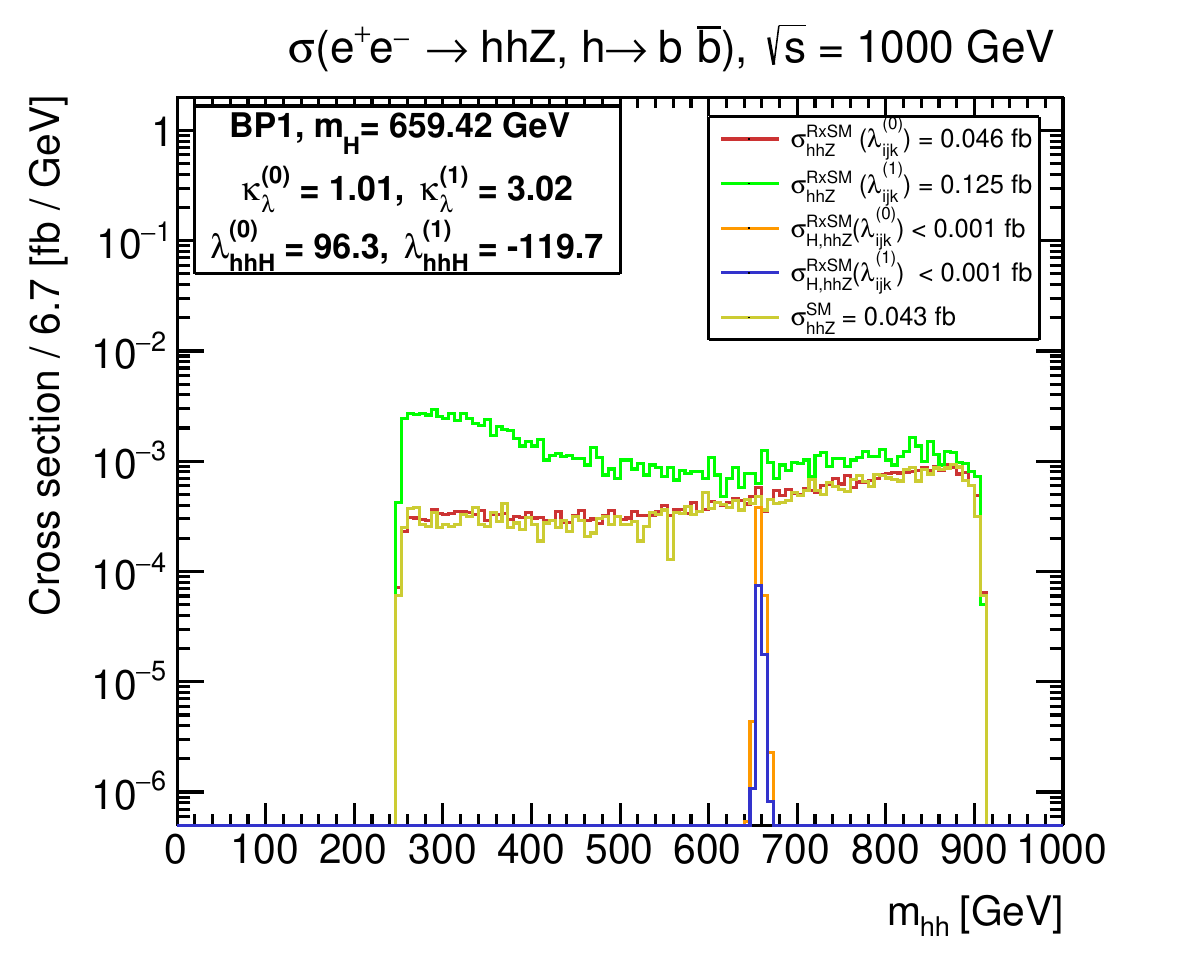}
    \includegraphics[width=0.7\linewidth]{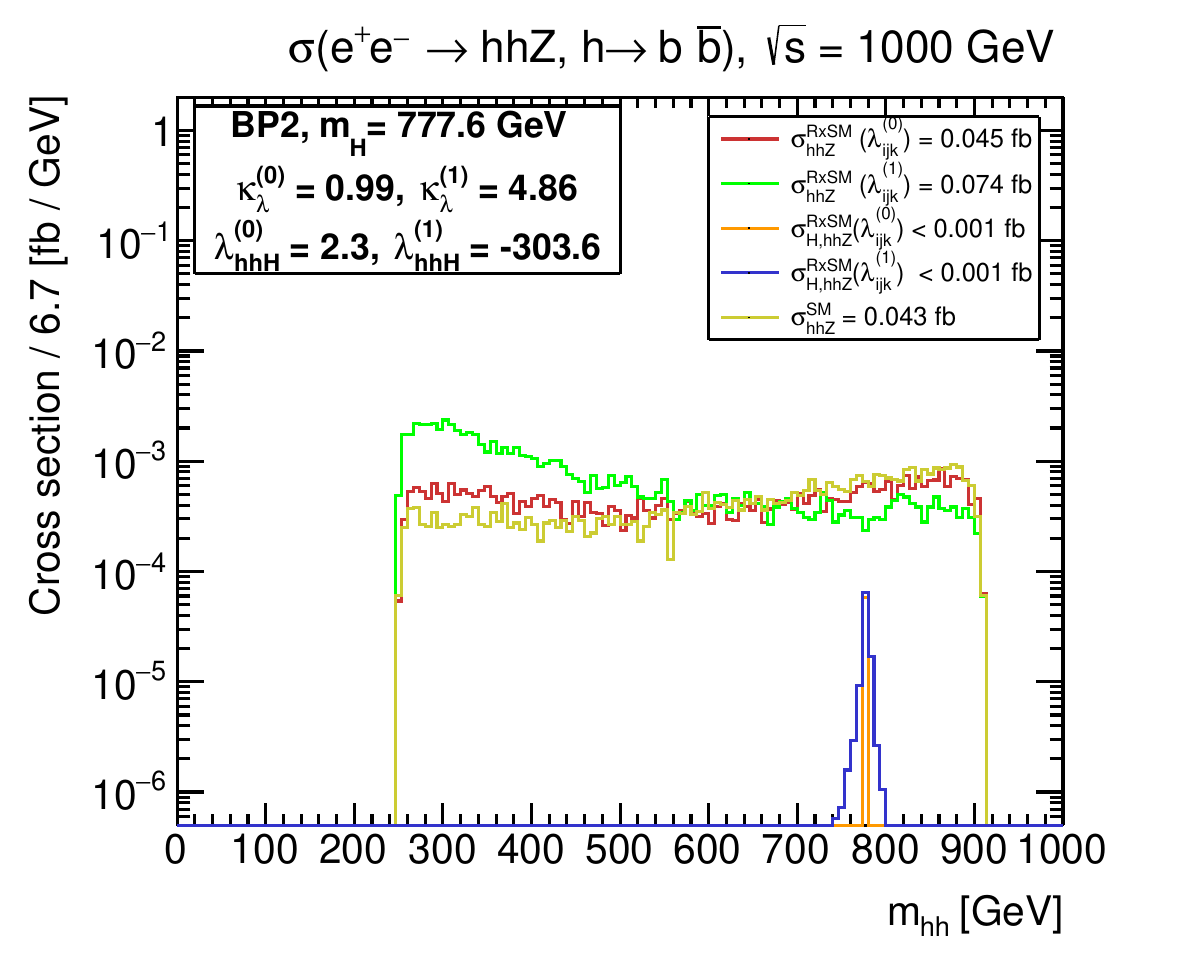}
    \caption{Differential polarised di-Higgs production cross-section distributions as a function of the di-Higgs invariant mass $m_{hh}$ for ILC1000. \textit{Top:} Results for BP1; \textit{bottom:} Results for BP2 from \cref{bps}. We plot the RxSM result using one-loop trilinear scalar couplings (green curve), the RxSM result using tree-level scalar couplings (red), the contribution of the diagram with $H$ in the $s$-channel using one-loop trilinear scalar couplings (blue curve), the contribution of the diagram with $H$ in the $s$-channel using tree-level trilinear scalar couplings (orange curve) and the SM result (yellow curve). Values for $\rlahhH{1}$ are shown in GeV.}
    \label{diffee1}
\end{figure}

\begin{figure}
    \centering
    \includegraphics[width=0.7\linewidth]{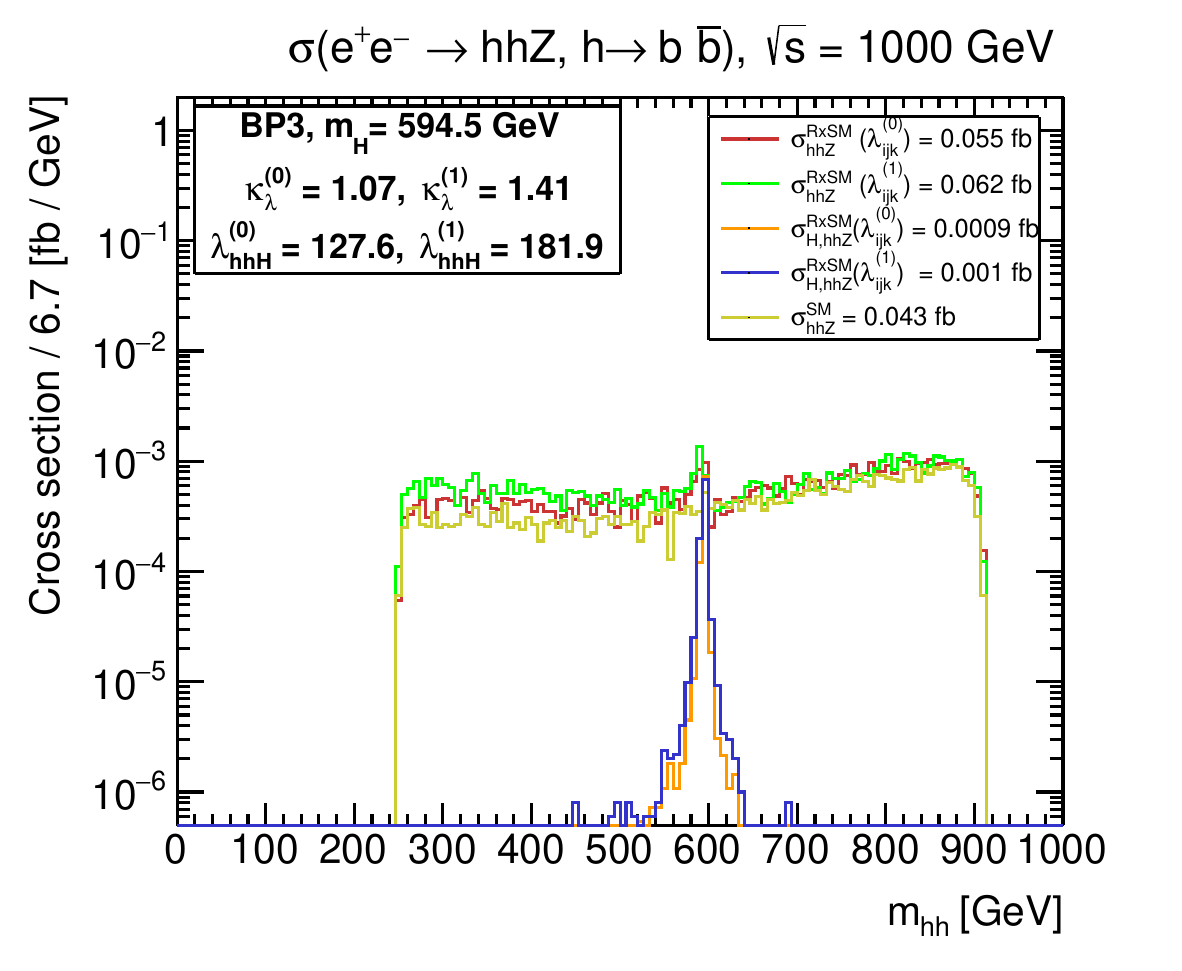}
    \includegraphics[width=0.7\linewidth]{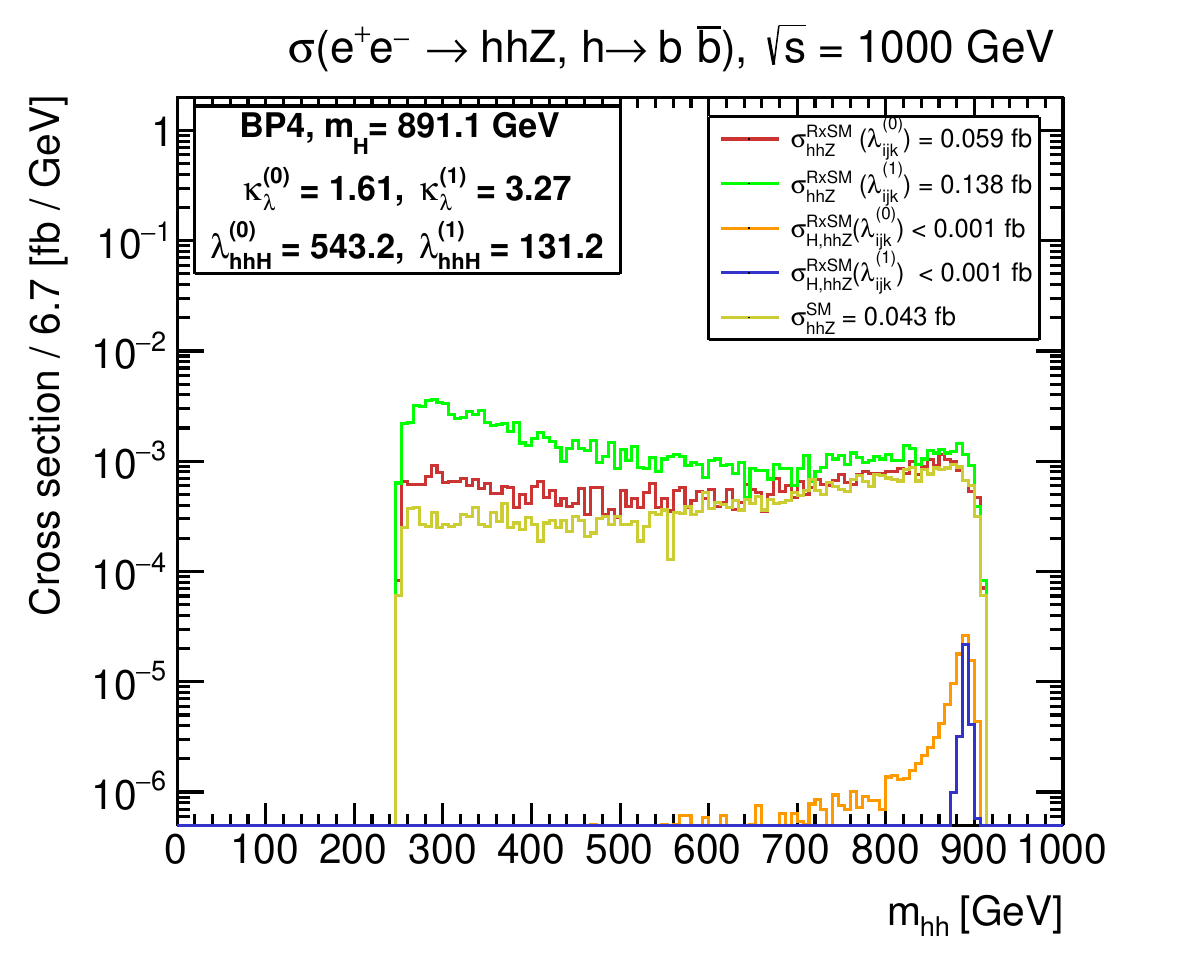}
    \caption{Differential polarised di-Higgs production cross-section distributions as a function of the di-Higgs invariant mass $m_{hh}$ for ILC1000. \textit{Top:} Results for BP3; \textit{bottom:} Results for BP4 from \cref{bps}. We plot the RxSM result using one-loop trilinear scalar couplings (green curve), the RxSM result using tree-level scalar couplings (red), the contribution of the diagram with $H$ in the $s$-channel using one-loop trilinear scalar couplings (blue curve), the contribution of the diagram with $H$ in the $s$-channel using tree-level trilinear scalar couplings (orange curve) and the SM result (yellow curve). Values for $\rlahhH{1}$ are shown in GeV.}
    \label{diffee2}
\end{figure}

\begin{figure}
    \centering
    \includegraphics[width=0.7\linewidth]{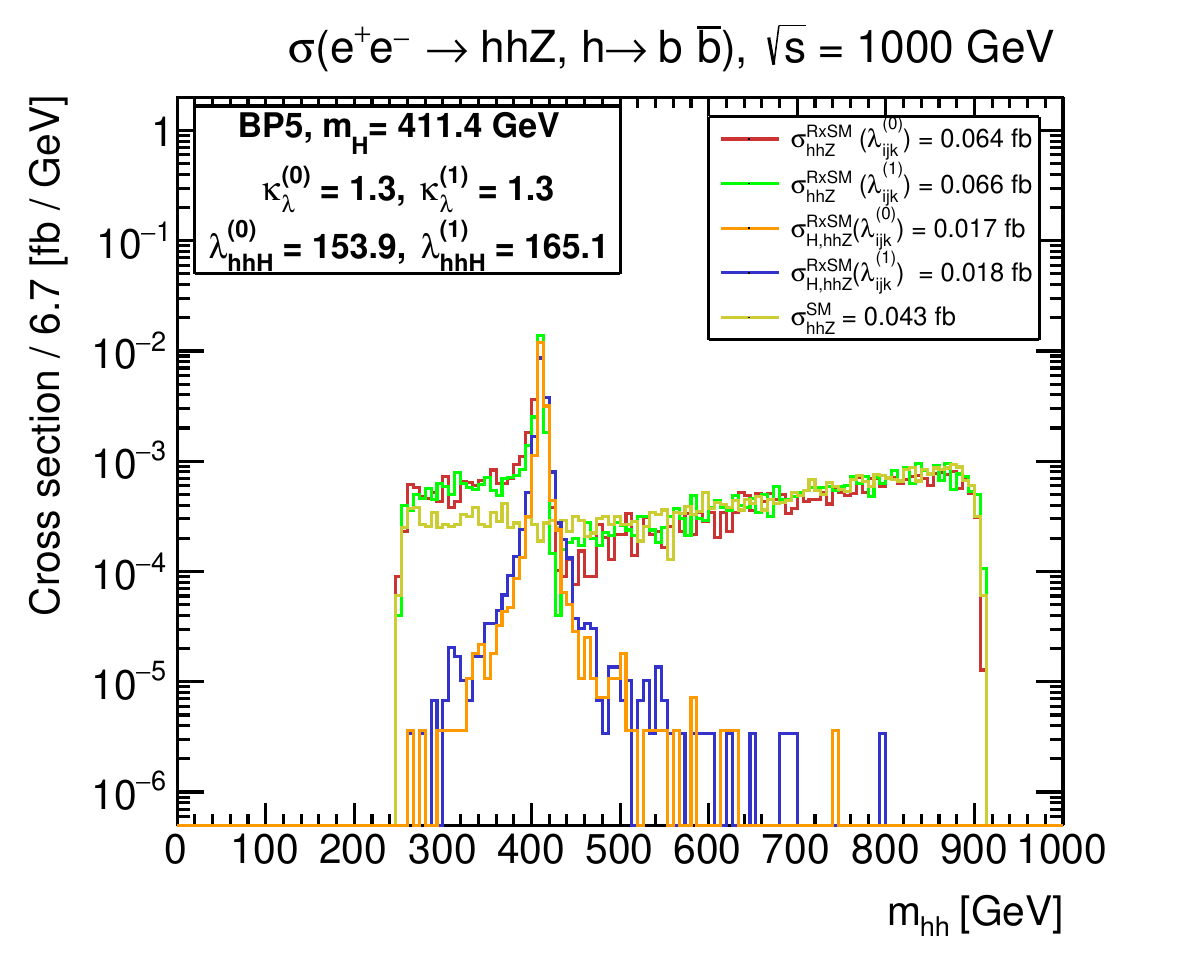}
    \includegraphics[width=0.7\linewidth]{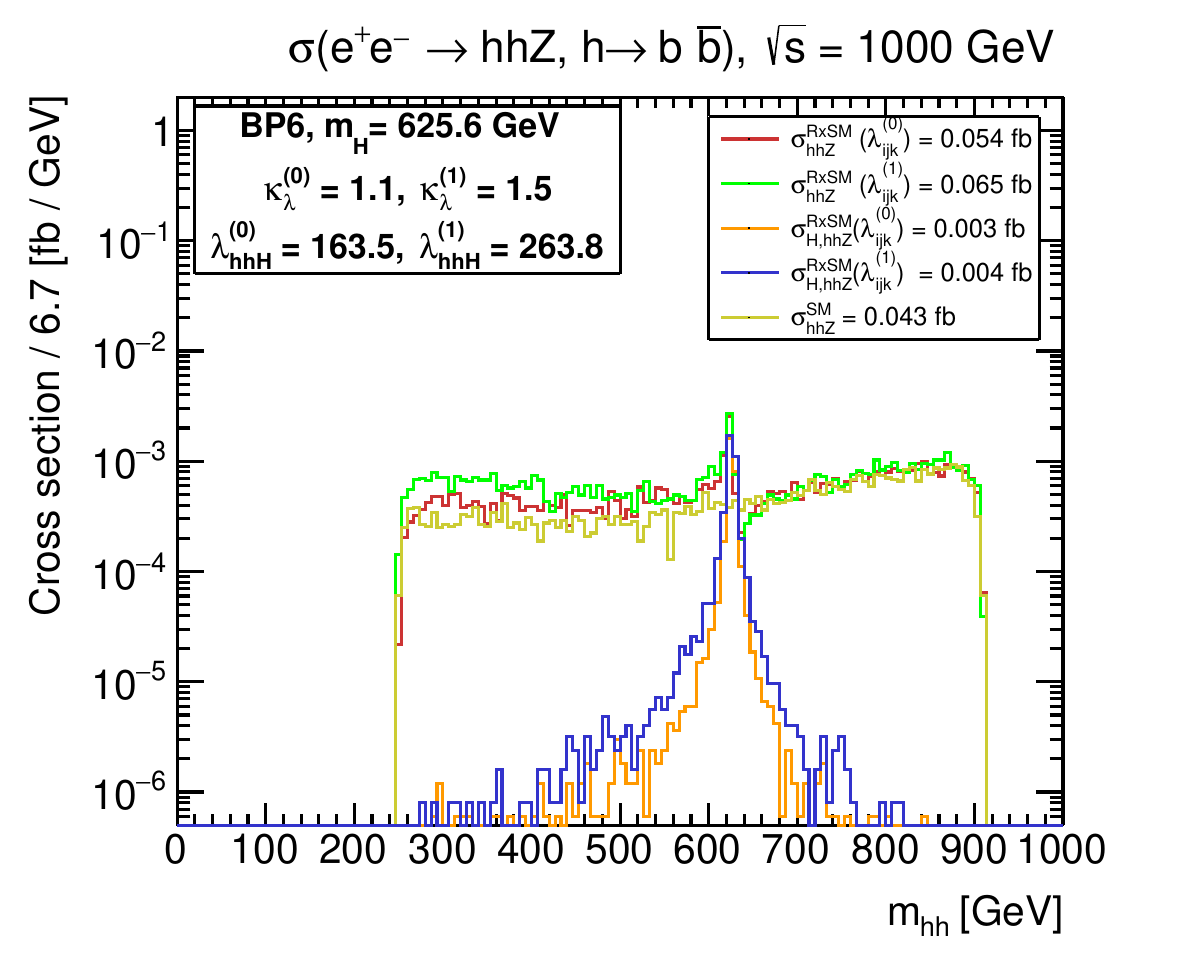}
    \caption{Differential polarised di-Higgs production cross-section distributions as a function of the di-Higgs invariant mass $m_{hh}$ for ILC1000. \textit{Top:} Results for BP5; \textit{bottom:} Results for BP6 from \cref{bps}. We plot the RxSM result using one-loop trilinear scalar couplings (green curve), the RxSM result using tree-level scalar couplings (red), the contribution of the diagram with $H$ in the $s$-channel using one-loop trilinear scalar couplings (blue curve), the contribution of the diagram with $H$ in the $s$-channel using tree-level trilinear scalar couplings (orange curve) and the SM result (yellow curve). Values for $\rlahhH{1}$ are shown in GeV.}
    \label{diffee}
\end{figure}

We estimate the experimental uncertainties affecting the differential cross-section distributions,  
following a similar scheme as for the study of the HL-LHC case. Firstly, we 
consider the decay channel of the SM-like Higgs boson to $b\bar{b}$ by multiplying the differential cross-section by 
$\br^2(h \to b\bar b)$ as well as by the efficiency of the detector in the reconstruction of the $b$-jets. 
This efficiency is defined in a different way than that at the HL-LHC, and we follow \cite{Arco:2025pgx}: 
we decompose it into the $b$-tagging efficiency of each $b$-jet, which we assume to be $\epsilon_b=80\%$, 
and the acceptance $\cA$ of the detector. We estimate this acceptance by applying the following pre-selection cuts to detect the final
$4b+Z$ events 
\cite{Arco:2025pgx}, 

\begin{equation}
E_b > 20\ \text{GeV}, \quad |\eta_b| < 2.5, \quad |\eta_Z| < 2.5, \quad y_{bb}>0.0010\,,
\label{eq:cuts-ee}
\end{equation}
where $E_b$ is the energy of the $b$-jets, $\eta_{b/Z}$ is the pseudo-rapidity of the $b/Z$ respectively and $y_{bb}$ is the difference in 
the pseudo-rapidity of the $b$-jets, which represents the distance between two $b$-jets. 
In order to estimate the acceptance $\cA$, we simulate the $e^+e^-\to  Zhh \to Z b\bar b b\bar b$ process 
with and  without the cuts of \cref{eq:cuts-ee} with \texttt{MadGraph} at the parton level. 
We obtain $\cA$ as the ratio of events with and without cuts. 
Taking also into account the luminosity $\cL = 8 \,\iab$, we can define the number of events $N$ as
\begin{equation}
    N = \sigma(e^+e^-\to Zhh) \times \cL \times \big(\br(h\to b\bar{b})\big)^2 \times \cA \times \epsilon_b^4.
\end{equation}
Next, we define the statistical significance very similar as for the HL-LHC case in \cref{z}. 
We calculated the significances for both polarisations, denoted as $Z_{-+}$ for \cref{sigma-pol-mp} and $Z_{+-}$ for \cref{sigma-pol-pm}.
The luminosity assumed for each polarisation is 40\% of the total luminosity, i.e.\ $\cL_{-+} = \cL_{+-} = 0.4 \times 8\,\iab = 3.2\,\iab$.
The overall significance is then obtained as $Z := \sqrt{Z_{-+}^2 + Z_{+-}^2}$.
The results 
for the significances 
are calculated with tree-level and one-loop trilinear scalar couplings, 
which we denote respectively $Z^{(0)}$ and $Z^{(1)}$. 
The results for the six benchmark points are summarised in \cref{bpscrossee}. 
The corresponding results for the differential distributions are presented in \cref{diffee1,diffee2,diffee}.
The red (green) curves show the distributions for $\siZhh$ using $\laijkz$ ($\rlaijk1$), the blue (orange) curves show the
distributions for $\siHZhh$, i.e.\ only taking into account the heavy Higgs resonance diagram, using $\laijkz$ ($\rlaijk1$), and
the yellow curves show the corresponding SM distribution for $\siZhhSM$.

\begin{table}[h]
\centering
\begin{tabular}{@{}ccccccccccc@{}}
\toprule
BP  & $\kala^{(0)}$ & $\kala^{(1)}$ & $\lahhH^{(0)}$ & $\lahhH^{(1)}$ & $\siZhh(\laijk^{(0)})$ & $\siZhh(\rlaijk1)$ & 
$Z^{(0)}$ & $Z^{(1)}$ & $Z_{\mathrm{peak}}^{(0)}$ & $Z_{\mathrm{peak}}^{(1)}$\\
 & & & [GeV] & [GeV] & [fb] & [fb] & & \\ \midrule
1 & 1.01                      & 3.02   & 96.30  & $-119.70$ & 0.060   & 0.120  & 0.0&27.23& 0.0&0.0    \\
2 & 0.99                      & 4.86 & 2.30  & $-303.60$  & 0.059   & 0.210   & 0.0 & 35.38 & 0.0&0.0    \\
3 & 1.07                   & 1.41 & 127.60 & 181.90  & 0.062    & 0.07  &0.0&0.0   & 0.0&0.0    \\
4 & 1.61                      & 3.27 & 543.20  & 131.20  & 0.075   & 0.13 &0.0 & 20.69   & 0.0&0.0    \\
5 & 1.30                      & 1.30& 153.90   & 165.10   & 0.077    & 0.079  &13.18&17.25   & 16.63&18.55   \\
6 & 1.10      & 1.50 & 163.5  & 263.8  & 0.061     & 0.073 &    3.54&3.95     & 3.29&4.82   \\\bottomrule
\end{tabular}
\caption{Predictions for $\kala$ and $\lahhH$ at tree level and one loop, for the total $Zhh$ production cross-section at a 1 TeV $e^+e^-$ collider using tree-level or one-loop trilinear couplings and statistical significances to distinguish the di-Higgs invariant mass differential distributions (with tree-level or one-loop trilinear couplings) from the SM, or from the case $\lahhH = 0$, for the benchmark points defined in \cref{bps}.}
\label{bpscrossee}
\end{table}

In \cref{diffee1}, we observe that for BP1 the RxSM distribution with tree-level trilinear scalar couplings (red curve) 
is practically indistinguishable from the SM distributions (yellow curves), while for BP2 both curves appear slightly different, although the statistical analysis shows that they are statistically indistinguishable. 
In terms of the statistical significance (see \cref{bpscrossee}), BP1 and BP2 both have $Z^{(0)}=0.0$,
meaning that a difference in the number of events cannot be detected. 
On the other hand, when we consider the distributions with one-loop trilinear scalar couplings (green curves) and 
compare them with the SM ones (yellow curves), we find that there is a large enhancement for the RxSM case due to the 
large loop corrections to $\kala$.  
We find values of $Z^{(1)}=27.23$ and $Z^{(1)}=35.38$ for BP1 and BP2, respectively, which means that 
these points can be distinguished from the SM when including one-loop corrections to the trilinear scalar couplings.

Turning next to BP3 and BP4, as shown in \cref{diffee2}, the RxSM distributions using tree-level trilinear scalar couplings (red curves) look slightly different from the SM curve (yellow curve). However, when we compute the significances we find that they are equal to zero due to the small difference in the number of events after taking into account the Higgs-boson decays and the cuts. 
For BP3 we do not find large loop corrections for the trilinear scalar couplings and therefore when we consider them in the calculation of the cross-section (green curve), we do not observe any improvement in the significance, see \cref{bpscrossee}. For BP4 there is a positive correction of \order{100\%} in $\kala$, which causes a general enhancement of the distribution with respect to the SM one. 
In turn, the statistical significance including one-loop corrections increases from $Z^{(0)}=0.00$ to $Z^{(1)}=20.69$. 
Consequently, one can distinguish the result of the RxSM from the SM one for BP4 provided that we take into account one-loop corrections to the trilinear scalar couplings.

Finally, for BP5 and BP6, as shown in \cref{diffee}, the RxSM distributions using tree-level trilinear scalar couplings (red curve) 
can already clearly be distinguished from the SM distributions (yellow curves), because of the 
resonance peak structure from the heavy Higgs boson. For the BP5 we have a significance using tree-level trilinear scalar couplings of $Z^{(0)}=13.18$, which means that we have evidence of new physics, while for BP6 since the resonance is smaller the significance is only $Z^{(0)}=3.54$. For these benchmark points when considering one-loop corrected trilinear scalar couplings we can see that the distributions (green curves) are not significantly modified because the loop corrections are not very large. However, we can see that taking into account the loop corrections still does improve the significances for discriminating the RxSM distributions from the SM ones. 
We have also evaluated for all the benchmark points $Z_{\mathrm{peak}}^{(0)}$ and $Z_{\mathrm{peak}}^{(1)}$, defined analogously to the HL-LHC
case --- see \cref{sec:mhh-HLLHC}. The results are shown the rightmost two columns in \cref{bpscrossee}. While for BP1-4 there is no sensitivity to distinguish the resonant peaks, the numbers for BP5 and BP6 indicate that
the main part of the significance to distinguish the RxSM from the SM in in these scenarios originates from the resonance peak, and not from the continuum.
For these two benchmark points the statistical significance indicates that the 
resonance peak structure can possibly be distinguished from the continuum, which may enable a measurement of $\lahhH$, see
\citere{Frank:2025zmj}.



\section{Conclusions}
\label{sec:conclusions}

In this work, we explored the phenomenology of di-Higgs production in the general Higgs singlet extension of the SM, the RxSM, at the 
HL-LHC and a possible future high-energy $e^+e^-$ collider. In particular, we investigated the impact of the radiative corrections to the trilinear 
scalar couplings $\lahhh$ and $\lahhH$, which are relevant for the di-Higgs processes. At the technical level, we computed the loop-corrected
trilinear scalar couplings using \texttt{anyH3}, and used these as inputs for calculations of di-Higgs cross-sections and distributions with
\texttt{HPAIR} (for $gg\to hh$) and \texttt{Madgraph5\_aMC} (for $e^+e^-\to Zhh$). We devised, for the first time, a complete set of on-shell
renormalisation conditions for all the parameters of the RxSM scalar sector,  including for the $\mathbb{Z}_2$-breaking Lagrangian 
trilinear couplings $\kappa_{SH}$ and $\kappa_S$ that were so far renormalised $\overline{\text{MS}}$ in the existing literature.

We performed an extensive scan of the RxSM parameter space and determined the possible values of $\lahhh$, $\lahhH$, $\sihh$ and $\siZhh$ 
that can be realised for points allowed by state-of-the-art theoretical and experimental constraints. 
For $\lahhh$, we find that enhancements up to $\kala \simeq 6$ are possible in the region where $\cos\alpha\gtrsim 0.998$, $m_H\in [800,\ 900]\gev$
and $v_S\lesssim 50\gev$. In the same region of parameter space $\lahhH$ can receive large negative corrections that can even flip its sign.
For this parameter region, the concurrent increase of $\lahhh$ and decrease of $\lahhH$ both contribute to enhancing $\sihh$,
where values up to $\sihh\sim 4.5\,\sihhSM$ were found. We also compared the magnitude of the enhancements in $\sihh$ achievable 
from tree-level or one-loop modifications of the trilinear scalar couplings:
while effects are possible already at the tree level, the largest effects arise from radiative corrections. 

We next investigated the impact of the loop corrections to the trilinear scalar couplings on the di-Higgs invariant mass distributions. 
We devised six benchmark scenarios, representative of the RxSM parameter space, and we quantified the ability to distinguish these scenarios 
from the SM experimentally, using the statistical significance defined in~\citere{Frank:2025zmj}.
In particular, we showed that for a number of these scenarios (BPs 1, 2 and 4), only an analysis with one-loop trilinear couplings would allow
discriminating the RxSM from the SM. Conversely, we found that for some scenarios with resonant contributions for low $m_H$ (like BP5), 
the inclusion of loop-corrections in $\rlaijk{1}$ has a smaller impact on the distributions and the conclusions that can be drawn from them. 

We performed similar studies of the total and differential cross-sections for the process $e^+e^-\to Zhh$ at a high-energy $e^+e^-$ collider.
The largest enhancements of the cross-section are found for the same regions of the parameter space as for the di-Higgs process at the HL-LHC, and values up to $\siZhh \sim 4\,\siZhhSM$ can be attained. 
For the six benchmark points we have evaluated the differential \mhh\ distributions. We find that the loop corrections to the trilinear
scalar couplings can significantly enhance the significance for distinguishing the RxSM from the SM. The resonance peak structure, on the other
hand, only contributes significantly in two of the benchmark points. In that case the statistical significance indicates that the 
resonance peak structure can possibly be distinguished from the continuum, which may enable a measurement of $\lahhH$.

Our results demonstrate the crucial importance of including, whenever possible, higher-order BSM corrections to analyses of di-Higgs production, 
in order to draw reliable conclusions from the comparison of experimental data with theoretical predictions. 
This provides further motivation for the automation of di-Higgs precision calculations, 
see \citere{anyHH} for recent advancements.  
Additionally, we concentrated in this work on corrections to trilinear scalar couplings, as these are known (see e.g.\ Ref.~\cite{Bahl:2022jnx}) 
to be the dominant source of BSM corrections to di-Higgs productions. However, for scenarios with total and differential cross-sections
undistinguishable from the SM ones, the investigation of corrections to other Higgs couplings entering the process 
(e.g.\ Higgs-top or Higgs-$Z$ interactions) becomes relevant. Finally, this work has been focused on probes of RxSM scenarios at high-energy
colliders. A next important step is to include into the investigation the dynamics of the EWPT and possible production of cosmological 
relics of a SFOEWPT. 
We leave this for future work~\cite{Braathen:2025svl}.


\section*{Acknowledgements}
\sloppy{
We thank Margarete M\"{u}hlleitner for communications and for implementing the 
RxSM in \texttt{HPAIR}. We are also grateful to 
Carlos~Pulido, 
Michael~Ramsey-Musolf
and 
Rui~Santos 
for interesting discussions. 
J.B. and A.V.S are supported by the DFG Emmy Noether Grant No.\ BR 6995/1-1. 
J.B.\ and A.V.S.\ acknowledge support by the Deutsche Forschungsgemeinschaft (DFG, German Research Foundation) under Germany's Excellence Strategy --- EXC 2121 ``Quantum Universe'' --- 390833306. This work has been partially funded by the Deutsche Forschungsgemeinschaft (DFG, German Research Foundation) --- 491245950. 
The work of S.H.\ has received financial support from the
grant PID2019-110058GB-C21 funded by
MCIN/AEI/10.13039/501100011033 and by ``ERDF A way of making Europe'', 
and in part by by the grant IFT Centro de Excelencia Severo Ochoa CEX2020-001007-S
funded by MCIN/AEI/10.13039/501100011033. 
S.H.\ also acknowledges support from Grant PID2022-142545NB-C21 funded by
MCIN/AEI/10.13039/501100011033/ FEDER, UE.
}


\appendix


\section{Summary of counterterms for the renormalisation of the RxSM}
\label{app:ren}

For convenience, we summarise in this appendix the set of expressions for counterterms in our OS renormalisation scheme for the RxSM, which were discussed and derived in \cref{sec:ren}.

Beginning with one-point functions, we have for the tadpole counterterms,
\begin{align}
    \delta^\text{CT}t_{\phi}&=\sin\alpha\;\delta^{(1)} t_H-\cos\alpha\;\delta^{(1)} t_h\,,\nn\\
    \delta^\text{CT}t_{S}&=-\cos\alpha\;\delta^{(1)} t_H-\sin\alpha\;\delta^{(1)} t_h\,.
\end{align}

Turning next to renormalisation conditions derived from two-point functions, the OS counterterms for the scalar masses and mixing angle are
\begin{align}
    \delta^{CT} m^2_{h}=&\ \rm{Re}[\Sigma_{hh}(m_h^2)], \nn\\ 
    \delta^{CT} m^2_{H}=&\ \rm{Re}[\Sigma_{HH}(m_H^2)]\,,\nn\\
    \delta^{\mathrm{ CT}} \alpha =&\ \frac{\rm{Re}[\Sigma_{hH}(m_h^2)+\Sigma_{hH}(m_H^2)]}{2(m_H^2-m_h^2)}\,.
\end{align}
Additionally, the scalar field renormalisation constants are
\begin{align}
    \delta^{\mathrm{CT}} Z_{hh}&=-\mathrm{Re}\left[\frac{\partial \Sigma_{hh}(p^2)}{\partial p^2}\right]_{p^2=m_h^2} \,,\nn\\
    \delta^{\mathrm{CT}} Z_{HH}&=-\mathrm{Re}\left[\frac{\partial \Sigma_{HH}(p^2)}{\partial p^2}\right]_{p^2=m_H^2} \,,\nn\\
    \delta^{\mathrm{CT}} Z_{hH}&=\frac{\mathrm{Re}[\Sigma_{hH}(m_H^2)]}{m_h^2-m_H^2} \,,\nn\\ 
    \delta^{\mathrm{CT}} Z_{Hh}&=\frac{\mathrm{Re}[\Sigma_{Hh}(m_h^2)] }{m_H^2-m_h^2}
\end{align}

The EW and singlet VEV counterterms are given by
\begin{align}
\frac{\delta^\text{CT} v}{v}&=\frac{1}{2}&\left[\frac{s_W^2-c_W^2}{s^2_W}\frac{\rm{Re}\big[\Sigma_{WW}^T(m_W^2)\big]}{m_W^2}+\frac{c_W^2}{s_W^2}\frac{\rm{Re}\big[\Sigma_{ZZ}^T(m_Z^2)\big]}{m_Z^2}-\Pi_{\gamma\gamma}(0)
-\frac{2s_W}{c_W}\frac{\Sigma_{\gamma Z}^T(0)}{m_Z^2}\right]\,,\nn\\
  \delta^{\rm CT} v_S&=0\,,
\end{align}
We note that the counterterm for $v_S$ is the only counterterm in our scheme that is technically not an OS counterterm. 

Finally, the OS counterterms for the Lagrangian trilinear couplings, $\kappa_S$ and $\kappa_{SH}$, which we determined from conditions on three-point functions, read
\begin{align}
  \delta^{\mathrm{CT}}\kappa_S&= \frac{\frac{\partial\lambda_{HHH}^{(0)}}{\partial\kappa_{SH}}(\delta^{(1)}_{\text{gen.}+\text{wfr}}\lambda_{hHH}+\sum_x\delta^\text{CT}_x\lambda_{hHH})-\frac{\partial\lambda_{hHH}^{(0)}}{\partial\kappa_{SH}}(\delta^{(1)}_{\text{gen.}+\text{wfr}}\lambda_{HHH}+\sum_x\delta^\text{CT}_x\lambda_{HHH})}{\frac{\partial\lambda_{hHH}^{(0)}}{\partial\kappa_{SH}}\frac{\partial\lambda_{HHH}^{(0)}}{\partial\kappa_{S}}-\frac{\partial\lambda_{hHH}^{(0)}}{\partial\kappa_{S}}\frac{\partial\lambda_{HHH}^{(0)}}{\partial\kappa_{SH}}}\,,\nn\\
  \delta^{\mathrm{CT}}\kappa_{SH}&= \frac{\frac{\partial\lambda_{hHH}^{(0)}}{\partial\kappa_{S}}(\delta^{(1)}_{\text{gen.}+\text{wfr}}\lambda_{HHH}+\sum_x\delta^\text{CT}_x\lambda_{HHH})-\frac{\partial\lambda_{HHH}^{(0)}}{\partial\kappa_{S}}(\delta^{(1)}_{\text{gen.}+\text{wfr}}\lambda_{hHH}+\sum_x\delta^\text{CT}_x\lambda_{hHH})}{\frac{\partial\lambda_{hHH}^{(0)}}{\partial\kappa_{SH}}\frac{\partial\lambda_{HHH}^{(0)}}{\partial\kappa_{S}}-\frac{\partial\lambda_{hHH}^{(0)}}{\partial\kappa_{S}}\frac{\partial\lambda_{HHH}^{(0)}}{\partial\kappa_{SH}}}\,.
\end{align}

\section{Definition of the significance of the resonant peak}
\label{app:peak}
\begin{figure}[!]
    \centering
    \includegraphics[width=0.7\linewidth]{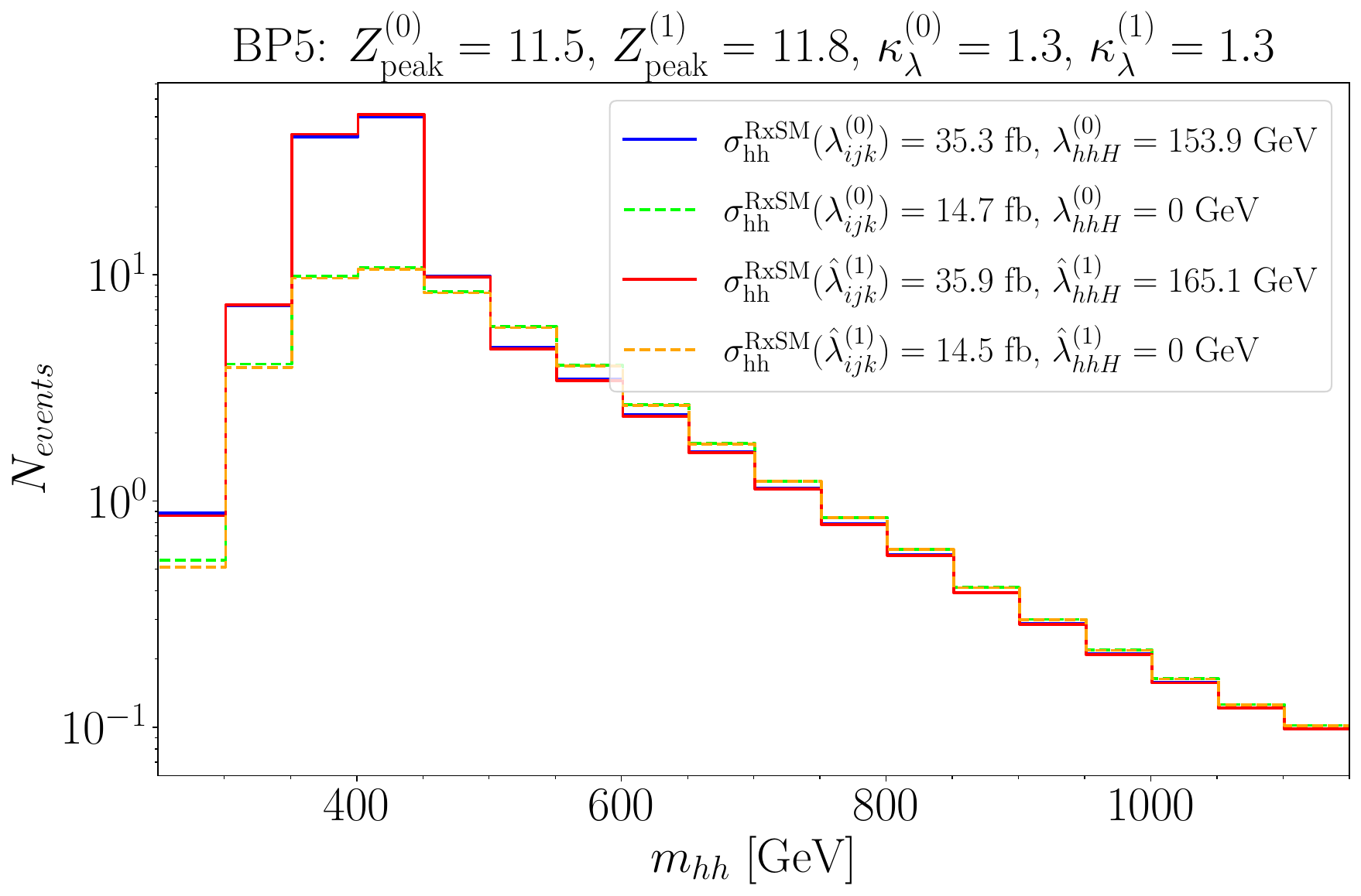}
    \includegraphics[width=0.7\linewidth]{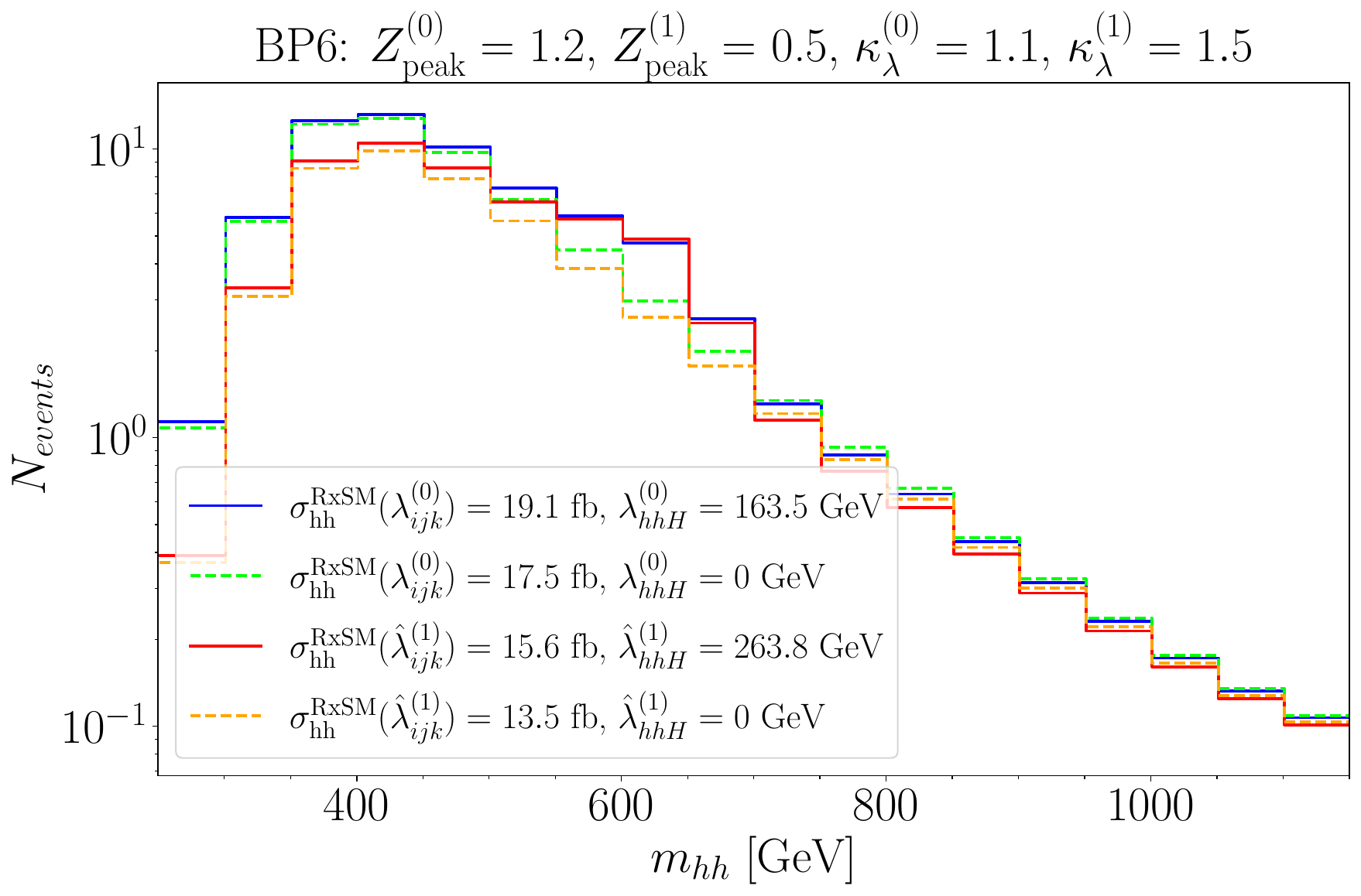}
    \caption{Distribution of the number of di-Higgs production events --- computed as in \cref{nevents}, taking into account the decay of the Higgs boson into $b\bar{b}$ as well as smearing and binning effects --- with respect to the invariant mass of the four reconstructed $b$ quarks $m_{b\bar{b}b\bar{b}}$. The solid curves show the result using the computed values of the trilinear scalar couplings and the dashed curves show the results of setting $\lahhH = 0$. The blue and green lines show the result using tree-level trilinear scalar couplings and the red and orange lines show the result using one-loop trilinear scalar couplings. \textit{Top:} results for BP5; \textit{bottom:} results for BP6.}
    \label{zpeak}
\end{figure}

In this appendix, we describe how we define the significance of 
the resonant peaks of BP5 and BP6 of \cref{bps} and \cref{bp456}. 
(An analogous definition is used for the $e^+e^-$ case.)
The definition of the significance is the same one as in \cref{z} but we modify the definition of the background $b_i$. 
As a signal we still use the number of events for the RxSM but for the background we use the result of the RxSM setting 
$\lahhH = 0$. In this way, we compare the resonant peak with a continuous distribution. The corresponding distributions are shown in \cref{zpeak}. 

\clearpage

\end{document}